\documentclass[sigconf]{acmart}
\usepackage[normalem]{ulem}
\usepackage{soul}
\usepackage{url}
\usepackage{xspace,balance,multirow}
\usepackage[ruled, vlined, linesnumbered]{algorithm2e}
\usepackage{xcolor}

\usepackage{colortbl}
\usepackage{bbold}
\usepackage{enumitem}
\usepackage[captionskip=-1pt]{subfig}
\usepackage{wrapfig}
\usepackage{flushend}
\usepackage{tikz}
\usepackage{pgfplots}
\usetikzlibrary{patterns}
\pgfplotsset{compat=1.16}
\usepackage{float}
\AtBeginDocument{%
  }

\setcopyright{acmlicensed}
\copyrightyear{2018}
\acmYear{2018}
\acmDOI{XXXXXXX.XXXXXXX}

\acmConference[Conference acronym 'XX]{Make sure to enter the correct
  conference title from your rights confirmation email}{June 03--05,
  2018}{Woodstock, NY}
\acmISBN{978-1-4503-XXXX-X/18/06}

\newcommand{\shorttitle}{}
\newcommand{\ie}{{i.e.,}\xspace}
\newcommand{\eg}{{e.g.,}\xspace}
\newcommand{\spara}[1]{\vspace{1mm}\noindent\textbf{#1.}}
\newcommand{\savespace}[1]{\ignorespaces}
\SetKw{KwAnd}{and}
\SetKw{KwOr}{or}
\SetKw{KwBreak}{break}
\SetKw{KwContinue}{continue}

\newcommand{\argmax}[1]{\underset{#1}{\operatorname{arg}\operatorname{max}}\;}
\newcommand{\ours}{\textsf{ThriftLLM}\xspace} % dynamic IM

\newcommand{\greedy}{\textsf{GreedyLLM}\xspace} 
\newcommand{\surgreedy}{\textsf{SurGreedyLLM}\xspace} 
 
\newcommand{\PA}{\xi\xspace} % dynamic IM
\newcommand{\roberta}{\textsf{RoBERTa}\xspace}
\newcommand{\ditto}{\textsf{Ditto}\xspace}

\newcommand{\problem}{{\sc Optimal Ensemble Selection}\xspace} % dynamic IM
\newcommand{\problemshort}{{\sc OES}\xspace}
\renewcommand{\P}{\mathcal{P}\xspace}
\newcommand{\A}{\mathcal{A}\xspace}

\newenvironment{squashitemize}{
  \begin{itemize}[noitemsep, topsep=0pt]
}{
  \end{itemize}
}

\newcommand{\T}{\mathbf{T}\xspace}

\newcommand{\x}{\mathbf{x}\xspace}

\renewcommand{\L}{\mathcal{L}\xspace}

\newcommand{\C}{\mathcal{C}\xspace}

\newcommand{\R}{\mathbb{R}\xspace}
\newcommand{\N}{\mathbb{N}\xspace}
\newcommand{\mathcalT}{\mathcal{T}\xspace}
\renewcommand{\S}{\mathcal{S}\xspace}

\newcommand{\Q}{\mathcal{Q}\xspace}

\newcommand{\e}{{\ensuremath{\mathrm{e}}}}

\newtheorem{definition}{Definition}%[section] XX
\newtheorem{theorem}{Theorem}%[section] XX
\newtheorem{lemma}{Lemma}%[section] XX
\newenvironment{proofsketch}{%
  \renewcommand{\proofname}{ Proof Sketch}\proof}{\endproof}

\newtheorem{fact}{Fact}%[section]

\newcommand{\sktch}[1]{#1}
\newcommand{\arxiv}[1]{}

\newcommand{\eat}[1]{}
\newcommand{\cover}[1]{}
\newcommand{\ravi}[1]{\textcolor{black}{#1}}

\newcommand{\hkk}[1]{}
\newcommand{\td}[1]{{\color{cyan}To do:}}

\newenvironment{customlegend}[1][]{%
    \begingroup
    \csname pgfplots@init@cleared@structures\endcsname
    \pgfplotsset{#1}%
}{%
    \csname pgfplots@createlegend\endcsname
    \endgroup
}%

\def\addlegendimage{\csname pgfplots@addlegendimage\endcsname}

\pgfplotsset{every tick label/.append style={font=\tiny}}
\definecolor{myblue}{RGB}{113,210,242}
\definecolor{myorange}{RGB}{249,205,173}
\definecolor{myred}{RGB}{183,0,162}
\definecolor{mycolor}{RGB}{0,0,0}
\def\st{0.85}
\def\gap{0.06}

\begin{document}

\settopmatter{printfolios=true}
%\newpage

\title{ThriftLLM: On Cost-Effective Selection of Large Language Models for Classification Queries}

\author{Keke Huang}
\email{kk.huang@ubc.ca}
\affiliation{%
  \institution{University of British Columbia}
  \country{}
}

\author{Yimin Shi}
\email{yiminshi@u.nus.edu}
\affiliation{%
  \institution{National University of Singapore}
  \country{}
}

\author{Dujian Ding}
\email{dujian.ding@gmail.com}
\affiliation{%
  \institution{University of British Columbia}
%  \institution{CNRS@CREATE, Singapore}
  \country{}
}

\author{Yifei Li}
\email{yfli@student.ubc.ca}
\affiliation{%
  \institution{University of British Columbia}
%  \institution{CNRS@CREATE, Singapore}
  \country{}
}

\author{Yang Fei}
\email{yfei11@u.nus.edu}
\affiliation{%
  \institution{National University of Singapore}
  \country{}
}

\author{Laks Lakshmanan}
\email{laks@cs.ubc.ca}
\affiliation{%
  \institution{University of British Columbia}
%  \institution{CNRS@CREATE, Singapore}
  \country{}
}

\author{Xiaokui Xiao}
\email{xkxiao@nus.edu.sg}
\affiliation{%
  \institution{National University of Singapore}
  \institution{CNRS@CREATE, Singapore}
  \country{}
}
%}

\begin{abstract}
Recently, large language models (LLMs) have demonstrated remarkable capabilities in understanding and generating natural language content, attracting widespread attention in both industry and academia. An increasing number of services  offer LLMs for various tasks via APIs. Different LLMs demonstrate expertise in different domains of queries (\eg text classification queries). Meanwhile, LLMs of different scales, complexities, and performance are priced diversely. Driven by this, several researchers are investigating  strategies for selecting an ensemble of LLMs, aiming to decrease overall usage costs while enhancing performance. However, to our best knowledge, none of the existing works addresses the problem, how to find an LLM ensemble subject to a cost budget, which maximizes the ensemble performance with guarantees.

In this paper, we formalize the performance of an ensemble of models (LLMs) using the notion of {correctness probability}, which we formally define. We develop an approach for aggregating responses from multiple LLMs to enhance ensemble performance. Building on this, we formulate the \problem (\problemshort) problem of selecting a set of LLMs subject to a cost budget that maximizes the overall correctness probability.  We show that the correctness probability function is non-decreasing and non-submodular {and provide evidence that the \problemshort problem is likely to be NP-hard.} By leveraging a submodular function that upper bounds {correctness probability},  we develop an algorithm, \ours, and prove that it achieves an instance-dependent approximation guarantee with high probability. Our framework functions as a data processing system that selects appropriate LLM operators to deliver high-quality results under budget constraints. It achieves state-of-the-art performance for text classification and entity matching queries on multiple real-world datasets against various baselines in our extensive experimental evaluation, while using a relatively lower cost budget, strongly supporting the effectiveness and superiority of our method. The source code and data have been made available at
\url{https://github.com/kkhuang81/thriftLLM}.
\end{abstract}

\maketitle

\begin{sloppy}
\section{Introduction}\label{sec:intro}

The evolution of large language models (LLMs) has significantly transformed the field of natural language processing. These models have enabled AI systems to generate human-like texts based on input instructions with remarkable accuracy. An increasing number of companies (e.g., OpenAI, Anthropic, and Google) have introduced a wide range of services powered by LLMs, such as GPT-4~\cite{GPT4},  Claude-3~\cite{claude32024}, and Gemini~\cite{Geminifamily}. These services, accessible {via} APIs, have achieved notable success across various applications, such as text classification, question answering, summarization, and natural language inference~\cite{SunL0WGZ023,LeiLHWYCK2023,SuXWXKLF19,ZhangLDLMH24}. In data management, various LLMs have been adopted for text-to-SQL generation~\cite{ZhangMFMG0LL24, GaoWLSQDZ24,FanGZZCCLMDT24}, query performance\footnote{We use the generic term ``performance'' to refer to any metrics used in the literature to gauge the performance of models {such as accuracy, BLEU score, etc.}}  
optimization~\cite{Trummer24,ZhuCXLSZSTL24,ZhaoWKGKC24}, schema matching \cite{parciak2024schemamatchinglargelanguage}, and entity matching \cite{10.1007/978-3-031-70421-5_6, PeetersSB25, Ditto20}. {In those applications, we can regard each LLM as a standalone operator that takes unstructured data as input and computes derived attributes (\eg sentiment, stance). From this perspective, an LLM ensemble acts as a data processing system to deliver high-quality results under budget constraints. More broadly, this strategy can be seen as extending the capabilities of traditional DBMSs by leveraging the power of LLM operators over unstructured data. This is particularly relevant for modern data management workloads, where unstructured data, \eg text, images, increasingly plays a central role.}

While LLMs offer significant improvements in performance, they come with substantial costs, particularly for high-throughput applications. 
For example, GPT-4 charges $\$30$ per million input tokens  and $\$60$ per million output tokens~\cite{GPT4Price}, while Gemini-1.5~\cite{GeminiPrice} costs $\$7$ and $\$21$ per million input and output tokens, respectively. Typically, models with higher performance charge a higher price. However, it is well recognized that expensive models with larger {number of} parameters do not necessarily dominate the cheaper models %\st{overwhelmingly in a wide range of} 
{across all} applications~\cite{ChenZZ2023,ChenLi2024v4,XiaK0GRK024,ShekharDubey2024,SunHSXYDTC24}. Different models may excel in different domains, 
and it has been observed that smaller LLMs can surpass larger LLMs on specific tasks~\cite{ChenZZ2023,FuLaskarChenTN2024,ding2024hybrid}.

This phenomenon raises an important question: \textit{can we leverage a set of cost-effective {LLMs}, within a specified budget, to achieve performance comparable to that of a more expensive LLM?} The literature offers two main strategies to address this: {\em LLM ensemble}~\cite{ChenZZ2023, Jiang0L23,ShekharDubey2024} and {\em tiny LM}~\cite{HillierGuertlerTan2024, chen2024role,TangLiuNiTian2024}. The ensemble approach aims to optimize performance by combining outputs from a carefully selected set of LLMs. In contrast, the tiny LM approach applies advanced techniques to reduce the parameter counts of models without significantly compromising the performance. Specifically, FrugalGPT~\cite{ChenZZ2023}, {a recent LLM ensemble work,} derives an LLM cascade from a ground set optimized for given queries under a cost budget constraint. 
As a {representative from the tiny LM family,} Octopus v2~\cite{ChenLi2024} fine-tunes the small model Gemma-2B to exploit fine-tuned functional tokens for function calling, achieving comparable performance with GPT-4 in accuracy and latency. Among the frameworks for LLM usage cost reduction and LLM performance enhancement, FrugalGPT is the most pertinent to our research. However, when FrugalGPT applies the derived LLM cascade to queries, it adopts the single output from the \textit{last executed} model in the sequence rather than taking advantage of {all generated} responses from {the model cascade} to produce an optimal solution. {Moreover,} the budget constraint is applied to queries in an \textit{expected} sense, allowing  
{the incurred costs on individual queries to exceed the budget in practice, a phenomenon confirmed by our experiments.}
In addition, {FrugalGPT} does not provide any performance guarantee for the {selected LLM} ensemble. 

\begin{figure}[!t]
\centering
\includegraphics[width=\linewidth]{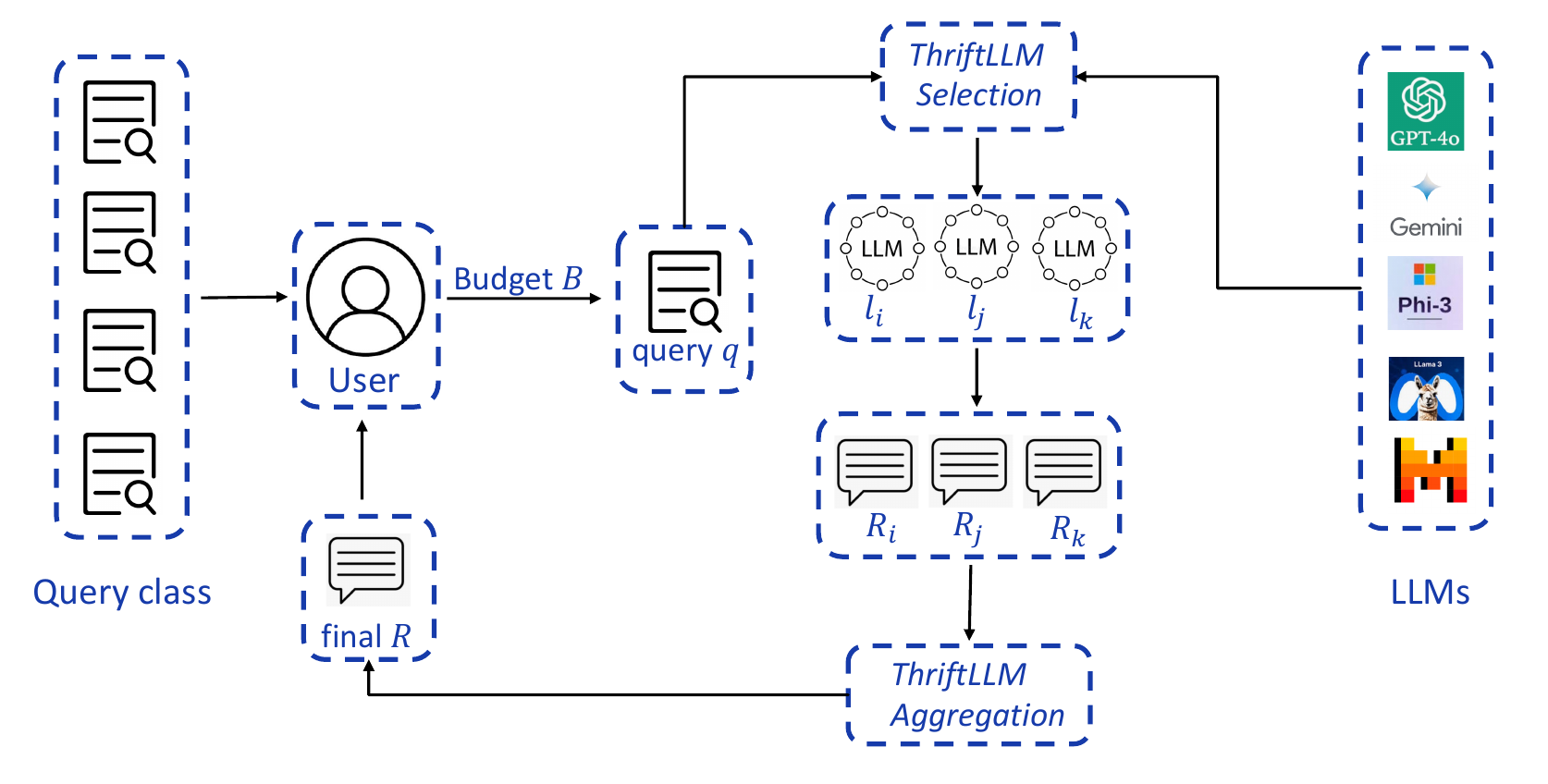}\vspace{-2mm}
\caption{Overview of \ours: $R_i, R_j, R$ denote responses.}
\label{fig:thriftllm}
\end{figure}

Inspired by these observations, we aim to {combine} the responses generated by a collection of LLMs in a non-trivial manner to deliver high-quality output, with a focus on classification queries. To this end, we devise a novel aggregation approach and quantify the quality of the aggregated response via a new metric of {\textit{correctness probability}} by leveraging likelihood maximization.
Building on this, we formalize an optimization problem, dubbed \textit{\problem} (\problemshort for short): given a set of LLMs, a specific budget, and a query, the objective is to identify a subset of LLMs whose total cost is within the budget while the aggregated {correctness probability} on the query is maximized. To address this challenge, we propose an adaptive LLM selection algorithm \ours, designed specifically for this budget-constrained LLM selection scenario. As illustrated in Figure~\ref{fig:thriftllm}, when receiving a query and budget from a user, \ours selects an appropriate subset of LLMs from the LLM candidates without exceeding the budget. These LLMs independently process the query, and their responses are subsequently aggregated by \ours to produce a final answer {of high quality}, which is then returned to the user. {Take a traditional data management task—entity matching—as an example. A global consulting firm may collect product information from different regions in varied formats and descriptions for market analysis. For example, Product 1: \textit{Samsung Galaxy S21 Ultra, Phantom Silver, Factory Unlocked} and Product 2: \textit{Samsung smartphone, 256GB storage, high-end camera, silver color}. They both refer to the same high-end Samsung Galaxy S21. To ensure reliable insights, it must identify records referring to the same real-world product across sources. In our framework, each LLM serves as an operator that takes a product pair and predicts whether they refer to the same entity. These LLM operators vary in capability. Given a fixed budget, \ours selects an optimal subset of them to form an ensemble that ensures high-quality matching.} %\lvsl{i think you should add a real example similar to the samsung one in our response to the second reviewer, D3. however, pls see my comment there too.}\kk{updated!}

{We conduct an in-depth theoretical analysis of the {correctness probability} function in the \problemshort problem and establish that it is non-decreasing and non-submodular. We also provide evidence of the hardness of the problem.}  {Nevertheless, we leverage a surrogate objective function that upper bounds the {correctness probability}, show that it is non-decreasing and submodular, and devise an algorithm for \problemshort that achieves an instance-dependent approximation guarantee with high probability, when LLM success probabilities are known. We show how our algorithms and approximation guarantees can be extended to the case when the success probabilities are unknown and are estimated within confidence intervals. We compare \ours with $3$ baselines on $5$ real-world datasets on various text classification queries across different domains as well as with $2$ SOTA baselines across $5$ real-world datasets on entity matching queries. The experimental results strongly demonstrate the superior performance of \ours in terms of accuracy respecting cost budgets and delivering high-quality results under given budgets.} 

{In sum, we make the following contributions. 
\begin{squashitemize}
\item We propose a new aggregation scheme for combining individual LLM responses and formalize the ensemble prediction quality using a novel notion of {\textit{correctness probability}} (Sections~\ref{sec:pre} and~\ref{sec:aggre}). We show that {correctness probability} is non-decreasing but non-submodular and offer evidence of {hardness} of \problem (Section~\ref{sec:PA}). %Furthermore, we formalize the {\em \problem} problem and reveal its NP-hard complexity.
\item We develop a greedy algorithm, \greedy, and show that it has an unbounded approximation factor. We then develop the \ours algorithm by leveraging a submodular upper bound function as a surrogate objective, and show that {it offers an instance dependent approximation to the optimum with high probability (%Section~\ref{sec:appro}
Sections~\ref{sec:surgreedy}-\ref{sec:adaptiveselection}). When the success probabilities are unknown and are estimated within confidence intervals, our data-dependent approximation guarantees continue to hold (Section~\ref{sec:interval}).}
\item We conduct extensive experiments on text classification tasks over $5$ real-world datasets across diverse domains against $5$ baselines and on entity matching over $5$ real-world datasets against $4$ baselines. Experiments show that \ours achieves comparable or superior performance on the tested datasets while respecting the %minimum 
budget constraints (Section~\ref{sec:exp}).
\end{squashitemize}
}
Proofs of most of our technical results can be found in Appendix~\ref{app:proofs}. \arxiv{For brevity, we sketch some of the proofs. 
The full details can be found in \cite{thriftLLM-full-2025}.} 

\sktch{
\begin{table}
    \centering
    \caption{Frequently used notations}\label{tbl:notations}
    %\vspace{-3mm}
    \setlength{\tabcolsep}{0.2em} % for the horizontal padding
    \renewcommand{\arraystretch}{1}% for the vertical padding
    \renewcommand{\aboverulesep}{0pt}
    \renewcommand{\belowrulesep}{0pt}
%	\begin{tabular}{|p{1.6cm}|p{6cm}|}\hline%{|p{1.5cm}|p{0.75cm}|p{0.75cm}|p{1.5cm}|p{2cm}|}
	\begin{tabular}{@{}c|m{5.6cm}@{}}\toprule %{@{}c|m{6.5cm}@{}}
 %{|p{1.5cm}|p{0.75cm}|p{0.75cm}|p{1.5cm}|p{2cm}|}
        \textbf{Notation} & \multicolumn{1}{c}{\textbf{Description}} \\ \midrule
        $\L, L, \S$ & the set of LLMs, the set size, and a subset of $\L$, \ie $|\L|=L$ and $\S\subseteq \L$ \\ \midrule        
        $\Q,q$ & the query class and one random query  \\ \midrule
        $R(l)$ & the response of model $l\in \L$ \\ \midrule
        $\C = \{C_1, \ldots, C_K\}$, $C_q$ & the set of classes and the ground-truth class of query $q$ with $C_q\in \C$ \\ \midrule
        $\P=\{p_1,p_2, \cdots, p_L\}$ & the set of success probabilities of LLMs on a query class \\ \midrule
        $B$, $b$ & the total budget and a model cost on a query\\ \midrule
        $\Omega, \phi$ & the observation space of {possible} responses and a specific observation \\ \midrule
        $\PA(\cdot)$ & the function of {correctness probability}  \\ \midrule
        $f(C_k \mid \phi)$ & the likelihood function of $C_k$ being the ground truth given observation $\phi$ \\ \midrule
        $h(C_k\mid \phi)$ & the belief function of the predicted class $C_k$ given observation $\phi$ \\ \bottomrule
    \end{tabular}
    %\vspace{-1mm}
\end{table}
}

\section{Preliminaries}\label{sec:pre}

We denote matrices, vectors, and sets with bold uppercase letters (\eg $\T$), bold lowercase letters (\eg $\x$), and calligraphic letters (\eg $\S$), respectively. The $i$-th row (resp.\ column) of matrix $\T$ is denoted $\T[i,\cdot]$ (resp.\ $\T[\cdot, i]$). We use $[n]$ to denote the set $\{1,2,\cdots, n\}$. %Frequently used notations are summarized in Table~\ref{tbl:notations}.

We refer to textual tasks submitted to LLMs as {\em queries}. Queries typically contain contexts, called {\em prompts},  preceding the actual questions. In general, we assume queries include necessary prompts. Let $\Q$ be a query class representing a specific category of queries in the real world and $\L=\{l_1,l_2,\cdots,l_L\}$ be a set of distinct LLMs ($L>0$). Given a random query $q\in \Q$ (with its corresponding prompt), the query processing cost of a model consists of two components: the input and output costs, which are directly determined by the {number} of input and output tokens, respectively. We let $b_i(q)\in \R_+$ denote the total {incurred} cost of model $l_i$ for {processing} query $q$. 
When the model $l_i$ is deterministic{, its cost $b_i(q)$ is solely determined by the query $q$}. 
For simplicity, we refer to this cost as $b_i$ when the query is clear from the context. The performance of the models from  $\L$ on the query class $\Q$ is represented by the set of {\em success probabilities} $\P=\{p_1,p_2, \cdots, p_L\}$, where each $p_i$ denotes the probability that the model $l_i$ generates a correct response to a query selected uniformly at random from $\Q$.

Specifically, for a classification\footnote{In this study, we focus on classification tasks due to space constraints. We aim to extend our methodology to regression tasks in our future work.} 
query $q\in \Q$, let $\C = \{C_1, \ldots, C_K\}$ denote the set of $K>0$ possible classes. When model  $l_i\in \L$ is applied to query $q$, it returns a prediction response denoted $R(l_i)\in \C$. For a subset $\S \subseteq \L$ of LLMs, their prediction responses yield an {\em observation} $\phi_\S=(R(l) \mid  l\in \S)$, a prediction sequence of the LLMs from $\S$. The set of all possible observations for $\S$ on the query class $\Q$ is termed as {\em observation space $\Omega_\S$}%\footnote{We sometimes drop the subscript in $\Omega_\S$ when $\S$ is clear from the context.}. 
We denote the prediction derived from observation $\phi_\S \in \Omega_\S$ as $C(\phi_\S)$. The derivation procedure is elaborated in Section~\ref{sec:aggregation}.

Given a set of {LLMs} $\S$ and considering a random query $q$ from $\Q$ with an associated ground-truth class $C_q$, the probability of observing $\phi_\S$, denoted as $\Pr[\phi_\S]$, is computed as follows. Let $\S^T = \{l \in \S \mid R(l) = C_q\}$ (resp. $\S^F = \{l \in \S \mid R(l) \neq  C_q\}$) be the subset of models whose prediction agrees (resp. disagrees) with $C_q$. Then $\Pr[\phi_\S]$ is 
\begin{equation}\label{eqn:observeprob}
\textstyle \Pr[\phi_\S] = \prod_{l_i\in \S^T}p_i\prod_{l_j \in \S^F}\tfrac{1-p_j}{K-1}.  
\end{equation}
Notice that a model making a wrong prediction could predict any one of the rest $K-1$ wrong classes. {For example, given a query class $\Q$, let the LLM set $\S=\{l_1, l_2, l_3\}$ with corresponding success probabilities $\P=\{0.9, 0.8, 0.8\}$ and the class set $\C = \{C_1, C_2, C_3\}$. Figure~\ref{fig:observation} demonstrates the observation space $\Omega_\S$. Suppose we sample a random query $q$ from $\Q$ uniformly. When the ground-truth class $C_q=C_1$, the probability of observation $\phi_3=(C_1,C_1,C_3)$ is $\Pr[\phi_3]=0.9\times 0.8\times \tfrac{1-0.8}{2}=0.072$. Similarly, the probability is updated to $\Pr[\phi_3]=0.0005$ or $\Pr[\phi_3]=0.004$ if $C_q=C_2$ or $C_q=C_3$, respectively. %Therefore, according to our response aggregation (details in Section~\ref{sec:aggregation}), we have $C(\phi_3)=C_1$. If  $C_1$ is the ground truth class, we have $\phi_3 \in \Omega^T_{\S}$ and the probability of  $\Pr[\phi_3]=0.072$ is added into $\PA_\P(\S)$; otherwise, $\phi_3$ is skipped. \lvsl{is it really necessary to refer to Section~\ref{sec:aggregation} at this stage? notice that $\PA_\P(\S)$ is not yet defined. also, even if the ground truth class is $C_3$, $\Pr[\phi_3] = 0.004$ will be added into $\PA_\P(\S)$. it's only when the ground truth class is $C_2$, $\phi_3$ is skipped. agree?}
By following this procedure, the probability $\PA_\P(\S)$ is %integrated 
{aggregated} over the entire observation space $\Omega_{\S}$.
}

\begin{figure}[!t]
\centering
\includegraphics[width=0.8\linewidth]{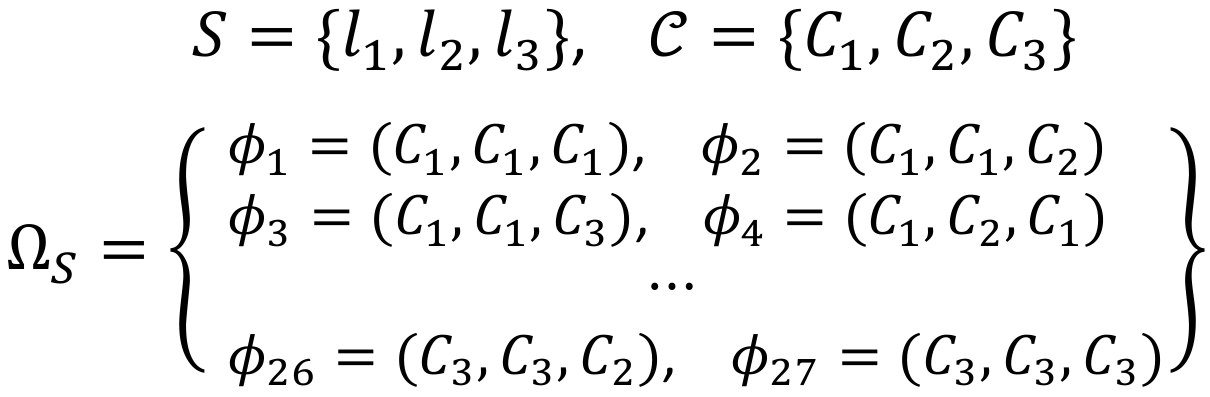}
\caption{{Example of an observation space $\Omega_\S$.}}
\label{fig:observation}
\end{figure}

By leveraging the realization probability $\Pr[\phi_\S]$,  the fundamental notion of {\em {correctness probability}} $\PA(\S)$ of $\S$, {i.e., the probability that $\S$ makes the correct prediction on a random query from  query class $\Q$,} is formalized as follows. 

\begin{definition}[{Correctness Probability}]\label{def:CP}
Given a random query $q$ sampled uniformly from class $\Q$ and a subset $\S$ of LLMs, let $C_q$ be the ground-truth response. Let $\Omega^T_\S \subset \Omega_\S$ be the subset of observations with $C(\phi_\S)=C_q$. The {correctness probability} on query class $\Q$ is {$\PA_\P(\S)=\sum_{\phi_\S\in \Omega^T_\S}\Pr[\phi_\S]$.} 
\end{definition} 

{When the success probabilities $\P$ are clear from the context, we will drop the subscript and denote {correctness probability} as $\PA(.)$. We next formally state the problem we study in the paper.}  
\begin{definition}[\problem]\label{def:problem}
Consider a query class $\Q$, a set $\L$ of LLMs, and a cost budget $B\in \R_+$. The \problem (\problemshort) problem is to find a subset $\S^\circ\subseteq \L$ {whose total cost} is under $B$, such that the {correctness probability} $\PA(\S^\circ)$ on $\Q$ is maximized, \ie 
\begin{equation}\label{eqn:objective}
\textstyle \S^\circ=\argmax{\S:\S\subseteq \L, c(\S)\le B}\PA(\S),    
\end{equation}
where {$c(\S) := \sum_{l_i\in\S} b_i$ is the total cost of LLMs in $\S$.}
\end{definition}

\section{Probability Estimation and Response Aggregation}\label{sec:aggre}

{A key ingredient in our optimization problem is the success probability of LLMs.} In this section, we first discuss our approach for estimating the success probability of an LLM  from historical data. We then elaborate on the methodology for deriving predictions based on observations. For ease of exposition, we assume the responses of different LLMs are mutually independent, and our analysis primarily concentrates on classification tasks.

\subsection{Estimation of Success Probability}\label{sec:estimation}

We assume that queries from the same query class $\Q$ exhibit semantic similarity. The success probability of a model $l$ on query class $\Q$ is the probability that $l$ returns the correct response for a query $q$ randomly sampled from $\Q$. This probability is crucial in addressing the \problemshort problem. However, success probabilities of LLMs are not known {\textit{a priori}} but can be estimated from historical data. 

Specifically, consider an input table $\T$ that records the historical performance of the $L$ LLMs on $N$ queries in a matrix format with $L, N\in \N_+$. The entries of $\T$ vary based on the type of query $q$. For classification tasks,  $\T$ contains boolean entries, i.e.,  $\T\in \{0,1\}^{N\times L}$. Conversely, for generation or regression tasks, it contains real values in the interval $[0,1]$, i.e., $\T\in [0,1]^{N\times L}$, with $\T_{\ell,k}$ indicating the quality or accuracy score of the response from the $\ell$-th model on the $k$-th query. In this paper, we focus on classification queries.  

To accurately estimate the success probability for each query class, we first cluster the queries in $\T$ into distinct groups based on their semantic similarity. To this end, we convert all queries into embeddings by leveraging the embedding API~\cite{OpenAIEmbedding}
%\footnote{\url{https://api.openai.com/v1/embeddings}} 
provided by OpenAI. Subsequently, we employ the DBSCAN~\cite{GanT15} algorithm to cluster the embeddings. The success probability $p_l$ of the $l$-th model on one resultant cluster $\Q_k$ is estimated as $p_l=\tfrac{1}{|\Q_k|}\sum_{q_i\in \Q_k}\T[i,l]$.

\subsection{Response Aggregation}\label{sec:aggregation}

Given a set $\S$ of LLMs and an observation $\phi_\S=(R(l_1),R(l_2),\cdots,R(l_{|\S|}))$, the aggregated prediction $C(\phi_\S)$ of $\phi_\S$ is derived by combining all responses in observation $\phi_\S$. Since the ground-truth $C_q$ of query $q$ is unknown, we take the class {with maximum likelihood} as the aggregated prediction $C(\phi_\S)$. Specifically, each response in observation $\phi_\S$ corresponds to a class from $\C=\{C_1, C_2, \cdots, C_K\}$,  which   includes the (unknown) ground truth $C_q$. We consider each of the $K$ classes in turn as the ground truth and compute the likelihood of observing $\phi_\S$. The class with the highest likelihood is selected as the prediction $C(\phi_\S)$.

Initially, the set $\S$ is partitioned into disjoint subsets as $\S=\{\S(C_1),\S(C_2),\cdots,\S(C_K)\}$ where $\S(C_k)=\{l:R(l)=C_k,l\in \S\}$ represents the subset of models that predicted class $C_k$. The likelihood function $f(C_q=C_k \mid \phi_\S)$, which measures the probability of $C_k$ being the ground truth, is defined as 
\begin{equation}\label{eqn:likelihood}
\textstyle f(C_q=C_k \mid \phi_\S)=\prod_{l_i\in \S(C_k)}p_i\prod_{l_j \in \S\setminus\S(C_k)}\tfrac{1-p_j}{K-1}
\end{equation}
for $k\in [K]$. Based on this likelihood, we can derive the prediction $C(\phi_\S)$ from observation $\phi_\S$ as $C(\phi_\S)=\arg\max_{C_k\in \C}f(C_q=C_k \mid \phi_\S)$. If there are multiple classes with the same maximal likelihood, we break the tie randomly.

Computing $f(C_q=C_k \mid \phi_\S)$ for all $k\in[K]$ is expensive. However, by identifying redundant calculations, we can speed it up. Specifically, computing $f(.)$  involves a substantial amount of repeated calculations, as $\prod_{l_j \in \S(C_k)}\tfrac{1-p_j}{K-1}$ for $k\in[K]$ is repeated $K-1$ times, leading to unnecessary overheads. We can {choose}  $C(\phi_\S)$ as the class with the highest likelihood by ranking the $K$ likelihoods $f(C_q=C_k \mid \phi_\S)$ \textit{without calculating their exact values}.

Specifically, it holds that
\begin{align*}    
f(C_q=C_k \mid \phi_\S)  
= &\textstyle \prod_{l_i\in \S(C_k)}p_i\prod_{l_j \in \S\setminus\S(C_k)}\tfrac{1-p_j}{K-1} \\
= &\textstyle \prod_{l_i\in \S(C_k)}\tfrac{p_i(K-1)}{1-p_i}\prod_{l_j \in \S}\tfrac{1-p_j}{K-1}.
\end{align*}
The factor $\prod_{l_j \in \S}\tfrac{1-p_j}{K-1}$ is independent of $k$ because it is shared across all likelihood functions $f(C_q=C_k \mid \phi_\S), \forall k\in[K]$. Consequently, it is the term $\prod_{l_i\in \S(C_k)}\tfrac{p_i(K-1)}{1-p_i}$ that {determines} the prediction of a given observation. For clarity, we define 
\begin{equation}\label{eqn:belief}
h(C_k\mid \phi_\S)=\textstyle\prod_{l_i\in \S(C_k)}\tfrac{p_i(K-1)}{1-p_i}    
\end{equation}
as the {\em belief} in $C_k$ being the ground truth,  conditioned on the observation $\phi_\S$ for $k\in [K]$. We use this belief in place of the likelihood when deriving the prediction from an observation. Without loss of generality, we set $h(C_k\mid \phi_\S)=\tfrac{p_{\min}}{2(1-p_{\min})}$ heuristically if $\S(C_k)=\emptyset$ where $p_{\min}=\min\{p_1,p_2,\cdots,p_L\}$. Let $H_k(\phi_\S)$ be the $k$-th highest belief value for $k\in[K]$ among all classes. This leads to the following easily proved fact.

{
\begin{fact}\label{fact:belief}
Given an observation $\phi_\S$, the class $C_k$ corresponding to $H_1(\phi_\S)$ is the prediction derived from $\phi_\S$. In formal terms,   $C(\phi_\S) = \arg\max_{C_k\in\C} h(C_k\mid \phi_\S)$.     
\end{fact}
}

Therefore, when deriving the prediction from observations, it is sufficient to examine the belief value of the subset $\S(C_k)$ rather than evaluating the likelihood across the entire set $\S$. One complication is that the ground truth $C_q$ is unknown. Our next result shows that we do not need to know the ground truth class $C_q$ to calculate  $\PA(\S)$.
\begin{proposition}\label{prob:accuracy}
The {correctness probability} $\PA(\S)$ is independent of the ground-truth class $C_q$ of the random query $q$.
\end{proposition}
The rationale is that {correctness probability} $\PA(\S)$ is determined by the {set of success probabilities} $\P$ on query class $\Q$. As long as $\P$ is fixed, varying $C_q$ of a random query $q \in \Q$ does not affect the overall {correctness probability} $\PA(\S)$ on $\Q$, i.e., it does not matter what the actual ground truth class is! {The intuition lies in the fact that observations are symmetrically distributed with respect to the underlying ground-truth labels.} Hence, we can assume any class in $\C$ to be the ground truth, without affecting the calculation of $\PA(\S)$.

\section{Adaptive LLM Selection}\label{sec:adaptive}

\sktch{In this section, we first study the properties of the {correctness probability} function $\PA(\cdot)$. Subsequently, we examine the hardness of the \problemshort problem. Following this, we propose a novel {\em surrogate greedy} strategy for solving the problem.}

\subsection{{Correctness Probability} and Problem Complexity}\label{sec:PA}

The {correctness probability} function $\PA(\S)$ (see Definition~\ref{def:CP}) determines the {probability of correctness} of a given set of LLMs $\S$ on a query class $\Q$, {\ie} the probability of obtaining the correct aggregated prediction  from any observation when applying $\S$ on a random query $q\in \Q$. The ultimate objective of \problemshort is to maximize this {probability} by identifying a subset of LLMs. To aid our analysis, we first analyze the properties of function $\PA(\cdot)$.

Let $\P = \{p_1, ..., p_L\}$ and $\P' = \{p'_1, ..., p'_L\}$ be two sets of success probabilities of the models in $\L$. We write $\P \preceq \P'$ iff $p_i \leq p'_i, i\in[L]$. We have the following lemmas.
\begin{lemma}\label{lem:nondecreasing}
The {correctness probability} function $\PA(\cdot)$ is non-decreasing. Specifically, (i) for any subset of models $\S \subseteq \L$ and success probability sets $\P, \P'$ such that $\P\preceq \P'$, we have $\PA_\P(\S) \leq \PA_{\P'}(\S)$; (ii) for any sets of models $\S \subset \S' \subseteq \L$, and any success probability set $\P$, $\PA_\P(\S) \leq \PA_\P(\S')$. 
\end{lemma}

{Unfortunately, our next result is negative:}  

\begin{lemma}\label{lem:nonsubmodular}
The {correctness probability} function $\PA(\cdot)$ is non-submodular.    
\end{lemma}
{Given this, we cannot directly find an (even approximately) optimal solution for the \problem problem.} Before we proceed, we establish the following proposition.
\begin{proposition}\label{pro:2models}
For a set $\S=\{l_1, l_2\}$ consisting of two LLMs, we have $\PA(\S)=\max\{p_1,p_2\}$ where $p_1$ and $p_2$ are the success probabilities of $l_1$ and $l_2$ respectively.
\end{proposition}  
The intuition is that when only two models are employed and they give two distinct predictions, the one with the higher success probability results in a higher belief value, as Equation~\eqref{eqn:belief} indicates, dominating the weaker one. Hence, the {correctness probability} is equal to the higher success probability between the two. {Our proof of Lemma~\ref{lem:nonsubmodular} (see Appendix~\ref{app:proofs}) builds on this idea.}

\spara{Hardness of \problem}
We offer evidence of the hardness of the \problemshort problem. We can rephrase the \problemshort problem as, ``select for each query, a subset of LLMs (items) so as to maximize the {sum of success probabilities} (value) while adhering to a pre-defined cost budget (weight limit)'', which is a variant of the 0-1 Knapsack problem~\cite{martello1987algorithms}, a well-known NP-hard problem. It is also worth noting that, for a given subset of LLMs, its {correctness probability} sums over exponentially many possible observations, which is clearly harder to compute than the ``total value'' in the 0-1 Knapsack problem. This suggests that the \problemshort problem should be at least as hard as the 0-1 Knapsack problem. While proving a formal reduction is challenging due to the complex calculation of $\PA(.)$, the above argument offers some evidence that \problemshort is likely to be computationally intractable.

\subsection{Our Surrogate Greedy Strategy}\label{sec:surgreedy}

As proved above, \problem is essentially a {\em budgeted non-submodular maximization} problem, which is substantially challenging. In the literature~\cite{BuchbinderFNS12, BianB0T17, Huang0XSL20, ShiL24}, the {\em greedy strategy} is recognized as a canonical approach for addressing  combinatorial optimization problems involving  submodular and non-submodular set functions. In the sequel, we first present the {\em vanilla greedy} strategy and demonstrate its inability to  solve  budgeted non-submodular maximization with  theoretical guarantees. To remedy this deficiency, we propose a novel {\em surrogate greedy}, which can provide a data-dependent approximation guarantee.

\spara{Vanilla Greedy} Algorithm~\ref{alg:greedy}, dubbed \greedy, presents the pseudo-code of the greedy strategy applied to the \problemshort problem. In general, \greedy selects models from the ground set $\L$ that achieve the highest ratio of marginal {correctness} gain to the associated cost in each iteration (see Line~\ref{line:marginalratio}). {When there is a tie, i.e., $\S^\prime$ contains more than one model with the same maximum ratio, the tie is broken in Line~\ref{line:tie-break} by choosing the model $l_\ast\in S^\prime$with the largest probability/cost ratio, which is then added to $\S$ if the remaining budget allows;} otherwise, $l_\ast$ is omitted, and \greedy proceeds to the next iteration. The process terminates if either the budget is exhausted or the set $\L$ becomes empty.

The selection mechanism of the greedy strategy is straightforward, making the algorithm efficient. However,  the greedy strategy for this budgeted non-submodular maximization problem does not come with any approximation guarantees and can indeed lead to arbitrarily bad results. For example, consider a set of LLMs $\L=\{l_1,l_2\}$ subject to a budget $B$, each with associated probabilities $\{p_1, p_2\}$ and costs $\{b_1, b_2\}$. Assume that $b_1 = B$, $b_2 \ll B$, and $p_1 \gg p_2$, yet the ratio $\tfrac{p_1}{b_1} < \tfrac{p_2}{b_2}$. In this case, $\{l_1\}$ is the optimal solution while \greedy would myopically choose $\{l_2\}$ as the solution. As a consequence, this misselection results in an approximation guarantee $\tfrac{p_2}{p_1}$, which can be arbitrarily small.

\begin{algorithm}[t]
\begin{small}
\caption{LLM Selection in Greedy - \greedy}\label{alg:greedy}
\KwIn{LLM set $\L$, success probability set $\P=\{p_1, p_2, \cdots, p_L\}$, cost set $\{b_1, \cdots, b_L\}$, budget $B$, and the set function $\PA(\cdot)$}
\KwOut{Subset $\S$}
$\S \gets \emptyset$\;
% Reorder the set $\L$ of LLMs on success probability in a non-increasing order\;
\While{$B > 0$ \KwAnd $\L\neq \emptyset$\label{line:neq}}
{    
    $\S^\prime \gets \arg\max_{l_i\in \L} \tfrac{\PA(\S\cup \{l_i\})-\PA(\S)}{b_i}$\label{line:marginalratio}\;  
    $l_\ast \gets \arg\max_{l_i\in \S^\prime}\tfrac{p_i}{b_i}$\label{line:tie-break}, $\L \gets \L \setminus\{l_\ast\}$\;
    \If{$B < b_\ast$}{\KwContinue\;}    
    $\S \gets \S \cup \{l_\ast\}$, $B\gets B-b_\ast$\;
}
\Return $\S$\;
\end{small}
\end{algorithm}

\spara{Our Surrogate Greedy} To derive a plausible approximation guarantee for this budgeted non-submodular maximization problem, we propose a {\em surrogate greedy} strategy. The idea is as follows. We design a surrogate set function $\gamma(\S)$ to approximate the {correctness probability} function $\PA(\S)$. This surrogate function is devised such that i) $\gamma(\S)$ is {\em submodular} and ii) $\gamma(\S) \ge \PA(\S)$ holds for every subset $\S\subseteq \L$. Subsequently, we establish an approximation guarantee for $\gamma(\S)$ concerning budgeted submodular maximization. Built on this foundation, we derive a data-dependent approximation guarantee for $\PA(\S)$ for the \problem problem.

Specifically, we define the surrogate set function as 
\begin{equation}\label{eqn:surrogatefunction}
\gamma(\S):=1-\textstyle\prod_{l_k \in \S}(1-p_k)    
\end{equation}
Accordingly, we have the following result. 

\begin{lemma}\label{lem:surrogatefunction}
The set function $\gamma(\S)$ in Equation~\eqref{eqn:surrogatefunction} is submodular and $\gamma(\S) \ge \PA(\S)$ holds for $\forall \S \subseteq \L$.
\end{lemma}
\begin{proof}[Proof of Lemma~\ref{lem:surrogatefunction}]
Consider two subsets $\S_1\subseteq \S_2 \subseteq \L$ and an LLM $l_i \in \L\setminus\S_2$. We calculate {the ratio of marginal gains}  
\begin{align*}
& \tfrac{\gamma(\S_1 \cup \{l_i\})-\gamma(\S_1)}{\gamma(\S_2 \cup \{l_i\})-\gamma(\S_2)} 
= \tfrac{1-\prod_{l_k \in \S_1 \cup \{l_i\}}(1-p_k) - 1 +\prod_{l_k \in \S_1}(1-p_k)}{1-\prod_{l_k \in \S_2 \cup \{l_i\}}(1-p_k) - 1 +\prod_{l_k \in \S_2}(1-p_k)} \\
= & \tfrac{\prod_{l_k \in \S_1}(1-p_k)-\prod_{l_k \in \S_1 \cup \{l_i\}}(1-p_k)}{\prod_{l_k \in \S_2}(1-p_k)-\prod_{l_k \in \S_2 \cup \{l_i\}}(1-p_k)} 
= \tfrac{p_i\prod_{l_k \in \S_1}(1-p_k)}{p_i\prod_{l_k \in \S_2}(1-p_k)} \\
= & \tfrac{1}{\prod_{l_k \in \S_2\setminus \S_1}(1-p_k)}
\ge  1.
\end{align*}
It follows that  $\gamma(\S_1 \cup \{l_i\})-\gamma(\S_1) \ge \gamma(\S_2 \cup \{l_i\})-\gamma(\S_2)$ for $\S_1\subseteq \S_2$, showing that $\gamma(.)$ is submodular. 

To compare $\PA(\S)$ with $\gamma(\S)$, we analyze the difference between the failure probabilities $1 - \PA(\S)$ and $1- \gamma(\S) = \prod_{l_i \in \S}(1-p_i)$. The probability $1 - \PA(\S)$ captures all cases where the aggregated prediction is incorrect. Those cases fall into two disjoint categories: {\bf Category I} contains cases where at least one model in $\S$ provides a correct prediction, whereas {\bf Category II} comprises cases where all models in $\S$ yield incorrect predictions. Consequently, we can express $1 - \PA(\S) = \Pr[\textbf{Category I}] + \Pr[\textbf{Category II}]$ and $1-\gamma(\S) = \prod_{l_i \in \S}(1-p_i)=\Pr[\textbf{Category II}]$. Given that $\Pr[\textbf{Category I}] \ge 0$, it follows that $1 - \PA(\S) \ge \prod_{l_i \in \S}(1-p_i)$, implying $\gamma(\S) \ge \PA(\S)$, which completes the proof.  
\end{proof}

\begin{algorithm}[t]
\begin{small}
\caption{Surrogate Greedy -  \surgreedy}\label{alg:surgreedy}
\KwIn{LLM set $\L$, success probability set $\P=\{p_1, p_2, \cdots, p_L\}$, cost set $\{b_1, \cdots, b_L\}$, budget $B$, {correctness} function $\PA(\cdot)$, and surrogate function $\gamma(\cdot)$}
\KwOut{Subset $\S$}
$l^\ast \gets \arg\max_{l_i\in \L, b_i \leq B} p_i$\; 
%$p^\ast \gets \PA(\{l^\ast\})$\; 
$\S_1 \gets $\greedy($\L,\P$, $\{b_1, \cdots, b_L\}$, $B$, $\PA(\cdot)$)\;
$\S_2 \gets $\greedy($\L,\P$, $\{b_1, \cdots, b_L\}$, $B$, $\gamma(\cdot)$)\;
$\S^\ast \gets \arg\max\{\{l^\ast\},\PA(\S_1), \PA(\S_2)\}$\;
\Return $\S^\ast$\;
\end{small}
\end{algorithm}

{Khuller et al.~\cite{KhullerMN99} study budgeted maximum set cover, a well-known submodular optimization problem, and propose a modified greedy algorithm that has a $(1-\tfrac{1}{\sqrt{\e}})$-approximation guarantee\footnote{They also propose a greedy algorithm with a $(1 - 1/\e)$ approximation guarantee, but its prohibitive $O(n^5)$ complexity makes it impractical for use.
}. Basically, the algorithm selects a single set $\S_1$ whose coverage is maximum within the budget; if the coverage of $\S_1$ is more than that of the vanilla greedy solution $\S_2$, then return $\S_1$; otherwise, return $\S_2$. By leveraging this as a building block, we now introduce our surrogate greedy approach, dubbed \surgreedy, for the budgeted non-submodular LLM selection problem in Algorithm~\ref{alg:surgreedy}. } 

In particular, \surgreedy first identifies the model $l^\ast$ with the highest success probability under the budget $B$. Next, it derives solution sets  $\S_1$ and $\S_2$ by leveraging \greedy on set functions  $\PA(\cdot)$ and $\gamma(\cdot)$ respectively. The one with the highest {correctness probability} among the three sets $\{\{l^\ast\},\S_1,\S_2\}$ is returned as the final solution. The following theorem shows that \surgreedy provides a data-dependent approximation guarantee. 

\begin{theorem}\label{thrm:surapproximation}
Consider sets $\{l^\ast\}$, $\S_1$, $\S_2$, and $\S^\ast$ derived from \surgreedy. It holds that
\begin{equation}\label{eqn:surapproximation}
\PA(\S^\ast) \ge \tfrac{\max\{\PA(S_1),\ \PA(S_2), \ p^\ast\}}{\max\{\gamma(\S_2),\ p^\ast\}}(1-\tfrac{1}{\sqrt{\e}})\cdot \PA(\S^\circ),    
\end{equation}
where $p^\ast$ is the success probability of $l^\ast$ and $\S^\circ$ is the optimal solution of \problem.
\end{theorem}

\subsection{Surrogate Greedy: Further Optimizations}\label{sec:adaptiveselection}

{In this section, we seek further optimizations on  Algorithm~\ref{alg:surgreedy}. Specifically, we shall first show that it is possible to further optimize the solution $\S^\ast$ returned by Surrogate Greedy by identifying models in $\S^\ast$ that can be safely pruned without changing the final prediction. Eliminating models from $\S^\ast$ helps cut down the cost incurred by the final solution. The intuition why this works is because when we apply models in $\S^\ast$ successively on a given task, by observing the predictions obtained so far we may be able to determine that the remaining models cannot influence the final aggregated prediction.}  

{Secondly, we note that calculating the {correctness} function $\PA(\cdot)$ exactly is expensive. Therefore, for practical implementation, we estimate $\PA(\cdot)$ using $\theta$ Monte Carlo simulations, where $\theta$ is determined by input parameters $\epsilon, \delta\in (0,1)$. The principle for selecting appropriate values for $\epsilon$ and $\delta$ is discussed in Section~\ref{sec:appro}.} 

\subsubsection{{Adaptive Selection}} \label{sec:adaptive}
After obtaining $\S^\ast$ from Algorithm~\ref{alg:surgreedy}, it can be further reduced at model invocation time, in an adaptive manner tailored for practical scenarios. Specifically, when we apply models from $\S^\ast$ in sequence on a given query, a tipping point arises upon which the aggregated prediction from the models applied  so far cannot be not influenced by subsequent models from $\S^\ast$ not yet used. At this juncture, the aggregated prediction can be deemed final and returned in response to the query. This procedure enables the derivation of a subset of LLMs $\S$ from $\S^\ast$ with a reduced cost  by leveraging real-time observational feedback, while ensuring the same prediction as that of $\S^\ast$. Building on this pivotal insight, we have developed an adaptive LLM selection strategy and introduce \ours, detailed in Algorithm~\ref{alg:LLMSurGreedy}.

{We first initialize $\mathcalT^\ast=\S^\ast$ obtained from Algorithm~\ref{alg:surgreedy} and select models from $\mathcalT^\ast$, add them to $\S$ while removing them from $\mathcalT^\ast$. Let $\S$ be the current set of  selected LLMs from $\mathcal{T}^\ast$, \ie $\S=\S^\ast\setminus \mathcal{T}^\ast$, and $\phi_\S$ be any real-time observation of $\S$ on input random query $q$. In each iteration before the selection, the algorithm  checks the termination condition $F(\mathcal{T}^\ast)H_2(\phi_\S) > H_1(\phi_\S)$ where 
$F(\mathcal{T}^\ast):=\textstyle\prod_{l_i\in \mathcal{T}^\ast}\tfrac{p_i(K-1)}{1-p_i}$ is the potential belief value of set $\mathcal{T}^\ast$. In particular, the potential belief $F(\S)$ of a set $\S$ represents the maximum possible belief value that the set $\S$ could contribute to any class. Note that the set $\mathcal{T}^\ast$ is updated (Line~\ref{line:update} in Algorithm~\ref{alg:LLMSurGreedy}) during each selection and contains the remaining unselected LLMs.}  

%%%%%%%%%%%%%%%%%%%%%%%%%%%%%%%% 
\begin{algorithm}
\begin{small}
\caption{Adaptive LLM Selection - \ours}\label{alg:LLMSurGreedy}
\KwIn{Set $\L$ of LLMs, set $\P$ of success probability, cost $b_1, \cdots, b_L$, budget $B$, parameters $\epsilon$, $\delta$, and a random query $q\in \Q$}
\KwOut{Subset $\S$ and prediction on $q$}
$p^\ast \gets \max\{p_i \colon l_i\in \L, b_i \leq B\} $, $\theta := \frac{8+2\epsilon}{\epsilon^2p^\ast}\ln(\frac{2L^2}{\delta})$\;
$\S^\ast \gets $ Apply Algorithm~\ref{alg:surgreedy} with $\theta$ Monte Carlo simulations for $\PA(\cdot)$ estimation\;
$\mathcal{T}^\ast \gets \S^\ast$, $\S \gets \emptyset$, $\phi_\S \gets \emptyset$, $H_2(\phi_\S)\gets 1 $, $ H_1(\phi_\S)\gets 1$\;
\While{$\mathcal{T}^\ast \neq \emptyset$\label{line:neq}}
{    
    \If{$F(\mathcal{T}^\ast)H_2(\phi_\S) > H_1(\phi_\S)\label{line:twocondition}$}
    {
       $l^\ast \gets \arg\max_{l_i\in \mathcal{T}^\ast} p_i$\;       
       $\S \gets \S \cup \{l^\ast\}$, $\mathcal{T}^\ast \gets \mathcal{T}^\ast \setminus\{l^\ast\}$\label{line:update}\; 
       Apply $l^\ast$ on query $q$ and update observation $\phi_\S$\;
    }
    \Else 
    {\KwBreak\;\label{line:break}}  
}
\Return $\S$ and the prediction with belief $H_1(\phi_\S)$\;
\end{small}
\end{algorithm}%\vspace{-2mm}
%%%%%%%%%%%%%%%%%%%%%%%%%%%%%%%% 

If the condition is satisfied (Line~\ref{line:twocondition}), this suggests the possibility that the application of the residual set $\mathcal{T}^\ast$ to query $q$ can yield a prediction that differs from the existing prediction {associated with the belief value} $H_1(\phi_\S)$. We formalize this observation in Proposition~\ref{prop:termination}. In this scenario, we persist in picking the model $l^\ast$ with the largest success probability from $\mathcal{T}^\ast$ into set $\S$. Subsequently, $l^\ast$ is applied to query $q$, and observation $\phi_\S$ is updated accordingly. This procedure terminates if the condition is not met or $\T^\ast$ is empty.

\begin{proposition}\label{prop:termination}
{If the condition $F(\mathcal{T}^\ast)H_2(\phi_\S) \le H_1(\phi_\S)$ holds in Algorithm~\ref{alg:LLMSurGreedy}, the prediction by set $\S$ is the same as the prediction by set $\S^\ast$.}
\end{proposition}

\subsubsection{Approximation Guarantee and Time Complexity}\label{sec:appro}

In Algorithm~\ref{alg:LLMSurGreedy}, {correctness probability} values of all subsets examined in SurGreedyLLM are estimated using a sufficient number of Monte Carlo simulations. We quantify the confidence of the estimation interval in the following result,  derived using Hoeffding's inequality.  

\begin{lemma}\label{lem:montecarlo}
Consider an arbitrary set $\S\subseteq \L$ with {correctness probability} $\PA(\S)$. The {correctness probability} estimation $\tilde{\PA}(\S)$ in Algorithm~\ref{alg:LLMSurGreedy} satisfies: 
\begin{equation}\label{eqn:montecarlo}
\Pr\left[|\PA(\S)-\tilde{\PA}(\S)| \le \tfrac{\epsilon p^\ast}{2} \right] \ge 1-\delta/L^2,   
\end{equation}
where $\epsilon, \delta \in (0,1)$, $p^\ast = \max\{p_i \colon l_i\in \L, b_i \leq B\}$, and $\tilde{\PA}(\S)$ is averaged from $\theta=\frac{8+2\epsilon}{\epsilon^2p^\ast}\ln(\frac{2L^2}{\delta})$ Monte Carlo simulations.
\end{lemma}

Lemma~\ref{lem:montecarlo} directly follows from Lemma 3 in~\cite{TangXS14}. Building on Lemma~\ref{lem:montecarlo}, we establish the approximation guarantee of the subset of LLMs from \ours.

\begin{theorem}\label{thrm:appro}
{Given parameters $\epsilon,\delta\in(0,1)$, let $\S^\ast$ be the solution returned by \ours and $\S^\circ$ be the optimal solution to the \problem problem. Then we have}    
\begin{equation}\label{eqn:approx}
\Pr\left[\PA(\S^\ast) \ge (\tfrac{\max\{\PA(S_1),\ \PA(S_2), \ p^\ast\}}{\max\{\gamma(\S_2),\ p^\ast\}}-\epsilon)(1-\tfrac{1}{\sqrt{\e}})\cdot \PA(\S^\circ) \right] \ge 1-\delta,
\end{equation}
where $\S_1$, $\S_2$, and $p^\ast$ are derived from \surgreedy.
\end{theorem}

\spara{Time Complexity} The time complexity of Algorithm~\ref{alg:LLMSurGreedy} is dominated by the selection process with set function $\PA(\cdot)$. Specifically, there are at most $O(L^2)$ evaluations of {correctness probability}. Each evaluation invokes $\theta$ Monte Carlo simulations, and each Monte Carlo simulation conducts $O(L)$ evaluations. Thus, the overall time complexity is  $O(\theta L^3)=O(\frac{L^3}{\epsilon^2 p^\ast}\ln(\frac{2L^2}{\delta}))=O(\frac{L^3}{\epsilon^2}\ln(\frac{L}{\delta}))$.

\subsection{Extension to Probability Interval Estimates}\label{sec:interval}

{Our algorithms  as well as the approximation analysis assume the precise ground truth success probabilities of the models $\L$ are available. In practice, they are unavailable and must be estimated, e.g., using the historical query responses of these models as illustrated in Section~\ref{sec:estimation}. These estimates have an associated confidence interval. Specifically, given arbitrary sample sizes, confidence intervals can be derived by leveraging concentration inequalities~\cite{BoucheronLB03}.
We next show how our algorithms (and analysis) can be extended to work with confidence intervals associated with estimates of success probabilities. For clarity, denote the estimate of ground truth success probability $p_l$ as $\hat{p}_l$. Let this estimate $\hat{p}_l$ have an associated confidence interval $[p^\bot_l, p^\top_l]$ at a confidence level of $1-\delta_l$, where $\delta_l \in (0,1)$. 
In the following, we explore the approximation guarantee of \ours when these intervals are provided as inputs.} 

Let $\P_{\textrm{low}}=\{p^\bot_1,p^\bot_2,\cdots,p^\bot_L\}$, $\hat{\P}=\{\hat{p}_1,\hat{p}_2,\cdots,\hat{p}_L\}$, and $\P_{\textrm{up}}=\{p^\top_1,p^\top_2,\cdots,p^\top_L\}$ {be the sets of lower bounds, estimated values, and upper bounds corresponding to model success probabilities, respectively.} The corresponding {correctness} functions $\PA(\cdot)$, $\PA_l(\cdot)$, and $\PA_u(\cdot)$ are defined based on {\em ground-truth} probabilities, $\P_{\textrm{low}}$, and $\P_{\textrm{up}}$, respectively. Although the same observation space is shared by the four scenarios involving ground-truth probabilities, $\P_{\textrm{low}}$, $\hat{\P}$ and $\P_{\textrm{up}}$, the corresponding probability distributions of observations intrinsically differ. {Run Algorithm~\ref{alg:LLMSurGreedy} with each of the success probability sets $\P_{\textrm{low}}$, $\hat{\P}$, and $\P_{\textrm{up}}$ and let $\S^\ast_l$, $\S^\ast$, and $\S^\ast_u$ be the solution returned by the algorithm on these inputs respectively. Based on this, we can establish the following theorem.} 

\begin{theorem}\label{thrm:approinterval}
{Consider a set $\L=\{l_1,l_2,\cdots,l_L\}$ of LLMs and a random query $q$ from class $\Q$. Suppose the  success probability of model $l$ on $q$ is estimated as $\hat{p}_l$ with confidence interval $[p^\bot_l, p^\top_l]$ at a confidence level of $1-\delta_l$ for $\delta_l \in (0,1)$. Given approximation parameters $\epsilon,\delta\in(0,1)$, we have}  
\begin{align*}
& \Pr\left[ \tfrac{\PA(\S^\ast)}{\PA(\S^\circ)} \ge \tfrac{\PA_l(\S^\ast_l)}{\PA_u(\S^\ast_u)}(\tfrac{\max\{\PA_u(S_{u1}),\ \PA_u(S_{u2}), \ p^\ast_u\}}{\max\{\gamma_u(\S_{u2}),\ p^\ast_u\}}-\epsilon)(1-\tfrac{1}{\sqrt{\e}})\right] \\ 
&\ge 1-(\delta+L^2\textstyle \sum^L_{l=1}\delta_l),   
\end{align*}
where $\gamma_u(\cdot)$ is the surrogate set function, $S_{u1}$ and $S_{u2}$ are selected by \surgreedy on $\P_{\textrm{up}}$, respectively. 
\end{theorem}

To ensure this data-dependent approximation guarantee holds with high probability, a failure probability $\delta+L^2\textstyle \sum^L_{l=1}\delta_l \ll 1$ is necessary. To this end, the term $L^2\textstyle \sum^L_{l=1}\delta_l$ is supposed to be in the scale of $\delta$, \ie $L^2\textstyle \sum^L_{l=1}\delta_l=\Theta(\delta)$. 
In the following, we discuss how to ensure a small failure probability.

\spara{Diminishing failure probability} When substantial samples are available {for query clusters,} the failure probability (confidence level) in Theorem~\ref{thrm:approinterval} can be significantly improved by enlarging the sample sizes. However, this may not be possible due to the lack of samples in real-world applications. In this scenario, we can calibrate the failure probability by repeatedly estimating the confidence intervals $[p^\bot_l, p^\top_l]$ with estimation $\hat{p}_l$ until a targeted failure probability (confidence level) is reached. Specifically, we sample a certain number of queries from each cluster and derive the confidence interval at a certain confidence level. We then repeat this procedure a sufficient number of times and return the median value among all estimates. By doing this, we can diminish the failure probability to a desired level. Formally, we establish the following Lemma.

\begin{lemma}\label{lem:failboosting}
{Let $\delta_l \in (0,1)$ be the failure probability of confidence interval $[p^\bot_l, p^\top_l]$ with estimate $\hat{p}_l$ of success probability $p_l$ derived by a sampling procedure  $\A$, such that \[\Pr\left[ p^\bot_l \le p_l \le p^\top_l \right] \ge 1 -\delta_l.\]  
By independently repeating $\A$ a total of $\Lambda_l$ times, and taking the interval with the corresponding estimates being the median among the $\Lambda_l$ estimates, denoted as $[\bar{p}^\bot_l, \bar{p}^\top_l]$, we have} 
\begin{equation}
    \Pr[\bar{p}^\bot_l \le p_l \le \bar{p}^\top_l] \ge 1-\exp(-\tfrac{\Lambda_l(1-2\delta_l)^2}{2}).
\end{equation}
\end{lemma}

Theoretically, we aim to limit probability $L^2\textstyle \sum^L_{l=0}\delta_l=\Theta(\delta)$ in Theorem~\ref{thrm:approinterval}. Lemma~\ref{lem:failboosting} proves that $\delta_l$ can be {diminished} to $\exp(-\tfrac{\Lambda_l(1-2\delta_l)^2}{2})$. In this regard, we can simply ensure $L^2 \exp(-\tfrac{\Lambda_l(1-2\delta_l)^2}{2}) \, {\leq} \, \tfrac{\delta}{L}$. Therefore, we have $\Lambda_l=\tfrac{6\log(L/\delta)}{(1-2\delta_l)^2}$. %Empirically, we can set $\Lambda_l=\Theta(L)$ for all $l\in[L]$.

\section{Related Work}\label{sec:related}

\spara{LLM Ensemble} FrugalGPT~\cite{ChenZZ2023} aims to reduce the utilization cost of large language models (LLMs) while improving the overall performance. In particular, it derives an LLM cascade from a candidate set tailored for queries under a budget constraint. However, its performance is suboptimal as only the response from the last {executed} model {in the cascade} is adopted, without exploiting previous responses. Furthermore, it generates one LLM cascade for the whole dataset, {i.e., for all query classes,} which can lead to inferior performance due to the inherent diversity of the datasets {and query classes.} LLM-Blender~\cite{Jiang0L23} is another recently proposed LLM ensemble approach, but it does not incorporate any budget constraint. Instead, it considers a set of existing LLMs from different mainstream providers. It first applies all models to the given query and selects the top-$K$ responses using a ranking model called PairRanker. The $K$ responses concatenated with the query are fed into a fine-tuned T5-like model, namely the GenFuser module, to generate the final response. In the development of LLMs, their performance grows gradually over time due to the scaling law and fine-tuning.
{LLM-Ensemble~\cite{llmEnsemble24} learns aggregation weights for each LLM and forms an ensemble by weights for the final response.
}
{Similarly, LLM-Topla~\cite{llmTopla24} selects a subset of LLMs optimized for diversity using a genetic algorithm. 
}
To predict the increasing performance and capture the convergence point along time, \citet{XiaK0GRK024} develop a time-increasing bandit algorithm TI-UCB. Specifically, TI-UCB identifies the optimal LLM among candidates regarding development trends with theoretical logarithmic regret upper bound. Different from TI-UCB on a single optimal LLM, C2MAB-V proposed by \citet{DaiLiLiuYuLui2024} seeks the optimal LLM combinations for various collaborative task types. It employs combinatorial multi-armed bandits with versatile reward models, aiming to balance cost and reward. Recently, \citet{ShekharDubey2024} try to reduce the usage costs of LLM on document processing tasks. In particular, they first estimate the ability of each individual LLM by a BERT-based predictor. By taking these estimations as inputs, they solve the LLM selection as a linear programming (LP) optimization problem and propose {the QC-Opt algorithm.} Instead of selecting combinations of well-trained LLMs, \citet{BansalSamanta2024} propose to compose the internal representations of several LLMs by leveraging a cross-attention mechanism, enabling new capabilities. Recently, Octopus-v4~\cite{ChenLi2024v4} has been proposed as an LLM router. It considers multiple LLMs with expertise in different domains and routes the queries to the one with the most matched topic. However, (i) it does not consider the budget constraint but only LLM expertise when routing, and (ii) it employs one single model instead of an LLM ensemble for enhanced performance. Among the above works, FrugalGPT \cite{ChenZZ2023} and LLM-Blender \cite{Jiang0L23} have similar goals and are most relevant to our work. We experimentally compare with them in Section~\ref{sec:exp}.  

\spara{Entity Matching} Entity Matching (EM) aims to identify whether two records from possibly different tables refer to the same real-world entity. It is also known as record linkage or entity resolution~\cite{GetoorM12}.
According to~\cite{BarlaugG21}, EM consists of five subtasks: data preprocessing, schema matching, blocking, record pair comparison, and classification. Magellan~\cite{KondaP16} serves as a representative end-to-end EM system, though it requires external human programming.
An emerging and fruitful line of work~\cite{BarlaugG21, EbraheemTSJOT17, MudgalLRDPKDAR18, ZhaoH19, CappuzzoPT20, YaoGCJL22} proposes applying deep learning-based methods to improve the classification accuracy and automate the EM process.
DeepMatcher~\cite{MudgalLRDPKDAR18} utilizes an RNN architecture to aggregate record attributes and perform comparisons based on the aggregated representations. DeepER~\cite{EbraheemTSJOT17} employs GloVe to generate word embeddings and trains a bidirectional LSTM-based EM model to obtain record embeddings.
AutoEM~\cite{ZhaoH19} introduces a methodology to fine-tune pre-trained deep learning-based EM models using large-scale knowledge base data through transfer learning.
Another line of research, such as EmbDI~\cite{CappuzzoPT20} and HierGAT~\cite{YaoGCJL22}, leverages graph structures to improve EM by learning proximity relationships between records.
With the emergence of transformer-based language models like BERT~\cite{kenton2019bert} and RoBERTa~\cite{roberta19}, Ditto~\cite{Ditto20} fine-tunes these pre-trained models with EM corpora and introduces three optimization techniques to improve matching performance. 
As the state-of-the-art method, \citet{PeetersSB25} shows that LLMs with zero-shot learning outperform pre-trained language model-based methods, offering a more robust, general solution. Their work demonstrates that LLMs can effectively generalize to EM tasks without fine-tuning, making them ideal for cases where large training sets are costly or impractical.

\section{Experiments}\label{sec:exp}

\sktch{In this section, we evaluate the performance of our proposed \ours across a collection of real-world datasets against state-of-the-art LLM-ensemble models for two types of queries, namely {\em text classification} and {\em entity matching}. }

\begin{table}[!t]
\centering
\caption{Dataset details {for text classification}} \label{tbl:dataset}%\vspace{-1mm} 
\setlength{\tabcolsep}{0.4em}
%\vspace{-2mm}
\small
\resizebox{0.48\textwidth}{!}{%
\begin{tabular} {@{}l|ccccc@{}}
\toprule
{\bf Dataset}  & \multicolumn{1}{c}{Overruling} & \multicolumn{1}{c} {AGNews}  & \multicolumn{1}{c}{SciQ} &\multicolumn{1}{c}{Hellaswag\footnote{Due to its enormous size, we randomly select $30\%$ of the queries from this dataset for our experiments.}} & \multicolumn{1}{c}{Banking77} \\ \midrule 
{\bf{Domain}}  & Law & News & Science & Commonsense & Banking \\
{\bf{Sizes}}  & $2.1$ K & $7.6$ K &  $12.7K$ & $15$ K  & $13$ K\\
{\bf \#Classes}  & 2 & 4 &  4 & 4 & 77\\ \bottomrule
\end{tabular}}
 %\vspace{-1mm}
\end{table}

\begin{table}[!t]
\centering
\caption{Dataset details for entity matching} \label{tbl:dataset_entity}\vspace{-1mm} 
\setlength{\tabcolsep}{0.4em}
\renewcommand{\arraystretch}{0.9}% for the vertical padding
%\vspace{-2mm}
\small
\resizebox{0.48\textwidth}{!}{%
\begin{tabular} {@{}l|cc|cc@{}}
\toprule
\multirow{2}{*}{\bf Dataset}  & \multicolumn{2}{c|}{Training set} & \multicolumn{2}{c} {Test set} \\ 
 & \# Pos & \# Neg & \# Pos & \# Neg \\ \midrule
(WDC) - WDC Products & 500 & 2,000 & 250 & 989 \\
(A-B) - Abt-Buy & 616 & 5,127  & 206 & 1,000 \\
(W-A) - Walmart-Amazon & 576 & 5,568 & 193 & 1,000 \\
(A-G) - Amazon-Google & 699 & 6,175  & 234 & 1,000 \\
(D-S) - DBLP-Scholar & 3,207 & 14,016 &  250 & 1,000 \\
%(D-A) - DBLP-ACM & 1,332 & 6,085  & 250 & 1,000 \\ 
\bottomrule
\end{tabular}}
 %\vspace{-1mm}
\end{table}

\subsection{Experimental Setup}\label{sec:settings}

\spara{Datasets for text classification query} We conduct experiments on $5$ datasets across various real-world applications, namely Overruling, AGNews, SciQ, Hellaswag, and Banking77, as summarized in Table~\ref{tbl:dataset}. Specifically, 
% Headlines~\cite{sinha2021impact} is a news dataset that contains headlines of financial news. It is used to predict the trend of gold prices into one of four states -- \textit{up}, \textit{down}, \textit{neutral}, or \textit{none}. 
Overruling~\cite{ZhengGA0H21} is a legal document dataset designed to determine if a given sentence is an overruling. In particular, an overruling sentence {overrides} the decision of a previous case as a precedent. AGNews~\cite{zhang2015character} contains a corpus of news articles categorized into four classes: {\em World}, {\em Sports}, {\em Business}, and {\em Science/Technology}. Hellaswag~\cite{ZellersHBFC19} consists of unfinished sentences for commonsense inference. Specifically, given an incomplete sentence, models are required to select the most likely follow-up sentence from $4$ candidates. SciQ~\cite{WelblLG17} is a collection of multiple-choice questions from science exams, with a unique correct option. Finally, Banking77~\cite{Casanueva2020} dataset consists of online banking queries from customer service interactions, with each query assigned a label corresponding to one of 77 fine-grained intents. %This makes Banking77 a more fine-grained intent classification. benchmark compared to others.

\begin{table}[!t]
 \caption{Summary of commercial LLM APIs.} \label{tbl:LLMAPI}\vspace{-1mm}
 \setlength{\tabcolsep}{0.4em} % for the horizontal padding
%\renewcommand{\arraystretch}{0.9}% for the vertical padding
%\vspace{-3mm}
 \small
\resizebox{0.48\textwidth}{!}{%
\begin{tabular}{@{}clcccc@{}}
\toprule
 \multirow{2}{*}{\bf Company} & \multirow{2}{*}{\bf LLM APIs} & \multirow{2}{*}{\bf Size (B)} & \multicolumn{2}{c}{\bf Cost/$1$M tokens (usd)}   \\ 
  {} & {} & {} &  {\bf Input} & {\bf Output} \\ \midrule %\cline{3-10} 
\multirow{2}{*}{\rotatebox[origin=c]{0}{\bf OpenAI}}
& GPT-4o-mini	   & N.A.	&	0.15	&	0.6 \\
& GPT-4o           &  N.A  & 5.0      & 15.0    \\ \midrule
\multirow{3}{*}{\rotatebox[origin=c]{0}{\bf Google}}
& Gemini-1.5 Flash   & N.A.  &  0.075  & 0.3 \\
& Gemini-1.5 Pro    & N.A &  3.5 & 10.5    \\
& Gemini-1.0 Pro    & N.A. & 0.5 &  1.5     \\ \midrule
\multirow{4}{*}{\rotatebox[origin=c]{0}{\bf Microsoft}}
& Phi-3-mini	    & 3.8	& 0.13	&	0.52 \\
& Phi-3.5-mini		&  3.8  & 0.13 & 0.52 \\
& Phi-3-small		&  7 &	0.15	&	0.6	\\
& Phi-3-medium	    & 14	&	0.17	&	0.68 \\ \midrule
\multirow{2}{*}{\rotatebox[origin=c]{0}{\bf Meta}}
& Llama-3 8B	    & 8	& 0.055	&	0.055		\\
& Llama-3 70B		&  70  & 0.35  &	0.4			\\ \midrule
\multirow{1}{*}{\rotatebox[origin=c]{0}{\bf Mistral AI}}
& Mixtral-8x7B		& 46.7   &	0.24	&	0.24	\\
\bottomrule
\end{tabular}
}%\vspace{-2mm}
\end{table}

\spara{Datasets for entity matching query}
For this task, we used five datasets with the same setup as in~\cite{PeetersSB25}. Dataset names and detailed statistics are presented in Table~\ref{tbl:dataset_entity}. On each dataset,  \eat{represents a binary classification task, where} the model is required to determine whether two real-world entity descriptions \ravi{(records)} refer to the same entity, answering either {\em yes} or {\em no}. Among these five datasets, DBLP-Scholar is a bibliographic entity dataset, while the remaining four are e-commerce datasets.

\spara{LLMs} We incorporate $12$ commercial LLM APIs provided by five leading companies -- OpenAI, Google, Microsoft, Meta, and Mistral AI. The details of these LLMs are summarized in Table~\ref{tbl:LLMAPI}. As shown, we select state-of-the-art LLMs from five companies as candidates. Additionally, we present the costs per $1$ million input and $1$ million output tokens, which vary from $\$0.055$ to $\$15.0$. 
Typically, larger models incur higher costs. To reduce the internal model randomness, we set the lowest temperature for all LLM candidates, thereby ensuring more deterministic responses.

\spara{Baselines} For {\em text classification}, we compare \ours with \greedy and {four related} models, namely FrugalGPT~\cite{ChenZZ2023}, LLM-Blender~\cite{Jiang0L23}, {LLM-Topla~\cite{llmTopla24}, and LLM-Ensemble~\cite{llmEnsemble24}}. FrugalGPT, {LLM-Ensemble, LLM-Topla,} and LLM-Blender are described in detail in Section~\ref{sec:related}. {Since LLM-Ensemble didn't address cost constraint in their original work, we apply a straightforward approach by greedily selecting the top-K weighted LLMs until the budget is met.} For LLM-Blender, we utilize the latest models for PairRanker and GenFuser, \eat{in LLM-Blender,}  following the {original parameter settings recommended by the authors.} For {\em entity matching}, we include the SOTA methods \roberta~\cite{roberta19} finetuned by~\cite{PeetersSB25} and \ditto~\cite{Ditto20}\sktch{(details in Section~\ref{sec:related})}.

\spara{Parameter settings} In the experiments, we split the datasets for text classification into $80\%/20\%$ as {historical and} test sets, {which form two disjoint query sets.} As with datasets for entity matching, we follow the train/test splits as those in~\cite{PeetersSB25}. For \ours, we fix parameters $\epsilon = 0.1$ and $\delta = 0.01$. According to the actual query costs, we set a series of budgets $B=\{1.0, 5.0, 10, 50, 100\}\times 10^{-5}$ (USD) such that only subsets of LLMs in Table~\ref{tbl:LLMAPI} are feasible given those budget constraints. {Empirically, for a single query, when $B=1.0 \times 10^{-5}$ USD, no more than 2–3 models are typically selected; whereas for $B=1.0 \times 10^{-3}$ USD, up to 9–10 models are chosen.}

\spara{Running environment} Our experiments are conducted on a Linux machine with an NVIDIA RTX A5000 (24GB memory), Intel Xeon(R) CPU (2.80GHz), and 500GB RAM.

\begin{figure}[!t]
\centering
\arxiv{\includegraphics[height=0.5\linewidth]{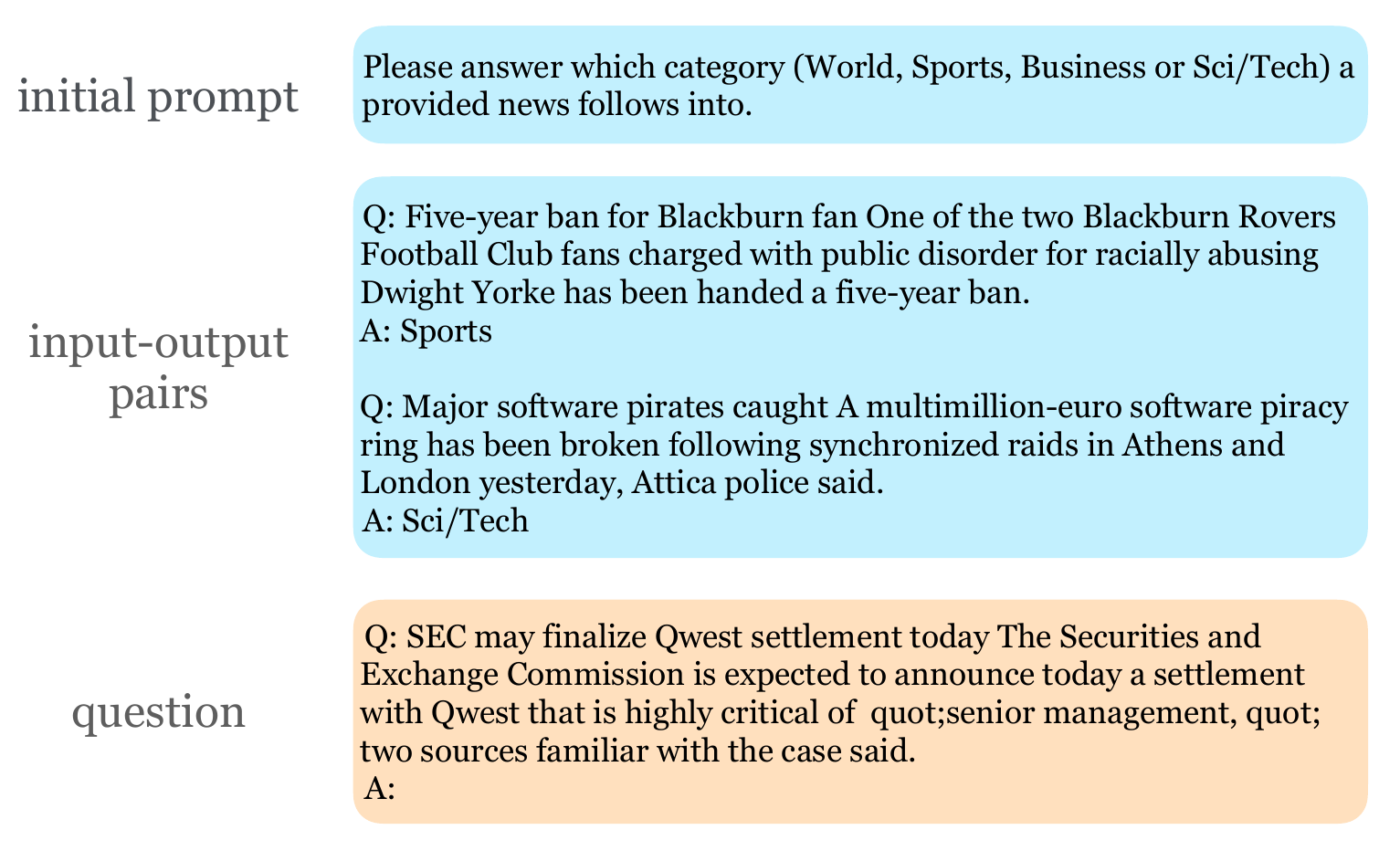}}
\sktch{\includegraphics[height=0.5\columnwidth]{figs/prompt_template.pdf}}
\vspace{-4mm}
\caption{Prompt template for AGNews dataset.}
\label{fig:prompt}\vspace{-2mm}
\end{figure}

\spara{Prompt Engineering}
We design two-shot prompting templates for text-classification datasets to ensure models generate outputs in the desired format. For AGNews, we adopt the prompt template from FrugalGPT~\cite{ChenZZ2023} but limit input-output examples to two, as shown in Figure~\ref{fig:prompt}. The blue text blocks contain the prompt and examples, while the orange block contains the target question. This structure is applied across all datasets. For other datasets in text classification, we randomly select two training records as input-output pairs. For datasets in entity matching, we follow the procedure outlined in~\cite{PeetersSB25} and use a zero-shot prompt with two templates: {\em domain-complex} for DBLP-Scholar and Amazon-Google, and {\em general-complex} for the others.

\begin{figure*}[!t]
\centering
\begin{small}
\hspace{-1mm}
\begin{tikzpicture}
\begin{customlegend}[legend columns=5,legend style={align=center,draw=none,column sep=1ex, font=\scriptsize},
        legend entries={\ours, \greedy, FrugalGPT, LLM-Ensemble, LLM-Topla}]
        \addlegendimage{mark=asterisk,red}
        \addlegendimage{mark=oplus,cyan}
        \addlegendimage{mark=triangle,blue}
        \addlegendimage{mark=diamond,orange}
        \addlegendimage{mark=+,violet}
        \end{customlegend}
\end{tikzpicture}\vspace{-3mm}

\subfloat[{Overruling}]{
\begin{tikzpicture}
\begin{axis}[
    height=\columnwidth/2.5,
    width=\columnwidth/2.2,            
    enlarge x limits=true,
    ymin=0.65, ymax=0.98,
    xmin=0.5, xmax=100.5,         
    ylabel={{\em Accuracy}},
    xlabel={\em Cost ($\times 10^{-5}$)},
    xlabel style={yshift=0.1cm},
    ylabel style={yshift=-0.1cm},    
    xtick={1,5,10,50,100},
    xticklabels={1,5,10,50,100},
    xmode=log,
    log basis x={10},
    xticklabel style={font=\scriptsize},
    %every axis y label/.style={at={(current axis.north west)},right=12mm,above=0mm},   
    %legend style={at={(0.05,0.95)},anchor=north west,font=\scriptsize},
    %legend columns=3,
    ]
    \addplot[only marks, mark size=1pt, mark=asterisk,red] coordinates {(0.5868229166666662,0.9513888888888888)
    (2.4283460648148167,0.9537037037037037)
    (4.4832361111111025,0.9537037037037037)
    (24.83767592592593,0.9537037037037037)
    (38.00079050925928,0.9560185185185185)}; % ThriftLLM  
    \addplot[only marks, mark size =1pt, mark=oplus,cyan] coordinates {(0.70159837962963,0.9467592592592593)
    (4.2812986111111126,0.9513888888888888)
    (7.260997685185179,0.9513888888888888)
    (34.17497453703705,0.9537037037037037)
    (47.612497685185193,0.9537037037037037)}; % greedy
    \addplot[only marks, mark size =1pt, mark=triangle,blue] coordinates {(0.66, 0.680555556) (3.07, 0.793981481)(3.57, 0.949074074)(5.97, 0.731481481481481)(72.77,0.84722)}; % FrugalGPT
    %\addplot[mark=diamond,cyan] coordinates {(1, 0.9490740740740741) (2, 0.9513888888888888)(3, 0.9513888888888888)(4, 0.9467592592592593)(5,0.9467592592592593)}; % Octopus
    \addplot[only marks, mark size =1pt, mark=diamond,orange] coordinates {(0.489, 0.696) (3.005, 0.94675)(3.85, 0.9467)(36.66,0.95138)(53.084,0.9513)}; % LLM-ensemble
    \addplot[only marks, mark size =1pt, mark=+,violet] coordinates {(1.3, 0.951388) (4.9, 0.939814)(4.9, 0.9213)}; % LLM-topla
\end{axis}
\end{tikzpicture}
\label{subfiga:overrunling}
}\hspace{2mm}
\subfloat[{AGNews}]{
\begin{tikzpicture}
\begin{axis}[
    height=\columnwidth/2.5,
    width=\columnwidth/2.2,            
    enlarge x limits=true,
    ymin=0.8, ymax=0.9,
    xmin=0.5, xmax=100.5,         
    ylabel={{\em Accuracy}},
    xlabel={\em Cost ($\times 10^{-5}$)},
    xlabel style={yshift=0.1cm},
    ylabel style={yshift=-0.1cm},    
    xtick={1,5,10,50,100},
    xticklabels={1,5,10,50,100},
    xmode=log,
    log basis x={10},
    xticklabel style={font=\scriptsize},
    %every axis y label/.style={at={(current axis.north west)},right=12mm,above=0mm},   
    %legend style={at={(0.05,0.95)},anchor=north west,font=\scriptsize},
    %legend columns=3,
    ]
    \addplot[only marks, mark size =1pt, mark=asterisk,red] coordinates {
    (0.5884641447368416,0.8447368421052631)
    (2.649204605263158,0.8473684210526315)
    (5.244262828947375,0.8486842105263158)
    (28.65560855263157,0.8625)
    (49.76167401315792,0.8644736842105263)}; % ThriftLLM  
    \addplot[only marks, mark size =1pt, mark=oplus,cyan] coordinates {
    (0.7256115131578955,0.8328947368421052)
    (4.5747421052631564,0.8453947368421053)
    (8.737631578947339,0.8480263157894737)
    (38.30910723684203,0.8625)
    (57.35128684210521,0.8625)}; % greedy
    \addplot[only marks, mark size =1pt, mark=triangle,blue] coordinates {(0.796, 0.7717) (1.55, 0.8513158)(3.31, 0.8236842)(4.45, 0.825)}; % FrugalGPT
    %\addplot[mark=diamond,cyan] coordinates {(1, 0.8401315789473685) (2, 0.8421052631578947)(3, 0.8421052631578947)(4, 0.8901315789473684)(5,0.8901315789473684)}; % Octopus
    % \addplot[only marks, mark size =1pt, mark=diamond,orange] coordinates {(0.916, 0.6111) (1.928, 0.847)(1.949, 0.847)(39.4,0.848)(58.794,0.842)}; % LLM-ensemble non empty
    \addplot[only marks, mark size =1pt, mark=diamond,orange] coordinates {(0.010, 0.007) (1.917, 0.842)(1.948, 0.846)(39.4,0.848)(58.794,0.842)}; % LLM-ensemble 
    \addplot[only marks, mark size =1pt, mark=+,violet] coordinates {(1.2, 0.833552) (4.7, 0.8941)}; % LLM-topla
\end{axis}
\end{tikzpicture}
\label{subfiga:agnews}
}\hspace{2mm}
\subfloat[{SciQ}]{
\begin{tikzpicture}
\begin{axis}[
    height=\columnwidth/2.5,
    width=\columnwidth/2.2,            
    enlarge x limits=true,
    ymin=0.8, ymax=1,
    xmin=0, xmax=100.5,         
    ylabel={{\em Accuracy}},
    xlabel={\em Cost ($\times 10^{-5}$)},
    xlabel style={yshift=0.1cm},
    ylabel style={yshift=-0.1cm},    
    xtick={1,5,10,50,100},
    xticklabels={1,5,10,50,100},
    xmode=log,
    log basis x={10},
    xticklabel style={font=\scriptsize},
    %every axis y label/.style={at={(current axis.north west)},right=12mm,above=0mm},   
    %legend style={at={(0.05,0.95)},anchor=north west,font=\scriptsize},
    %legend columns=3,
    ]
    \addplot[only marks, mark size =1pt, mark=asterisk,red] coordinates {
    (0.15048481861198773,0.9834384858044164)
    (2.8386851340693945,0.9905362776025236)
    (5.185557965299676,0.9909305993690851)
    (13.28856723186122,0.9921135646687698)
    (26.328808951104146,0.9925078864353313)
    }; % ThriftLLM  
    \addplot[only marks, mark size =1pt, mark=oplus,cyan] coordinates {
    (0.565797121451106,0.8497634069400631)
    (4.211140378548891,0.9917192429022083)
    (8.944871451104126,0.9905362776025236)
    (37.1398888012618,0.9925078864353313)
    (74.78353509463713,0.9929022082018928)}; % greedy
    \addplot[only marks, mark size =1pt, mark=triangle,blue] coordinates {(0, 0) (2.86, 0)(7.02, 0.8769161)(7.98, 0.95977918)(7.98, 0.95977918)}; % FrugalGPT
    %\addplot[mark=diamond,cyan] coordinates {(1, 0.8525236593059937) (2, 0.988564668769716)(3, 0.9889589905362776)(4, 0.9893533123028391)(5,0.9909305993690851)}; % Octopus
    % \addplot[only marks, mark size =1pt, mark=diamond,orange] coordinates {(0, 0) (0, 0)(0, 0)(37.775,0.992769)(68.639,0.994)}; % LLM-ensemble (non empty)
    \addplot[only marks, mark size =1pt, mark=diamond,orange] coordinates {(0, 0) (0, 0)(0, 0)(26.782,0.703)(63.903,0.925)}; % LLM-ensemble
    \addplot[only marks, mark size =1pt, mark=+,violet] coordinates {(4, 0.986198) (18.5, 0.9885646)(10.9, 0.9641)}; % LLM-topla
\end{axis}
\end{tikzpicture}
\label{subfiga:sciq}
}\hspace{2mm}
\subfloat[{Hellaswag}]{
\begin{tikzpicture}
\begin{axis}[
    height=\columnwidth/2.5,
    width=\columnwidth/2.2,            
    enlarge x limits=true,
    ymin=0.82, ymax=0.92,
    xmin=0, xmax=100.5,         
    ylabel={{\em Accuracy}},
    xlabel={\em Cost ($\times 10^{-5}$)},
    xlabel style={yshift=0.1cm},
    ylabel style={yshift=-0.1cm},    
    xtick={1,5,10,50,100},
    xticklabels={1,5,10,50,100},
    xmode=log,
    log basis x={10},
    xticklabel style={font=\scriptsize},
    %every axis y label/.style={at={(current axis.north west)},right=12mm,above=0mm},   
    %legend style={at={(0.05,0.95)},anchor=north west,font=\scriptsize},
    %legend columns=3,
    ]
    \addplot[only marks, mark size =1pt, mark=asterisk,red] coordinates {
    (2.823668222740919,0.8579526508836278)
    (3.798212404134714,0.8829609869956653)
    (6.160533344448156,0.8842947649216405)
    (21.502863121040377,0.8942980993664554)
    (38.62696048682907,0.8986328776258753)
    }; % ThriftLLM  
    \addplot[only marks, mark size =1pt, mark=oplus,cyan] coordinates {
    (0.29839613204401476,0.4274758252750917)
    (3.6556410470156624,0.8579526508836278)
    (8.547986162054007,0.8842947649216405)
    (37.18816055351784,0.8946315438479493)
    (54.57482210736922,0.8982994331443814)}; % greedy
    %\addplot[mark=triangle,blue] coordinates {(1, 0.1) (2, 0.2)(3, 0.3)(4, 0.4)(5,0.5)}; % FrugalGPT
    %\addplot[mark=diamond,cyan] coordinates {(1, 0.448149383127709) (2, 0.887295765255085)(3, 0.887295765255085)(4, 0.9023007669223074)(5,0.924974991663888)}; % Octopus
    % \addplot[only marks, mark size =1pt, mark=diamond,orange] coordinates {(0, 0) (0, 0)(0, 0)(41.494,0.8306)(85.298,0.8526)}; % LLM-ensemble (non empty)
    \addplot[only marks, mark size =1pt, mark=diamond,orange] coordinates {(0, 0) (0, 0)(0, 0)(9.311,0.186)(35.325,0.353)}; % LLM-ensemble
     \addplot[only marks, mark size =1pt, mark=+,violet] coordinates {(4, 0.843614) (9, 0.887295)(14.9, 0.8113)}; % LLM-topla
\end{axis}
\end{tikzpicture}
\label{subfiga:hellaswag}
}\hspace{2mm}
\subfloat[{Banking77}]{
\begin{tikzpicture}
\begin{axis}[
    height=\columnwidth/2.5,
    width=\columnwidth/2.2,            
    enlarge x limits=true,
    ymin=0.7, ymax=0.8,
    xmin=0.5, xmax=100.5,         
    ylabel={{\em Accuracy}},
    xlabel={\em Cost ($\times 10^{-5}$)},
    xlabel style={yshift=0.1cm},
    ylabel style={yshift=-0.1cm},    
    xtick={1,5,10,50,100},
    xticklabels={1,5,10,50,100},
    xmode=log,
    log basis x={10},
    xticklabel style={font=\scriptsize},
    %every axis y label/.style={at={(current axis.north west)},right=12mm,above=0mm},   
    %legend style={at={(0.05,0.95)},anchor=north west,font=\scriptsize},
    %legend columns=3,
    ]
    \addplot[only marks, mark size =1pt, mark=asterisk,red] coordinates {
    (0.5752521971723287,0.7279327474207108)
    (2.095943064577756,0.7260221627818112)
    (3.386388995032475,0.7302254489873902)
    (13.371494459304555,0.7363393198318685)
    (13.4589249140237,0.7363393198318685)
    }; % ThriftLLM  
    \addplot[only marks, mark size =1pt, mark=oplus,cyan] coordinates {
    (0.8582737867787458,0.7264042797095911)
    (3.1174235766144444,0.7264042797095911)
    (6.39955177684374,0.7298433320596103)
    (19.707959686664236,0.7363393198318685)
    (20.13988001528479,0.7382499044707681)}; % greedy
    \addplot[only marks, mark size =1pt, mark=triangle,blue] coordinates {(0, 0) (0, 0)(0, 0)(33.259, 0.75888422)(33.259, 0.75888422)}; % FrugalGPT
    %\addplot[mark=diamond,cyan] coordinates {(1, 0.8375) (2, 0.8375)(3, 0.8375)(4, 0.8375)(5,0.8375)}; % Octopus
    % \addplot[only marks, mark size =1pt, mark=diamond,orange] coordinates {(0.927, 0.7536) (1.240, 0.724)(4.334, 0.732)(21.216,0.736)(21.593,0.736)}; % LLM-ensemble(non empty)
    \addplot[only marks, mark size =1pt, mark=diamond,orange] coordinates {(0.584, 0.474) (1.231, 0.718)(4.307, 0.727)(21.216,0.736)(21.593,0.736)}; % LLM-ensemble
     \addplot[only marks, mark size =1pt, mark=+,violet] coordinates {(3.4, 0.7332823) (6, 0.8112)(23.8, 0.7799)}; % LLM-topla
\end{axis}
\end{tikzpicture}
\label{subfiga:headlines}
}
%\vspace{-1mm}%\hspace{2mm}
\end{small}\vspace{-1mm}
\captionof{figure}{Accuracy vs cost for text classification query.} \label{fig:accuracycosttext}%\vspace{-3mm} 
\end{figure*}

\subsection{Performance on Text Classification Query}\label{sec:performance}

The tested methods are evaluated in terms of accuracy scores against LLM usage costs. Figure~\ref{fig:accuracycosttext} presents the results of the accuracy versus costs of the $3$ tested methods except for LLM-Blender on the five datasets.
In particular, FrugalGPT encounters the out-of-memory issue on our machine (24GB GPU memory) on dataset Hellaswag, so its performance is not reported in Figure~\ref{subfiga:hellaswag}. Moreover, FrugalGPT enforces budget constraints based on the {\em expected} training cost and does not respect the budget constraint strictly in a per-query manner, unlike \ours. For a fair comparison, we modify the budget constraint of FrugalGPT on testing queries to align with this per-query approach in our experiments.
LLM-Blender utilizes all $12$ LLM candidates for response collection, subsequently aggregating the most prominent responses to formulate the final solution. Given that LLM-Blender is not budget-aware, it is not appropriate to compare it with budget-constrained methods across different budget scenarios. As such, we report the performance of LLM-Blender and compare it with \ours separately in Table~\ref{tbl:LLMBlender}.

Figure~\ref{fig:accuracycosttext} demonstrates the accuracy scores with the corresponding utilized cost of each method on the $5$ datasets for text classification queries. As shown, \ours consistently outperforms all other baseline models with either superior accuracy at the lowest costs or the highest accuracy with lower costs on all datasets. In particular, it achieves the highest accuracy on $4$ out of the $5$ tested datasets, except on Banking77. {\ours may select several weaker models instead of stronger ones on Banking77. This selection discrepancy arises because the success probabilities of these weaker models are overestimated when queries are from a substantial number of distinct classes.} Compared with \greedy, \ours acquires comparable accuracy scores but consumes notably lower costs. This observation {demonstrates} the effectiveness of adaptive selection in \ours on cost saving without sacrificing the performance. On datasets AGNews, Hellaswag, and Banking77, where the accuracy scores do not approach $1$, \ours exhibits an ability to enhance accuracy further as costs increase. This indicates that \ours effectively harnesses the capabilities of the LLM ensemble and scales efficiently with increased budget allocation. {For small budgets, LLM-Ensemble suffers significant performance degradation because the single top-weighted model exceeds the budget.}
{LLM-Topla performs worse than \ours except on AGNews and Banking77. The performance gains stem from LLM-Topla being fine-tuned individually on each dataset, which incurs substantially higher computational cost.}

Overall, \ours demonstrates a steadily strong performance, outperforming \eat{both \greedy and FrugalGPT}{the baselines}. The analysis reveals a general trend where \ours provides higher accuracy at lower cost levels, indicating its efficiency in utilizing computational resources.

\begin{table}[!t]
\centering
\caption{Accuracy (\%) of \ours and LLM-Blender.} \label{tbl:LLMBlender}\vspace{-1mm} 
\setlength{\tabcolsep}{0.4em}
%\vspace{-2mm}
\small
\resizebox{0.48\textwidth}{!}{%
\begin{tabular} {@{}l|ccccc@{}}
\toprule
{\bf Dataset}  & \multicolumn{1}{c}{Overruling} & \multicolumn{1}{c} {AGNews}  & \multicolumn{1}{c}{SciQ}  &\multicolumn{1}{c}{Hellaswag}  &\multicolumn{1}{c}{Banking77} \\ \midrule 
{\ours} &  95.60   & 86.45    &  99.25 & 89.94 & 73.82\\ \midrule
{LLM-Blender}  &  89.35  & 83.02 & 90.06 & 53.18 & 52.08\\ 
\bottomrule
\end{tabular}}
 \vspace{-2mm}
\end{table}

\spara{Comparison with LLM-Blender} In Table~\ref{tbl:LLMBlender}, we compare the best accuracy scores of \ours across five different budget settings with the accuracy of LLM-Blender, which uses all model outputs as candidates for response selection. Despite this, it can be seen that on all datasets \ours clearly outperforms LLM-Blender by a significant margin. Distinct from the response aggregation mechanism in \ours, LLM-Blender relies on the GenFuser component, a generative model fine-tuned on the T5-like architecture, to generate the final outputs by fusing collected candidate responses with additional query interpretations, which, however, leads to suboptimal quality. These results reveal the superior performance of \ours over LLM-Blender in text classification even with smaller budgets.

\begin{figure*}[!t]
\centering
\begin{small}
\hspace{-1mm}
\begin{tikzpicture}
\begin{customlegend}[legend columns=5,legend style={align=center,draw=none,column sep=1ex, font=\scriptsize},
        legend entries={\ours, \roberta, FrugalGPT, \ditto, GPT-4o}]
        \addlegendimage{mark=asterisk,red}
        \addlegendimage{mark=oplus,cyan}
        \addlegendimage{mark=triangle,blue}
        \addlegendimage{mark=diamond,orange}
        \addlegendimage{mark=+,violet}
        \end{customlegend}
\end{tikzpicture}\vspace{-3mm}

\subfloat[{WDC Products}]{
\begin{tikzpicture}
\begin{axis}[
    height=\columnwidth/2.5,
    width=\columnwidth/2.2,            
    enlarge x limits=true,
    ymin=0.75, ymax=0.9,
    xmin=0.5, xmax=100.5,         
    ylabel={{\em F1 score}},
    xlabel={\em Cost ($\times 10^{-5}$)},
    xlabel style={yshift=0.1cm},
    ylabel style={yshift=-0.1cm},    
    xtick={1,5,10,50,100},
    xticklabels={1,5,10,50,100},
    xmode=log,
    log basis x={10},
    xticklabel style={font=\scriptsize},
    %every axis y label/.style={at={(current axis.north west)},right=12mm,above=0mm},   
    %legend style={at={(0.05,0.95)},anchor=north west,font=\scriptsize},
    %legend columns=3,
    ]
    \addplot[only marks, mark size=1pt, mark=asterisk,red] coordinates {
    (0.74652502017756454, 0.7943171688179822)
    (2.6298805488296988,0.8263838999575926)
    (5.296172316384183,0.8496269706542695)
    (30.586340597255844,0.8837031369548584)
    (57.94180669895068,0.8840656239188623)}; % ThriftLLM  
    \addplot[only marks, mark size =1pt, mark=oplus,cyan] coordinates {(2.04, 0.7753)}; % roberta
    \addplot[only marks, mark size =1pt, mark=diamond,orange] coordinates {(5.2,0.8490)}; % ditto
    \addplot[only marks, mark size =1pt, mark=triangle,blue] coordinates {(0.616, 0.769565217)(3.15, 0.768211921)(6.9, 0.780911063)(48.3058, 0.866019417)(50.2945, 0.854870775)
}; % FrigalGPT
    \addplot[only marks, mark size =1pt, mark=+,violet] coordinates {(33.70661824051653, 0.8767)}; % GPT-4o

\end{axis}
\end{tikzpicture}
\label{subfiga:wdc}
}\hspace{2mm}
\subfloat[{Abt-Buy}]{
\begin{tikzpicture}
\begin{axis}[
    height=\columnwidth/2.5,
    width=\columnwidth/2.2,            
    enlarge x limits=true,
    ymin=0.9, ymax=1.0,
    xmin=0.5, xmax=100.5,         
    ylabel={{\em F1 score}},
    xlabel={\em Cost ($\times 10^{-5}$)},
    xlabel style={yshift=0.1cm},
    ylabel style={yshift=-0.1cm},    
    xtick={1,5,10,50,100},
    xticklabels={1,5,10,50,100},
    xmode=log,
    log basis x={10},
    xticklabel style={font=\scriptsize},
    %every axis y label/.style={at={(current axis.north west)},right=12mm,above=0mm},   
    %legend style={at={(0.05,0.95)},anchor=north west,font=\scriptsize},
    %legend columns=3,
    ]
    \addplot[only marks, mark size =1pt, mark=asterisk,red] coordinates {
    (3.893199170124466,0.9394590265390076)
    (2.2903261410788372,0.929235632246122)
    (4.1900950207468775,0.9449864804685919)
    (30.644734854771836,0.9537074969274887)
    (44.859014522821337,0.9570028473959902)}; % ThriftLLM  
    \addplot[only marks, mark size =1pt, mark=oplus,cyan] coordinates {(4.67,0.9121)}; % roberta
    \addplot[only marks, mark size =1pt, mark=diamond,orange] coordinates {(11.33,0.9131)}; % ditto
    \addplot[only marks, mark size =1pt, mark=triangle,blue] coordinates {(0.68, 0.928746929)(3.36, 0.93398533)(3.36, 0.93398533)(40.3562, 0.939467312)(40.4545, 0.941747573)
}; % FrigalGPT
    \addplot[only marks, mark size =1pt, mark=+,violet] coordinates {(25.19045643153525, 0.9395)}; % GPT-4o
\end{axis}
\end{tikzpicture}
\label{subfiga:abtbuy}
}\hspace{2mm}
\subfloat[{Walmart-Amazon}]{
\begin{tikzpicture}
\begin{axis}[
    height=\columnwidth/2.5,
    width=\columnwidth/2.2,            
    enlarge x limits=true,
    ymin=0.84, ymax=0.95,
    xmin=0, xmax=100.5,         
    ylabel={{\em F1 score}},
    xlabel={\em Cost ($\times 10^{-5}$)},
    xlabel style={yshift=0.1cm},
    ylabel style={yshift=-0.1cm},    
    xtick={1,5,10,50,100},
    xticklabels={1,5,10,50,100},
    xmode=log,
    log basis x={10},
    xticklabel style={font=\scriptsize},
    %every axis y label/.style={at={(current axis.north west)},right=12mm,above=0mm},   
    %legend style={at={(0.05,0.95)},anchor=north west,font=\scriptsize},
    %legend columns=3,
    ]
    \addplot[only marks, mark size =1pt, mark=asterisk,red] coordinates {
    (0.5213130762782895,0.9142498162766991)
    (1.759007124895221,0.9150150633798814)
    (3.966659681475268,0.918312693059767)
    (28.21421374685669,0.9281735410652298)
    (57.49349958088859,0.9285680906941315)
    }; % ThriftLLM  
    \addplot[only marks, mark size =1pt, mark=oplus,cyan] coordinates {(4.67,0.8702)}; % roberta
    \addplot[only marks, mark size =1pt, mark=diamond,orange] coordinates {(11.33,0.8639)}; % ditto
    \addplot[only marks, mark size =1pt, mark=triangle,blue] coordinates {(0.727, 0.864450128)(3.37, 0.791907514)(3.64, 0.787172012)(8.5, 0.838356164)(8.6, 0.844686649)
}; % FrigalGPT
    \addplot[only marks, mark size =1pt, mark=+,violet] coordinates {(34.30301760268235, 0.8665)}; % GPT-4o
\end{axis}
\end{tikzpicture}
\label{subfiga:amazon}
}\hspace{2mm}
\subfloat[{Amazon-Google}]{
\begin{tikzpicture}
\begin{axis}[
    height=\columnwidth/2.5,
    width=\columnwidth/2.2,            
    enlarge x limits=true,
    ymin=0.69, ymax=0.88,
    xmin=0, xmax=100.5,         
    ylabel={{\em F1 score}},
    xlabel={\em Cost ($\times 10^{-5}$)},
    xlabel style={yshift=0.1cm},
    ylabel style={yshift=-0.1cm},    
    xtick={1,5,10,50,100},
    xticklabels={1,5,10,50,100},
    xmode=log,
    log basis x={10},
    xticklabel style={font=\scriptsize},
    %every axis y label/.style={at={(current axis.north west)},right=12mm,above=0mm},   
    %legend style={at={(0.05,0.95)},anchor=north west,font=\scriptsize},
    %legend columns=3,
    ]
    \addplot[only marks, mark size =1pt, mark=asterisk,red] coordinates {
    (0.41773390383047965,0.8274285714285714)
    (2.3381140994295004,0.6974678628202562)
    (3.36254034229829,0.8280233136117013)
    (21.418745721271432,0.8645165202801727)
    (23.81295762021186,0.8649015614392397)
    }; % ThriftLLM  
    \addplot[only marks, mark size =1pt, mark=oplus,cyan] coordinates {(4.86,0.7927)}; % roberta
    \addplot[only marks, mark size =1pt, mark=diamond,orange] coordinates {(11,0.8007)}; % ditto
    \addplot[only marks, mark size =1pt, mark=triangle,blue] coordinates {(0.624, 0.688034188)(2.15, 0.697892272)(2.33, 0.699763593)(2.33, 0.699763593)(2.33, 0.699763593)
}; % FrigalGPT
    \addplot[only marks, mark size =1pt, mark=+,violet] coordinates {(23.19, 0.7356)}; % GPT-4o
\end{axis}
\end{tikzpicture}
\label{subfiga:google}
}\hspace{2mm}
\subfloat[{DBLP-Scholar}]{
\begin{tikzpicture}
\begin{axis}[
    height=\columnwidth/2.5,
    width=\columnwidth/2.2,            
    enlarge x limits=true,
    ymin=0.89, ymax=0.95,
    xmin=0.5, xmax=100.5,         
    ylabel={{\em F1 score}},
    xlabel={\em Cost ($\times 10^{-5}$)},
    xlabel style={yshift=0.1cm},
    ylabel style={yshift=-0.1cm},    
    xtick={1,5,10,50,100},
    xticklabels={1,5,10,50,100},
    xmode=log,
    log basis x={10},
    xticklabel style={font=\scriptsize},
    %every axis y label/.style={at={(current axis.north west)},right=12mm,above=0mm},   
    %legend style={at={(0.05,0.95)},anchor=north west,font=\scriptsize},
    %legend columns=3,
    ]
    \addplot[only marks, mark size =1pt, mark=asterisk,red] coordinates {
    (0.5175240770465483,0.9242097738542632)
    (2.1408816211878053,0.9292063809882177)
    (4.448489566613205,0.9362730986456106)
    (29.15401364365977,0.9391283549450737)
    (56.86117616372392,0.9461198708206688)
    }; % ThriftLLM  
    \addplot[only marks, mark size =1pt, mark=oplus,cyan] coordinates {
    (3.53,0.9388)
    }; % roberta
    \addplot[only marks, mark size =1pt, mark=diamond,orange] coordinates {
    (8.24,0.9431)
    }; % ditto
    \addplot[only marks, mark size =1pt, mark=triangle,blue] coordinates {(0.594, 0.920152091)(2.32, 0.93957115)(2.33, 0.93957115)(33.6537, 0.925490196)(33.7074, 0.92519685)
}; % FrugalGPT
    \addplot[only marks, mark size =1pt, mark=+,violet] coordinates {(33.556179775280864, 0.8976)}; % GPT-4o
\end{axis}
\end{tikzpicture}
\label{subfiga:scholar}
}
%\vspace{-1mm}%\hspace{2mm}
\end{small}\vspace{-1mm}
\captionof{figure}{F1 score vs cost  for entity matching query.} \label{fig:accuracycostentity}%\vspace{-3mm} 
\end{figure*}

\subsection{Performance on Entity Matching Query}\label{sec:entity}

Figure~\ref{fig:accuracycostentity} displays the F1 scores against the utilized costs of the three tested models. \roberta and \ditto, both based on the BERT~\cite{kenton2019bert} architecture, are fine-tuned tailored for each tested dataset. Consistent with the experiment settings in~\cite{PeetersSB25}, we incorporate their reported results for \roberta and \ditto to ensure comparability. However, note that the corresponding utilization costs are not disclosed in~\cite{PeetersSB25}. To address this gap, we estimate the average cost per query by leveraging the reported fine-tuning time and the corresponding AWS pricing for computation time, given that their experiments were conducted on a p3.8xlarge AWS EC2 machine with 4 V100 GPUs (1 GPU per run).

As shown in Figure~\ref{fig:accuracycostentity} for entity matching queries, \ours persistently dominates the \eat{other two} baselines, yielding higher F1 scores but incurring lower costs on the four e-commerce datasets and acquiring the highest F1-score on DBLP-Scholar. In particular, \ours boosts the F1 scores with a notable improvement of $3.51\%$, $4.39\%$, $5.84\%$, $6.42\%$, and $0.30\%$ on the $5$ datasets respectively. Meanwhile, the observed performance pattern, where F1 scores increase with {increasing budgets}, signifies the strength of \ours to maximize the efficiency of allocated budgets thereby enhancing overall performance. This observation demonstrates that \ours aggregates less expensive LLMs in an effective manner, yielding superior performance at reduced costs.

{In general, \ours achieves higher accuracy as the budget increases. When the budget is low, around $1\sim 5\times 10^{-5}$ USD, \ours tends to select $3\sim 5$ weaker but inexpensive models that, when combined, offer improved ensemble performance. As the budget rises to $50\sim 100\times 10^{-5}$ USD, \ours typically selects $1\sim 2$ stronger yet costly models (\eg GPT-4o and Gemini-1.5 Pro), complemented by $5\sim 7$ cheaper ones, to form a more effective ensemble, which enables flexible budget-adaptive selection.}

\subsection{Ablation Study}\label{sec:abla}

\begin{table}
\centering
\caption{Accuracy (\%) across confidence intervals on AGNews.} \label{tbl:interval}%\vspace{-1mm} 
\setlength{\tabcolsep}{0.4em}
%\vspace{-2mm}
\small
%\resizebox{\columnwidth}{!}{%
\begin{tabular} {@{}c|ccccc@{}}
\toprule
{\bf $\alpha$} & 0 & 0.02 & 0.04 & 0.08 & 0.1 \\ \midrule 
{Acc. of $\P_{\textrm{low}}$} & 84.80 & 84.93 & 84.80 & 84.80 & 84.74  \\ \midrule
{Acc. of $\P_{\textrm{up}}$} & 84.80 & 84.80 & 84.87 & 84.73 & 84.80    \\ \bottomrule
\end{tabular}
%}
\vspace{-2mm}
\end{table}

\spara{Confidence interval on approximation guarantees} Let $\alpha$ represent the length of the confidence interval in Section~\ref{sec:interval}, \ie $\alpha = p^\top_l - p^\bot_l$ for $l\in[L]$. 
Specifically, given the current estimated probability $\hat{p}_l$, we set $p^\bot_l=\hat{p}_l-\tfrac{\alpha}{2}$ and $p^\top_l=\min\{\hat{p}_l+\tfrac{\alpha}{2}, 1.0\}$. By feeding the resultant probability sets $\P_{low}=\{p^\bot_1, p^\bot_2, \cdots, p^\bot_L\}$ and $\P_{up}=\{p^\top_1, p^\top_2, \cdots, p^\top_L\}$ to \ours respectively, we record the accuracy scores of the selected LLMs. %Please note that these recorded accuracy scores are calculated using the ground-truth labels of the testing queries and, therefore, based on ground-truth success probabilities. It is worth pointing out that these scores are distinct from the definitions of $\PA_l(\cdot)$ and $\PA_u(\cdot)$. 

We vary $\alpha=\{0,0.02,0.04,0.08,0.1\}$ and conduct this experiment on dataset AGNews with the budget $B=1\times 10^{-5}$. The results are reported in Table~\ref{tbl:interval}. The accuracy score of $84.80\%$ with $\alpha=0$ acts as the base case. Compared with this case, accuracy scores with $\alpha>0$ are either the same or approximately $84.80\%$ with slight variations incurred by the inherent randomness of accuracy estimation in model selection. This observation reveals that \ours is robust to the \ravi{estimation errors} in success probabilities. 

\begin{table}[!t]
\centering
\caption{Accuracy (\%) of \ours vs single LLMs.} \label{tbl:singleLLM}%\vspace{-1mm} 
\setlength{\tabcolsep}{0.4em}
%\vspace{-2mm}
\small
\resizebox{0.48\textwidth}{!}{%
\begin{tabular} {@{}l|ccccc@{}}
\toprule
{\bf Dataset}  & \multicolumn{1}{c}{Overruling} & \multicolumn{1}{c} {AGNews}  & \multicolumn{1}{c}{SciQ} &\multicolumn{1}{c}{Hellaswag} & \multicolumn{1}{c}{Banking77} \\ \midrule 
{\ours}   &  \bf{95.60}   & 86.45   &  \bf{99.25} & 89.86  & \underline{73.82} \\ \midrule
{GPT-4o}  &  94.68   &  \underline{86.71}   & \underline{99.17}  &  \bf{93.29} & \bf{75.05} \\  \midrule
{Gemini-1.5 Pro}  &  \underline{95.14}   &  \bf{89.01}  &  96.61 &  \underline{90.96} & 25.91 \\ \midrule
{Phi-3-medium}  &  \underline{95.14}   & 84.21   &  98.50 & 88.73  & 59.84 \\ \midrule
{Llama-3 70B}  &   94.68  & 81.31   &  98.86 & 86.53 &  69.66 \\ \midrule
{Mixtral-8x7B}  &  94.90   &  79.34 & 96.29  & N.A. &  N.A. \\ 
\bottomrule
\end{tabular}}
 %\vspace{-1mm}
\end{table}

\begin{table}
\centering
\caption{Accuracy (\%) across historical data on Overruling.} \label{tbl:varhistorical}%\vspace{-1mm} 
\setlength{\tabcolsep}{0.4em}
\renewcommand{\arraystretch}{0.3}
%\vspace{-2mm}
\small
\resizebox{0.48\textwidth}{!}{%
\begin{tabular} {@{}c|ccccc@{}}
\toprule
{\bf Budget} & $20\%$ & $40\%$ & $60\%$ & $80\%$ & Original \\ \midrule 
{$1.0 \times 10^{-5}$} & 95.37 & 94.91 & 96.06 & 94.90 & 95.14  \\ \midrule
{$5.0 \times 10^{-5}$} & 95.37 & 95.37 & 95.60 & 95.13 & 95.37    \\ \midrule
{$10.0 \times 10^{-5}$} & 95.37 & 95.37 & 95.37 & 95.60 & 95.37    \\ \midrule
{$50.0 \times 10^{-5}$} & 95.60 & 95.37 & 95.37 & 95.13 & 95.37    \\ \midrule
{$100.0 \times 10^{-5}$} & 95.60 & 95.37 & 95.37 & 95.60 & 95.60    \\ 
\bottomrule
\end{tabular}
}
 %\vspace{-1mm}
\end{table}

\spara{\ours vs Single LLM} To further validate the advantage of the LLM ensemble over an individual LLM, we compare the accuracy scores obtained by \ours with single models in Table~\ref{tbl:LLMAPI} on the $5$ tested datasets for text classification queries. For a convincing comparison, we select the most powerful {and most expensive} LLMs provided by each company, including GPT-4o from OpenAI, Gemini-1.5 Pro from Google, Phi-3-medium from Microsoft, Llama-3 70B from Meta, and Mixtral-8x7B from Mistral AI. The results are summarized in Table~\ref{tbl:singleLLM}. For clarity, we highlight the highest accuracy score in bold and underline the second-highest score for each dataset. As displayed, \ours achieves the best on $2$ out of $5$ datasets. \eat{and second best on one of the remaining datasets.} On the other $3$ datasets, \ours either achieves or closely approaches the second-highest accuracy scores with negligible gaps.
This evidence not only underscores the superior performance of \ours as an ensemble model across diverse topics but also suggests that individual powerful models do not consistently offer advantages across all domains.

{
\def\st{0.5}
\def\gap{0.55}
\begin{figure}
\centering
\begin{small}
\begin{tikzpicture}
    \begin{customlegend}[        
        legend columns=3,
        area legend,
        legend style={align=center,draw=none,column sep=1ex, font=\footnotesize},
        legend entries={{\ours},{\surgreedy},{GPT-4o}}]
        \addlegendimage{,fill=red}
        \addlegendimage{,fill=blue}
        \addlegendimage{,fill=cyan}
    \end{customlegend}
\end{tikzpicture}\vspace{-5mm}

\subfloat[{Accuracy}]{
\begin{tikzpicture}
\begin{axis}[
    ybar=2pt,
    height=\columnwidth/2.5,
    width=\columnwidth/1.7,            
    bar width=0.12cm,
    enlarge x limits=true,
    ymin=0.9, ymax=1,
    xmin=0.35, xmax=\st+4*\gap,         
    xlabel={\em Budget ($\times 10^{-5}$)},
    ylabel={{\em Accuracy}},
    xtick={{\st},{\st+\gap},{\st+2*\gap},{\st+3*\gap},{\st+4*\gap},{\st+5*\gap},{\st+6*\gap}},
    xticklabels={1,5,10,50,100}, 
    xticklabel style={font=\scriptsize},
    yticklabel style={font=\small}, 
    every axis y label/.style={at={(current axis.north west)},right=10mm,above=0mm},   
    ]
    {\addplot[fill=red] coordinates {
    (\st, 0.9421296296296297) (\st+\gap, 0.9513888888888888)(\st+2*\gap, 0.9537037037037037)(\st+3*\gap, 0.9537037037037037)(\st+4*\gap,0.9537037037037037)};}
    {\def\st{0.45}
    \addplot[fill=blue] coordinates {(\st, 0.9421296296296297) (\st+\gap, 0.9513888888888888)(\st+2*\gap, 0.9537037037037037)(\st+3*\gap, 0.9537037037037037)(\st+4*\gap,0.9537037037037037)};}
    {\def\st{0.4}
    \addplot[fill=cyan] coordinates {(\st, 0.9467592592592593) (\st+\gap, 0.9467592592592593)(\st+2*\gap, 0.9467592592592593)(\st+3*\gap, 0.9467592592592593)(\st+4*\gap,0.9467592592592593)};}
    %\legend{\ours,\NMF,\ConTinEst}
\end{axis}
\end{tikzpicture}
\label{subfig:accuracy}
}
\subfloat[{Cost}]{
\begin{tikzpicture}
\begin{axis}[
    ybar=2pt,
    height=\columnwidth/2.5,
    width=\columnwidth/1.7,            
    bar width=0.12cm,
    enlarge x limits=true,
    ymin=0, ymax=30,
    %ymode=log,
    xmin=0.35, xmax=\st+4*\gap,         
    xlabel={\em Budget ($\times 10^{-5}$)},
    ylabel={{\em Cost Proportion}},
    xtick={{\st},{\st+\gap},{\st+2*\gap},{\st+3*\gap},{\st+4*\gap}},
    xticklabels={1,5,10,50,100}, 
    xticklabel style={font=\scriptsize},
    yticklabel style={font=\small}, 
    every axis y label/.style={at={(current axis.north west)},right=10mm,above=0mm},   
    ]
    {\addplot[fill=red] coordinates {
    (\st, 6.352/10) (\st+\gap, 2.6830208/5)(\st+2*\gap, 5.03473032407407/10)(\st+3*\gap, 0.000307842/0.0005)(\st+4*\gap,0.440937326)};}
    {\def\st{0.45}
    \addplot[fill=blue] coordinates {(\st, 8.9121/10) (\st+\gap, 4.72367/5)(\st+2*\gap, 8.44258/10)(\st+3*\gap, 0.0003532392/0.0005)(\st+4*\gap,0.5128175)};} 
    {\def\st{0.4}
    \addplot[fill=cyan] coordinates {(\st, 27.643518518518533) (\st+\gap, 27.643518518518533/5)(\st+2*\gap, 27.643518518518533/10)(\st+3*\gap, 0.00027643518518518533/0.0005)(\st+4*\gap,0.00027643518518518533/0.001)};}
\end{axis}
\end{tikzpicture}
\label{subfig:cost}
}
\end{small}\vspace{-2mm}
\captionof{figure}{{Accuracy vs cost on Overruling.}} \label{fig:adapt}\vspace{-2mm}
\end{figure}}

\spara{Adaptive selection} We have proved {(see Proposition~\ref{prop:termination})} that the subset of LLMs selected by \ours makes the same prediction as that selected by \surgreedy, while utilizing  lower budgets. To verify this point and quantify the saved costs, we evaluate \ours and \surgreedy on dataset Overruling by following the budget $B$ setting. {For comparison, we also include one strong single model GPT-4o as a baseline}. The results are displayed in Figure~\ref{fig:adapt}. {As shown in Figure~\ref{subfig:accuracy}, \ours and \surgreedy achieve exactly the same accuracy scores, consistent with our result in Proposition~\ref{prop:termination}.}  Figure~\ref{subfig:cost} presents the comparison between their cost proportions relative to the given budgets. {It is {worth noting} that \ours achieves a saving of  $\sim 10\%\mbox{-}40\%$ of the allowed budget compared to \surgreedy.} Furthermore, as the budget decreases, the cost savings achieved by \ours compared to \surgreedy become more noticeable. {Both \ours and \surgreedy outperform GPT-4o when the budget is at least $5\times 10^{-5}$ USD per query. Overall, GPT-4o requires up to $\sim$30$\times$ higher cost to achieve only a marginal improvement over \ours and \surgreedy, as shown in Figure~\ref{subfig:cost}.}

\spara{Sensitivity to size of historical data} In text classification, $80\%$ of the dataset is used as historical data for {success} probability estimation. To evaluate the sensitivity of \ours to the size of historical data, we select $\{20\%, 40\%, 60\%, 80\%\}$ of the original historical data on Overruling respectively for probability estimation, and then evaluate the performance of \ours on test queries by varying the budget $B=\{1.0, 5.0, 10, 50, 100\}\times 10^{-5}$ (see  Table~\ref{tbl:varhistorical}). The performance of \ours is stable and robust relative to the proportion of available historical data. The stability is further enhanced with increased budget allocations. These results imply that \ours consistently performs well {across a wide range of sizes of available historical data.} 
 
\section{Conclusion}\label{sec:conclusion}

{We investigate the problem of finding an LLM ensemble under budget constraints for optimal query performance, with a focus on classification queries. We formalize this problem as the \problem problem. To solve this problem, we design a new aggregation scheme for combining individual LLM responses and devise a notion of {correctness probability} to measure the aggregation quality. We prove that {correctness probability} is non-decreasing and non-submodular. Despite this, we develop \ours, a surrogate greedy algorithm that offers an instance dependent approximation guarantee. We evaluate \ours on several real-world datasets on text classification and entity matching queries. Our experiments show that it achieves state-of-the-art performance while utilizing a relatively small  budget {compared to the baselines tested.} Extensions of  \ours to regression and generation tasks are intriguing directions for future work. }

\clearpage
\appendix
\section{Appendix: Proofs}\label{sec:app} \label{app:proofs}

\sktch{
\begin{proof}[Proof of Proposition~\ref{prob:accuracy}]
Consider a set of LLMs $\S$ and a query class $\Q$ with labels from the set $\C=\{C_1, \cdots, C_K\}$. Let $q$ be a random query uniformly sampled from $\Q$. According to the definition, the correctness probability of $\S$ %on query class $\Q$ 
is $\PA_\P(\S)=\sum_{\phi_\S\in \Omega^T_\S}\Pr[\phi_\S]$ where $\Omega^T_\S$ is the set of observations with correct prediction. For a specific observation $\phi_\S\in \Omega^T_\S$, $\S^T$ is the set of models in $\S$ predicting the correct class, and $\S^F=\S\setminus \S^T$ is the remaining LLMs predicting one of the $K-1$ incorrect classes. According to Equation~\eqref{eqn:observeprob}, we have $\PA_\P(\S)=\sum_{\phi_\S\in \Omega^T_\S}\prod_{l_i\in \S^T}p_i\prod_{l_j \in \S^F}\tfrac{1-p_j}{K-1}$. 

\ravi{Suppose the ground-truth class is $C_q$. The sets $\Omega^T_\S$, $\S^T$, and $\S^F$ clearly depend on the class $C_q$. Making this dependence explicit, let us denote those sets as $\Omega^T_\S(C_q)$, $\S^T(C_q)$, and $\S^F(C_q)$ respectively. Let us write the correctness probability of $\S$ on a random query $q\in \Q$, given that the ground-truth class of $q$ is $C_q$, as $\PA_\P(\S\mid C_q)$.} For a different ground-truth class $C^\prime_q \neq C_q$, we prove that $\PA_\P(\S \mid C_q) = \PA_\P(\S \mid C^\prime_q)$, \ie 
\begin{align}
&\textstyle \sum_{\phi_\S\in \Omega^T_\S(C_q)}\prod_{l_i\in \S^T(C_q)}p_i\prod_{l_j \in \S^F(C_q)}\tfrac{1-p_j}{K-1} \notag\\
= &\textstyle \sum_{\phi_\S\in \Omega^T_\S(C^\prime_q)}\prod_{l_i\in \S^T(C^\prime_q)}p_i\prod_{l_j \in \S^F(C^\prime_q)}\tfrac{1-p_j}{K-1}. \label{eq:gt-label-ind}
\end{align}
\ravi{Note that there exists a one-to-one mapping between $\Omega^T_\S(C_q) $ and $ \Omega^T_\S(C^\prime_q)$ such that for every observation $\phi_\S\in \Omega^T_\S(C_q)$, there is a distinct corresponding observation $\phi^\prime_\S \in \Omega^T_\S(C^\prime_q)$: $\Pr[\phi_\S] = \Pr[\phi^\prime_\S]$.  Specifically, for any observation $\phi_x \in \Omega^T_\S(C_q) \setminus \Omega^T_\S(C^\prime_q)$, we can always find an observation $\phi_y \in \Omega^T_\S(C^\prime_q) \setminus \Omega^T_\S(C_q)$ such that $\Pr[\phi_x(C_q)]=\Pr[\phi_y(C^\prime_q)]$. As a consequence,  Equation~\eqref{eq:gt-label-ind} holds true and hence the correctness probability $\PA_\P(\S\mid C_q)$ is independent of the underlying ground-truth label $C_q$, for a randomly chosen query $q\in\Q$. On the other hand, the correctness probability $\PA(\S)$ is determined by the {set of success probabilities} $\P$ on query class $\Q$. As long as $\P$ is fixed, varying $C_q$ of a random query $q \in \Q$ does not affect the value $\PA_\P(\S)$ on $\Q$.}     
\end{proof}
}

\arxiv{
\begin{proofsketch}[Proof Sketch of Lemma~\ref{lem:nondecreasing}]
Part (i) is trivial. For part (ii), consider a random query $q\in \Q $ with ground-truth class $C_q$, any LLM set $\S \subset \L$, and model $l\in \L\setminus \S$. Let $\S^\prime := \S\cup\{l\}$. We can prove that $\PA(\S^\prime)\ge \PA(\S)$. Part (ii) will follow from this.  
\end{proofsketch}}
%%%%%%%%%%%%%%%%%% skipping full proof 
\sktch{
\begin{proof}[Proof of Lemma~\ref{lem:nondecreasing}]
For part (i), note that as the success probabilities of models increase, it is trivial that the prediction accuracy is non-decreasing. We focus on part (ii) in what follows. 

Consider a random query $q\in \Q $ with ground-truth class $C_q$, any LLM set $\S \subset \L$, and model $l\in \L\setminus \S$. Let $\S^\prime := \S\cup\{l\}$. We prove that $\PA(\S^\prime)\ge \PA(\S)$. Part (ii) will follow from this. 

Let $\Omega$ and $\Omega^\prime$ be the observation spaces of $\S$ and $\S^\prime$ on query class $\Q$, respectively. Let $\Omega_T$ (resp.\ $\Omega^\prime_T$) be the subspace containing observations with the correct prediction {$C_q$} and $\Omega_F=\Omega\setminus\Omega_T$ (resp.\ $\Omega^\prime_F = \Omega^\prime \setminus \Omega^\prime_T$) be the remaining subspace with incorrect predictions for $\S$ (resp.\ $\S^\prime$). When invoking the additional LLM $l$ on $\Q$, the observation space $\Omega$ is converted into $\Omega^\prime$: each observation $\phi_{\S} \in \Omega$ is uniquely mapped to $K$ distinct observations $\phi_{\S^\prime} \in \Omega^\prime$ with $\phi_{\S^\prime}=\phi_\S\oplus R(l)$, where $K$ is the number of classes, $\oplus$ is the concatenation operation, and response $R(l)\in \{C_1, C_2, \cdots, C_K\}$. Note that $Pr[\phi_{{\S^\prime}}]=\sum_{R(l)\in \{C_1, C_2, \cdots, C_K\}}Pr[\phi_\S\oplus R(l)]$. 

Consider the scenarios involving $\phi_{\S}$ and $\phi_{\S^\prime}$: (1) observation $\phi_{\S} \in \Omega_F$ generates observations $\phi_{\S^\prime} \in \Omega^\prime_T$, and (2) observation $\phi_{\S} \in \Omega_T$ generates observations $\phi_{\S^\prime} \in \Omega^\prime_F$. We prove the probability of scenario (1) is no smaller than the probability of scenario (2). Notice that the other two scenarios, where observations in $\Omega_T$ (resp. $\Omega_F$) generate those in $\Omega^\prime_T$ (resp. in $\Omega^\prime_F$), do not contribute to changes in the prediction accuracy between  $\PA_\P(\S)$ and $\PA_\P(\S')$.

Let $\phi^\prime_x \in \Omega^\prime_F$ be an observation generated from $\phi_x \in \Omega_T$ corresponding to scenario (2). Suppose $R(l)=C_k$, so  $\phi^\prime_x=\phi_x \oplus C_k$. Since $\phi_x \in \Omega_T$ and $\phi^\prime_x \in \Omega^\prime_F$, we know $H_1(\phi_x)=h(C_q\mid \phi_x)$ and $H_1(\phi^\prime_x)=h(C_k\mid \phi^\prime_x)$ where $C_q \neq C_k$. This fact implies that the ranking of belief values of the $K$ classes is altered after applying the additional model $l$. Therefore, we can infer the inequality 
\begin{equation}\label{eqn:TtoF}
H_1(\phi^\prime_x) = h(C_k\mid \phi^\prime_x) = h(C_k\mid \phi_x)\tfrac{p_l(K-1)}{1-p_l} > h(C_q\mid \phi_x),
\end{equation}
where $p_l$ is the success probability of model $l$. 

We can derive an observation pair corresponding to scenario (1) from the observation pair above. To start with, observe the inherent symmetry of the observation space. To explain, for observation space $\Omega$, there exists an observation pair $(\phi, \phi^\circ)$ in $\Omega$ such that an arbitrary model outputs $C_q$ in $\phi$ if and only if it outputs $C_k$ in $\phi^\circ$ and vice versa, and for models outputting a class outside $\{C_q, C_k\}$, the two observations coincide. Denote by $\S(C_i\mid \phi)$ the set of models which predict the class $C_i$ as per the observation $\phi$. By virtue of the symmetry property, there exists an observation $\phi_y\in \Omega_F$ symmetric to $\phi_x$ from $\Omega_T$ such that  $\S(C_k\mid \phi_y)=\S(C_q\mid \phi_x)$, $\S(C_q\mid \phi_y)=\S(C_k\mid \phi_x)$, and the remaining LLMs $\S_r=\S\setminus (\S(C_q\mid \phi_x) \cup \S(C_k\mid \phi_x))$ make the same predictions {in both $\phi_x$ and $\phi_y$}. Therefore, it holds that $H_1(\phi_y)=h(C_k\mid \phi_y)$. In such a case, consider the case $R(l)=C_q$ and the corresponding $\phi^\prime_y=\phi_y \oplus C_q$. According to Equation~\eqref{eqn:TtoF}, we have
\begin{equation}\label{eqn:FtoT}
h(C_q\mid \phi_y)\tfrac{p_l(K-1)}{1-p_l} = h(C_k\mid \phi_x)\tfrac{p_l(K-1)}{1-p_l} > h(C_q\mid \phi_x) = h(C_k\mid \phi_y).
\end{equation}
As a result, $H_1(\phi^\prime_y)=h(C_q\mid \phi^\prime_y)=h(C_q\mid \phi_y)\tfrac{p_l(K-1)}{1-p_l}$, \ie $\phi^\prime_y \in \Omega^\prime_T$, and the observation pair $(\phi_y, \phi\prime_y)$ corresponds to scenario (1).

We now prove that the probability of scenario (1) is no smaller than that of scenario (2), \ie accuracy gain from scenario (1) should be no less than the accuracy loss due to scenario (2). In particular, for $Pr[\phi^\prime_y]$, we have
\begin{align*}
Pr[\phi^\prime_y] = & \prod_{l_i\in \S(C_q\mid \phi^\prime_y)}p_i\prod_{l_i\in \S(C_k\mid \phi^\prime_y)}\tfrac{1-p_i}{K-1}\prod_{l_i\in \S_r}\tfrac{1-p_i}{K-1} \\
= & \prod_{l_i\in \S(C_q\mid \phi_y)\cup \{ l \}}p_i\prod_{l_i\in \S(C_k\mid \phi_y)}\tfrac{1-p_i}{K-1}\prod_{l_i\in \S_r}\tfrac{1-p_i}{K-1} \\
= & \prod_{l_i\in \S(C_k\mid \phi_x)\cup \{\l\}}p_i\prod_{l_i\in \S(C_q\mid \phi_x)}\tfrac{1-p_i}{K-1}\prod_{l_i\in \S_r}\tfrac{1-p_i}{K-1}
\end{align*}
Similarly, for $Pr[\phi^\prime_x]$, we have \[Pr[\phi^\prime_x]=\prod_{l_i\in \S(C_q\mid \phi_x)}p_i\prod_{l_i\in \S(C_k\mid \phi_x)\cup \{l\}}\tfrac{1-p_i}{K-1}\prod_{l_i\in \S_r}\tfrac{1-p_i}{K-1}.\] When calculating the ratio of $\tfrac{Pr[\phi^\prime_y]}{Pr[\phi^\prime_x]}$, it holds that 
\begin{align*}
\tfrac{Pr[\phi^\prime_y]}{Pr[\phi^\prime_x]} = & \tfrac{\prod_{l_i\in \S(C_k\mid \phi_x)\cup \{\l\}}p_i\prod_{l_i\in \S(C_q\mid \phi_x)}\tfrac{1-p_i}{K-1}}{\prod_{l_i\in \S(C_q\mid \phi_x)}p_i\prod_{l_i\in \S(C_k\mid \phi_x)\cup \{l\}}\tfrac{1-p_i}{K-1}} \\
= & \tfrac{\prod_{l_i\in \S(C_k\mid \phi_x)\cup \{\l\}}p_i}{\prod_{l_i\in \S(C_k\mid \phi_x)\cup \{l\}}\tfrac{1-p_i}{K-1}} \tfrac{\prod_{l_i\in \S(C_q\mid \phi_x)}\tfrac{1-p_i}{K-1}}{\prod_{l_i\in \S(C_q\mid \phi_x)}p_i}\\
= & h(C_k\mid \phi_x)\tfrac{p_l(K-1)}{1-p_l} \tfrac{1}{h(C_q\mid \phi_x)} 
> 1.
\end{align*}
The last inequality follows from  Equation~\eqref{eqn:FtoT}. Let $\Omega^\prime_x$ and $\Omega^\prime_y$ be the set of such observations $\phi^\prime_x$ and $\phi^\prime_y$ respectively. Therefore $\PA(\S^\prime)-\PA(\S)=Pr[\Omega^\prime_y]-Pr[\Omega^\prime_x]$. As $Pr[\Omega^\prime_y]> Pr[\Omega^\prime_x]$, we have $\PA(\S^\prime)>\PA(\S)$. In particular, $\PA(\S^\prime)=\PA(\S)$ holds if both $\Omega^\prime_x$ and $\Omega^\prime_y$ are empty sets, which completes the proof. 
\end{proof}
} 
%%%%%%%%%%%%%%%%%%%%%%%%%%%%%%%%% 

%\begin{proofsketch}[Proof Sketch of Lemma~\ref{lem:nonsubmodular}]
\begin{proof}[Proof of Lemma~\ref{lem:nonsubmodular}]
We construct a counterexample to demonstrate that the function $\PA(\S)$ is not submodular.
Consider sets $\S=\{l_1\}$, $\mathcal{T}=\{l_1, l_2\}$, and a LLM $l_3$. W.l.o.g., we assume their success probabilities follow  the partial ranking of $p_1>p_2$, $p_1>p_3$, and $\tfrac{p_2 (K-1)}{1-p_2} \tfrac{p_3 (K-1)}{1-p_3} > \tfrac{p_1 (K-1)}{1-p_1}$, \ie $\tfrac{p_2p_3(K-1)}{(1-p_2)(1-p_3)}>\tfrac{p_1}{1-p_1}$. As $\tfrac{p}{1-p}\in (0,\infty)$ for $p\in (0,1)$ and $\tfrac{p_1}{1-p_1} > \tfrac{p_2}{1-p_2}$ and $\tfrac{p_1}{1-p_1} > \tfrac{p_3}{1-p_3}$ hold simultaneously, such $p_1, p_2, p_3$ always exist when fixing $K$. 
If set function $\PA(\cdot)$ is submodular, it should satisfy 
\begin{equation}\label{eqn:submodular}
\PA(\S\cup \{l_3\})-\PA(\S) \ge \PA(\mathcal{T}\cup \{l_3\})-\PA(\mathcal{T}).
\end{equation}

According to Proposition~\ref{pro:2models}, we have $\PA(\S)=\PA(\S\cup \{l_3\})=\PA(\T)=p_1$. For $\PA(\mathcal{T}\cup \{l_3\})$, since $p_1>p_2$, $p_1>p_3$, and $\tfrac{p_2 (K-1)}{1-p_2} \tfrac{p_3 (K-1)}{1-p_3} > \tfrac{p_1 (K-1)}{1-p_1}$ hold, the prediction accuracy $\PA(\mathcal{T}\cup \{l_3\})$ is the {total} probability of two cases: (i) $l_1$ makes the correct prediction while $\{l_2,l_3\}$ fail to concur on the same incorrect prediction, and (ii) $l_1$ makes the incorrect prediction while $l_2$ and $l_3$ both predict correctly. Therefore, $\PA(\mathcal{T}\cup \{l_3\})$ is calculated as $\PA(\mathcal{T}\cup \{l_3\}) = p_1-p_1(1-p_2)\tfrac{1-p_3}{K-1}+(1-p_1)p_2p_3$, 
\arxiv{where the two terms in the sum correspond to the two cases. It can be shown from this that $\PA(\mathcal{T}\cup \{l_3\})> p_1 = \PA(\mathcal{T})$,  violating Equation~\eqref{eqn:submodular}. } 
\sktch{where the factor $p_1-p_1(1-p_2)\tfrac{1-p_3}{K-1}$ is the probability of case (i) and the factor $(1-p_1)p_2p_3$ is the probability of case (ii).} 
\sktch{
As $\tfrac{p_2 (K-1)}{1-p_2} \tfrac{p_3 (K-1)}{1-p_3} > \tfrac{p_1 (K-1)}{1-p_1}$, we have $(1-p_1)p_2p_3-p_1(1-p_2)\tfrac{1-p_3}{K-1} >0$. Therefore $\PA(\mathcal{T}\cup \{l_3\})> p_1 = \PA(\mathcal{T})$. Hence Equation~\eqref{eqn:submodular} does not hold, which completes the proof.
} 
%\end{proofsketch}
\end{proof}

\begin{proof}[Proof of Theorem~\ref{thrm:surapproximation}]
\citet{KhullerMN99} prove that the modified greedy strategy yields a solution with a $(1-\tfrac{1}{\sqrt{\e}})$-approximation guarantee, \ie $\max\{\gamma(\S_2), p^\ast\} \ge (1-\tfrac{1}{\sqrt{\e}})\gamma(\S^\circ_\gamma)$ where $\S^\circ_\gamma$ is the optimal solution for the budgeted submodular maximization with set function $\gamma(\cdot)$. Consequently, we have 
\begin{align*}
& \tfrac{\PA(\S^\ast)}{\PA(\S^\circ)}    
\ge  \tfrac{\max\{\PA(S_1),\ \PA(S_2), \ p^\ast\}}{\gamma(\S^\circ)} 
\ge \tfrac{\max\{\PA(S_1),\ \PA(S_2), \ p^\ast\}}{\gamma(\S^\circ_\gamma)} = \\
&  \tfrac{\max\{\PA(S_1),\ \PA(S_2), \ p^\ast\}}{\max\{\gamma(\S_2), p^\ast\}}\tfrac{\max\{\gamma(\S_2), p^\ast\}}{\gamma(\S^\circ_\gamma)}
\ge \tfrac{\max\{\PA(S_1),\ \PA(S_2), \ p^\ast\}}{\max\{\gamma(\S_2),\ p^\ast\}}(1-\tfrac{1}{\sqrt{\e}}),
\end{align*}
which completes the proof.
\end{proof}

\begin{proof}[Proof of Proposition~\ref{prop:termination}]
Let $C_\ast$ be the class with $H_1(\phi_\S)$, which {implies}  $h(C_\ast\mid \phi_\S)=\textstyle\prod_{l_i\in \S(C_\ast)}\tfrac{p_i(K-1)}{1-p_i}$. The potential belief $F(\mathcal{T}^\ast)=\prod_{l_i\in \mathcal{T}^\ast}\tfrac{p_i(K-1)}{1-p_i}$ is the highest possible belief that the remaining LLMs in $\mathcal{T}^\ast$ can contribute to any class. If $F(\mathcal{T}^\ast)H_2(\phi_\S) \le H_1(\phi_\S)$ holds, we have \[F(\mathcal{T}^\ast)H_K(\phi_\S) \le \cdots \le F(\mathcal{T}^\ast)H_3(\phi_\S)\le F(\mathcal{T}^\ast)H_2(\phi_\S) \le H_1(\phi_\S).\] This inequality implies that the remaining models in $\mathcal{T}^\ast$ are not able to contribute the belief to any class except $C_\ast$ so as to achieve a belief higher than $H_1(\phi_\S)$. Therefore, Proposition~\ref{prop:termination} holds.
\end{proof}

\begin{proof}[Proof of Theorem~\ref{thrm:appro}]
By Lemma~\ref{lem:montecarlo}, the prediction accuracy $\PA(\S)$ of each inspected $\S\subseteq \L$ is estimated within error $\tfrac{\epsilon p^\ast}{2}$ with at least $1-\frac{\delta}{L^2}$ probability. Let $\S^\ast$ be the set returned from \surgreedy with $\theta$ Monte Carlo simulations of $\PA(\cdot)$. Therefore, for $\S^\ast= \arg\max\{p^\ast,\tilde{\PA}(\S_1), \tilde{\PA}(\S_2)\}$, it holds that $|\PA(\S^\ast)-\tilde{\PA}(\S^\ast)| \le \tfrac{\epsilon p^\ast}{2}$ with high probability. Since the set function $\gamma(\cdot)$ can be exactly computed in linear time, $\max\{\gamma(\S_2), p^\ast\} \ge (1-\tfrac{1}{\sqrt{\e}})\gamma(\S^\circ_\gamma)$ holds without involving estimation errors. As a consequence, we have 
\begin{align*}
& \tfrac{\PA(\S^\ast)}{\PA(\S^\circ)} \ge \tfrac{\tilde{\PA}(\S^\ast) - \epsilon p^\ast/2}{\PA(\S^\circ)}    
\ge  \tfrac{\max\{\PA(S_1),\ \PA(S_2), \ p^\ast\}-\epsilon p^\ast}{\gamma(\S^\circ)} \ge \\
&  \tfrac{\max\{\PA(S_1),\ \PA(S_2), \ p^\ast\}-\epsilon p^\ast}{\gamma(\S^\circ_\gamma)} =  \tfrac{\max\{\PA(S_1),\ \PA(S_2), \ p^\ast\} -\epsilon p^\ast}{\max\{\gamma(\S_2), p^\ast\}}\tfrac{\max\{\gamma(\S_2), p^\ast\}}{\gamma(\S^\circ_\gamma)} \ge\\
&  \tfrac{\max\{\PA(S_1),\ \PA(S_2), \ p^\ast\}-\epsilon p^\ast}{\max\{\gamma(\S_2),\ p^\ast\}}(1-\tfrac{1}{\sqrt{\e}})  \ge (\tfrac{\max\{\PA(S_1),\ \PA(S_2), \ p^\ast\}}{\max\{\gamma(\S_2),\ p^\ast\}}-\epsilon)(1-\tfrac{1}{\sqrt{\e}}).
\end{align*}
Considering that at most $L^2$ possible subsets are checked in \greedy, the failure probability is bounded by union bound $\frac{\delta}{L^2}\cdot L^2=\delta$. Therefore, it holds that \[\Pr\left[\PA(\S^\ast) \ge (\tfrac{\max\{\PA(S_1),\ \PA(S_2), \ p^\ast\}}{\max\{\gamma(\S_2),\ p^\ast\}}-\epsilon)(1-\tfrac{1}{\sqrt{\e}})\cdot \PA(\S^\circ) \right] \ge 1-\delta,\] which completes the proof.
\end{proof}

\begin{proof}[Proof of Theorem~\ref{thrm:approinterval}]
By taking $\P_{\textrm{low}}$, $\hat{\P}$, and $\P_{\textrm{up}}$ as inputs to \ours, let $\PA_l(\S^\circ_l)$, $\PA(\S^\circ)$, and $\PA_u(\S^\circ_u)$ be the accuracy scores of the corresponding optimal sets respectively, and $\PA_l(\S^\ast_l)$, $\PA(\S^\ast)$, and $\PA_u(\S^\ast_u)$ be the accuracy scores of the selected sets for $\P_{\textrm{low}}$, $\hat{\P}$, and $\P_{\textrm{up}}$ respectively. According to Theorem~\ref{thrm:appro}, we have \[\Pr\left[\tfrac{\PA(\S^\ast)}{\PA(\S^\circ)} \ge (\tfrac{\max\{\PA(S_1),\ \PA(S_2), \ p^\ast\}}{\max\{\gamma(\S_2),\ p^\ast\}}-\epsilon)(1-\tfrac{1}{\sqrt{\e}}) \right] \ge 1-\delta.\] 
\sktch{Similarly, for $\P_{\textrm{up}}$, we have }
\[\Pr\left[\tfrac{\PA_u(\S^\ast_u)}{\PA_u(\S^\circ_u)} \ge (\tfrac{\max\{\PA_u(S_{u1}),\ \PA_u(S_{u2}), \ p^\ast_u\}}{\max\{\gamma_u(\S_{u2}),\ p^\ast_u\}}-\epsilon)(1-\tfrac{1}{\sqrt{\e}}) \right] \ge 1-\delta,\]
where $\gamma_u(\cdot)$ is the surrogate set function, $S_{u1}$ and $S_{u2}$ are selected by \surgreedy on $\P_{\textrm{up}}$, respectively. 
\sktch{As $p^\bot_l \le p_l \le p^\top_l$ for $l\in[L]$ holds w.p. $\geq 1-\delta_l$, according to the non-decreasing property of $\PA(\cdot)$ (Lemma~\ref{lem:nondecreasing}), we have $ \PA(\S^\circ) \le \PA_u(\S^\circ_u)$ and $\PA_l(\S^\ast_l)\le \PA(\S^\ast)$. } Therefore, we have
\[\tfrac{\PA(\S^\ast)}{\PA(\S^\circ)} \ge \tfrac{\PA_l(\S^\ast_l)}{\PA_u(\S^\circ_u)} \ge \tfrac{\PA_l(\S^\ast_l)}{\PA_u(\S^\ast_u)}(\tfrac{\max\{\PA_u(S_{u1}),\ \PA_u(S_{u2}), \ p^\ast_u\}}{\max\{\gamma_u(\S_{u2}),\ p^\ast_u\}}-\epsilon)(1-\tfrac{1}{\sqrt{\e}}).\]
Meanwhile, as proved in Theorem~\ref{thrm:appro}, there are at most $L^2$ possible subsets for accuracy estimation. Therefore, each model is involved in estimation at most $L^2$ times. Therefore, the resulting failure probability is at most $\delta+L^2 \sum^L_{l=1}\delta_l$. Consequently, we have 
\begin{align*}
& \Pr\left[ \tfrac{\PA(\S^\ast)}{\PA(\S^\circ)} \ge \tfrac{\PA_l(\S^\ast_l)}{\PA_u(\S^\ast_u)}(\tfrac{\max\{\PA_u(S_{u1}),\ \PA_u(S_{u2}), \ p^\ast_u\}}{\max\{\gamma_u(\S_{u2}),\ p^\ast_u\}}-\epsilon)(1-\tfrac{1}{\sqrt{\e}})\right] \\ 
&\ge 1-(\delta+L^2\textstyle \sum^L_{l=1}\delta_l).   
\end{align*}
\sktch{which completes the proof.}
\end{proof}
\vspace{-4mm}

\begin{proof}[Proof of Lemma~\ref{lem:failboosting}]
Let $X_1,\cdots, X_{\Lambda_l}\in \{0,1\}$ be random variables such that $X_i=1$ if $p^{(i)\bot}_l \le p_l \le p^{(i)\top}_l $ holds where $[p^{(i)\bot}_l, p^{(i)\top}_l]$ is the confidence interval obtained in the $i$-th repetition; otherwise $X_i=0$. Let $X=\tfrac{1}{\Lambda_l}\sum^{\Lambda_l}_{i=1}X_i$ be the average. Therefore, we have $\mathbb{E}[X]\ge 1-\delta_l$ as $X_i = 1$ holds with at least $1-\delta_l$ probability. Let the value centered at the interval $[p^{(i)\bot}_l, p^{(i)\top}_l]$ be the corresponding estimation. After algorithm $\A$ is repeated $\Lambda_l$ times, let $[\bar{p}^\bot_l, \bar{p}^\top_l]$ be the confidence interval whose estimation is the median value among obtained estimations. The true probability $p_l$ does not belong to $[\bar{p}^\bot_l, \bar{p}^\top_l]$ if and only if at least half of the estimations fail, \ie $X_i=0$. In this regard, we have 
\begin{align*}
& Pr[p_l\notin [\bar{p}^\bot_l, \bar{p}^\top_l]] \le \Pr[X\le \tfrac{1}{2}] 
\le  \Pr[X \le \mathbb{E}[X] - \tfrac{1-2\delta_l}{2}] \\
\le &\exp(-2\Lambda_l(\tfrac{1-2\delta_l}{2})^2) 
= \exp(-\tfrac{\Lambda_l(1-2\delta_l)^2}{2}),   
\end{align*}
where the second inequality is due to the fact that $\mathbb{E}[X] - \tfrac{1-2\delta_l}{2} \ge \tfrac{1}{2}$ as $\mathbb{E}[X]\ge 1-\delta_l$, and last one\sktch{inequality} is due to Hoeffding's inequality~\cite{hoeffding1963}. \sktch{Thus Lemma~\ref{lem:failboosting} is proved.}
\end{proof}

\sktch{
\section{Additional Experiments}\label{app:exp}

\spara{Assumption verification} Our assumption states that LLMs have ``similar performance'' on queries \ravi{that are ``semantically similar''.}  To be specific, ``similar performance'' is measured by {\em success probability} (Sections~\ref{sec:pre} and ~\ref{sec:estimation}) of models in predicting the correct class labels for a given class of queries, and ``semantic similarity'' between two queries is measured by the similarity of their contextual query embeddings. To validate their inherent correlation, we conducted a carefully designed set of experiments, outlined below.

(1) \ravi{Given that the ground-truth success probabilities of the LLMs are not available,} we apply the K-means clustering on the test queries, forming  $K^\prime$ clusters, where $K^\prime$ is selected to ensure all cluster sizes are above an empirical threshold. On the $i$-th cluster, $i\in [K^\prime]$, we calculate the success probability vector $\mathcal{P}^\ast_i\in \R^L$ of all $L$ LLM candidates (Table~\ref{tbl:LLMAPI} in our revision) by following the procedure in Section~\ref{sec:estimation}. $\mathcal{P}^\ast_i$ is regarded as the ground truth success probabilities of the models  on test queries from the $i$-th cluster. Recall that the historical queries are disjoint from the test queries. Let $\mathbf{P}^\ast\in \R^{K^\prime \times L}$ denote  the ground-truth matrix where $\mathbf{P}^\ast[i,\cdot]:=\mathcal{P}^\ast_i$. (2) We estimate the ground-truth $\mathbf{P}^\ast$ by clustering historical queries into $k$ clusters. To track the estimation quality variations, we vary $k\in\{1,2,4,8,16,32,64,128\} \cup \{K^\prime\}$. For each specific $k$, we calculate the success probability vector $\mathcal{\tilde{P}}_j \in \R^L$ on the $j$-th cluster of historical queries, $j\in [k]$. (3) Subsequently, we map the $i$-th cluster of {\em test queries} to the historical query cluster with the highest semantic similarity, say the $j$-th cluster, and take vector $\mathcal{\tilde{P}}_j$ as the \textit{estimation} of $\mathcal{P}^\ast_i$ where $i\in [K^\prime], j\in [k]$. By following this procedure, for each specific $k$, we can construct a matrix $\mathbf{\tilde{P}}\in \R^{K^\prime \times L}$ where $\mathbf{\tilde{P}}[i,\cdot]:=\mathcal{\tilde{P}}_j$, as the estimation of $\mathbf{P}^\ast$. (4) The estimation error, denoted  $\epsilon$, is calculated as $\epsilon = \tfrac{1}{K^\prime}\sum^{K^\prime}_{k=1}\tfrac{1}{L}\sum^{L}_{l=1}|\mathbf{P}^\ast[k,l]-\mathbf{\tilde{P}}[k,l]|$. 

For convenience, we refer to the above mapping strategy between test clusters and historical clusters as {\em semantic similarity mapping} (SSM), \ie the strategy based on our assumption. For a more thorough evaluation, we adopt two extra mapping strategies, namely {\em random mapping} (RM) and {\em semantic dissimilarity mapping} (SDM). For the $i$-th cluster of testing queries, RM  assigns a random historical cluster, and SDM  assigns the historical cluster with the {\em lowest} semantic similarity with the test cluster. Similarly, we calculate the corresponding estimation errors as $\epsilon_{\textrm{RM}}$ and $\epsilon_{\textrm{SDM}}$, respectively. For clarity, we denote the error of semantic similarity mapping in our assumption as $\epsilon_{\textrm{our}}$ instead of as $\epsilon_{\textrm{SSM}}$. 

We select three out of five datasets for text classification and for entity matching, \ie AGNews, Hellaswag, Banking77 and Abt-Buy, Walmart-Amazon, Amazon-Google, respectively. The corresponding estimation errors are reported in Figure~\ref{fig:assumption}. In particular, the $K^\prime$ value is shown for each dataset. First, the estimation error $\epsilon_{\textrm{our}}$ is notably smaller than errors $\epsilon_{\textrm{RM}}$ and $\epsilon_{\textrm{SDM}}$ on all the $6$ tested datasets. In particular, the error $\epsilon_{\textrm{our}}$ consistently decreases as the number of clusters $k$ approaches $K^\prime$, but slightly increases when $k > K^\prime$. \ravi{The initial decrease reflects the effect of tighter clusters consisting of queries with greater pairwise similarity, while the later increase results from estimation errors caused by small-sized historical clusters.} For example, in the Walmart-Amazon dataset, $54.69\%$ of clusters are smaller than the threshold used to select $K^\prime$, which explains the sharp rise in error with larger clustering number $k$. These observations provide strong evidence to support our assumption that LLMs have similar performance on queries with semantic similarity.

\begin{figure*}[!ht]
\centering
\begin{small}
\hspace{-1mm}
\begin{tikzpicture}
\begin{customlegend}[legend columns=5,legend style={align=center,draw=none,column sep=1ex, font=\small},
        legend entries={$\epsilon_{\textrm{our}}$, $\epsilon_{\textrm{RM}}$, $\epsilon_{\textrm{SDM}}$}]
        \addlegendimage{mark=asterisk,red}
        \addlegendimage{mark=oplus,cyan}
        \addlegendimage{mark=triangle,blue}
        \end{customlegend}
\end{tikzpicture}

%\vspace{-3mm}
\subfloat[{AGNews $(K^\prime=51)$}]{
\begin{tikzpicture}
\begin{axis}[
    height=\columnwidth/2.5,
    width=\columnwidth/1.6,            
    enlarge x limits=true,
    ymin=0.05, ymax=0.3,
    xmin=2, xmax=9,         
    ylabel={{\em Error}},
    xlabel={\em Cluster numbers},
    xlabel style={yshift=0.1cm},
    ylabel style={yshift=-0.1cm},    
    xtick={2,3,4,5,6,7,8,9},
    xticklabels={2,4,8,16,32,51,64,128},
    xticklabel style={font=\scriptsize},
    %every axis y label/.style={at={(current axis.north west)},right=12mm,above=0mm},   
    %legend style={at={(0.05,0.95)},anchor=north west,font=\scriptsize},
    %legend columns=3,
    ]
    \addplot[line width=0.25mm, mark size=1pt, mark=asterisk,red] coordinates {(2, 0.131726857)(3, 0.120600227)(4, 0.099298512)(5, 0.083814328)(6, 0.080895279)(7, 0.075703567)(8, 0.077677186)(9, 0.078849779)
    }; % ThriftLLM  
    \addplot[line width=0.25mm, mark size =1pt, mark=oplus,cyan] coordinates {(2, 0.140767174)(3, 0.148011961)(4, 0.159726177)(5, 0.162934778)(6, 0.173361797)(7, 0.176261502)(8, 0.179166245)(9, 0.183658458)}; % greedy
    \addplot[line width=0.25mm, mark size =1pt, mark=triangle,blue] coordinates {(2, 0.148477324)(3, 0.150329355)(4, 0.151645)(5, 0.170983183)(6, 0.172834109)(7, 0.180040121)(8, 0.188120988)(9, 0.217295509)
}; % FrugalGPT
\end{axis}
\end{tikzpicture}
\label{subfiga:overrunling}
}\hspace{2mm}
\subfloat[{Hellaswag $(K^\prime=21)$}]{
\begin{tikzpicture}
\begin{axis}[
    height=\columnwidth/2.5,
    width=\columnwidth/1.6,            
    enlarge x limits=true,
    ymin=0.01, ymax=0.2,
    xmin=2, xmax=9,         
    ylabel={{\em Error}},
    xlabel={\em Cluster numbers},
    xlabel style={yshift=0.1cm},
    ylabel style={yshift=-0.1cm},    
    xtick={2,3,4,5,6,7,8,9},
    xticklabels={2,4,8,16,21,32,64,128},
    xticklabel style={font=\scriptsize},
    %every axis y label/.style={at={(current axis.north west)},right=12mm,above=0mm},   
    %legend style={at={(0.05,0.95)},anchor=north west,font=\scriptsize},
    %legend columns=3,
    ]
    \addplot[line width=0.25mm, mark size=1pt, mark=asterisk,red] coordinates {(2,0.0332715)(3,0.028902839)(4,0.027842401)(5,0.027531683)(6,0.026686858)(7,0.029456383)(8,0.03225985)(9,0.041982611)
    }; % ThriftLLM  
    \addplot[line width=0.25mm, mark size =1pt, mark=oplus,cyan] coordinates {(2,0.09626246)(3,0.097183351)(4,0.100793401)(5,0.094087647)(6,0.098333805)(7,0.096984779)(8,0.101616538)(9,0.09959524)}; % greedy
    \addplot[line width=0.25mm, mark size =1pt, mark=triangle,blue] coordinates {(2,0.162183911)(3,0.169256853)(4,0.169430911)(5,0.156000938)(6,0.155124308)(7,0.162411018)(8,0.157899717)(9,0.155938842)
}; % FrugalGPT
\end{axis}
\end{tikzpicture}
\label{subfiga:hellaswag}
}\hspace{2mm}
\subfloat[{Banking77 $(K^\prime=36)$}]{
\begin{tikzpicture}
\begin{axis}[
    height=\columnwidth/2.5,
    width=\columnwidth/1.6,            
    enlarge x limits=true,
    ymin=0.05, ymax=0.4,
    xmin=2, xmax=9,         
    ylabel={{\em Error}},
    xlabel={\em Cluster numbers},
    xlabel style={yshift=0.1cm},
    ylabel style={yshift=-0.1cm},    
    xtick={2,3,4,5,6,7,8,9},
    xticklabels={2,4,8,16,32,36,64,128},
    xticklabel style={font=\scriptsize},
    %every axis y label/.style={at={(current axis.north west)},right=12mm,above=0mm},   
    %legend style={at={(0.05,0.95)},anchor=north west,font=\scriptsize},
    %legend columns=3,
    ]
    \addplot[line width=0.25mm, mark size=1pt, mark=asterisk,red] coordinates {(2,0.158591432)(3,0.154926932)(4,0.150894936)(5,0.125930773)(6,0.09779538)(7,0.095469294)(8,0.101622665)(9,0.132643153)
    }; % ThriftLLM  
    \addplot[line width=0.25mm, mark size =1pt, mark=oplus,cyan] coordinates {(2,0.164449832)(3,0.168652138)(4,0.174649564)(5,0.187690808)(6,0.195265812)(7,0.206543504)(8,0.247721736)(9,0.271568938)}; % greedy
    \addplot[line width=0.25mm, mark size =1pt, mark=triangle,blue] coordinates {(2,0.170029267)(3,0.165914234)(4,0.193621642)(5,0.180959825)(6,0.214905436)(7,0.22270198)(8,0.308592618)(9,0.31249914)
}; % FrugalGPT
\end{axis}
\end{tikzpicture}
\label{subfiga:banking}
}\hspace{2mm}

\subfloat[{Abt-Buy $(K^\prime=43)$}]{
\begin{tikzpicture}
\begin{axis}[
    height=\columnwidth/2.5,
    width=\columnwidth/1.6,            
    enlarge x limits=true,
    ymin=0.05, ymax=0.3,
    xmin=2, xmax=9,         
    ylabel={{\em Error}},
    xlabel={\em Cluster numbers},
    xlabel style={yshift=0.1cm},
    ylabel style={yshift=-0.1cm},    
    xtick={2,3,4,5,6,7,8,9},
    xticklabels={2,4,8,16,32,43,64,128},
    xticklabel style={font=\scriptsize},
    %every axis y label/.style={at={(current axis.north west)},right=12mm,above=0mm},   
    %legend style={at={(0.05,0.95)},anchor=north west,font=\scriptsize},
    %legend columns=3,
    ]
    \addplot[line width=0.25mm, mark size=1pt, mark=asterisk,red] coordinates {(2, 0.065408568)(3, 0.05778351)(4, 0.055504823)(5, 0.051979378)(6, 0.051971047)(7, 0.050379492)(8, 0.051636268)(9, 0.05813885)
    }; % ThriftLLM  
    \addplot[line width=0.25mm, mark size =1pt, mark=oplus,cyan] coordinates {(2, 0.066748122)(3, 0.078029209)(4, 0.077825545)(5, 0.082830581)(6, 0.083060227)(7, 0.083570071)(8, 0.088035632)(9, 0.094367048)}; % greedy
    \addplot[line width=0.25mm, mark size =1pt, mark=triangle,blue] coordinates {(2, 0.067960652)(3, 0.105199205)(4, 0.087767594)(5, 0.14616159)(6, 0.152622906)(7, 0.148614025)(8, 0.153113565)(9, 0.264278706)
}; % FrugalGPT
\end{axis}
\end{tikzpicture}
\label{subfiga:AbtBuy}
}\hspace{2mm}
\subfloat[{Walmart-Amazon $(K^\prime=27)$}]{
\begin{tikzpicture}
\begin{axis}[
    height=\columnwidth/2.5,
    width=\columnwidth/1.6,            
    enlarge x limits=true,
    ymin=0.04, ymax=0.2,
    xmin=2, xmax=9,         
    ylabel={{\em Error}},
    xlabel={\em Cluster numbers},
    xlabel style={yshift=0.1cm},
    ylabel style={yshift=-0.1cm},    
    xtick={2,3,4,5,6,7,8,9},
    xticklabels={2,4,8,16,27,32,64,128},
    xticklabel style={font=\scriptsize},
    %every axis y label/.style={at={(current axis.north west)},right=12mm,above=0mm},   
    %legend style={at={(0.05,0.95)},anchor=north west,font=\scriptsize},
    %legend columns=3,
    ]
    \addplot[line width=0.25mm, mark size=1pt, mark=asterisk,red] coordinates {(2,0.061152077)(3,0.061013746)(4,0.053953578)(5,0.053829142)(6,0.042916912)(7,0.045942533)(8,0.048456303)(9,0.059751725)
    }; % ThriftLLM  
    \addplot[line width=0.25mm, mark size =1pt, mark=oplus,cyan] coordinates {(2,0.067236946)(3,0.066544296)(4,0.073022257)(5,0.081224718)(6,0.082048216)(7,0.086007637)(8,0.096516604)(9,0.096762002)}; % greedy
    \addplot[line width=0.25mm, mark size =1pt, mark=triangle,blue] coordinates {(2,0.073213138)(3,0.073375163)(4,0.073390689)(5,0.106201309)(6,0.085115893)(7,0.087882032)(8,0.136399218)(9,0.160981637)
}; % FrugalGPT
\end{axis}
\end{tikzpicture}
\label{subfiga:Walmart}
}\hspace{2mm}
\subfloat[{Amazon-Google $(K^\prime=32)$}]{
\begin{tikzpicture}
\begin{axis}[
    height=\columnwidth/2.5,
    width=\columnwidth/1.6,            
    enlarge x limits=true,
    ymin=0.04, ymax=0.2,
    xmin=2, xmax=8,         
    ylabel={{\em Error}},
    xlabel={\em Cluster numbers},
    xlabel style={yshift=0.1cm},
    ylabel style={yshift=-0.1cm},    
    xtick={2,3,4,5,6,7,8},
    xticklabels={2,4,8,16,32,64,128},
    xticklabel style={font=\scriptsize},
    %every axis y label/.style={at={(current axis.north west)},right=12mm,above=0mm},   
    %legend style={at={(0.05,0.95)},anchor=north west,font=\scriptsize},
    %legend columns=3,
    ]
    \addplot[line width=0.25mm, mark size=1pt, mark=asterisk,red] coordinates {(2, 0.084664519)(3, 0.071216633)(4, 0.065423125)(5, 0.054354304)(6, 0.047634473)(7, 0.054868169)(8, 0.055799216)
    }; % ThriftLLM  
    \addplot[line width=0.25mm, mark size =1pt, mark=oplus,cyan] coordinates {(2, 0.093913203)(3, 0.104419625)(4, 0.104824495)(5, 0.113893701)(6, 0.129965746)(7, 0.128445676)(8, 0.131207888)}; % greedy
    \addplot[line width=0.25mm, mark size =1pt, mark=triangle,blue] coordinates {(2, 0.102399814)(3, 0.107571986)(4, 0.105485174)(5, 0.122898526)(6, 0.14501399)(7, 0.141470404)(8, 0.16448416)
}; % FrugalGPT
\end{axis}
\end{tikzpicture}
\label{subfiga:Amazon}
}\hspace{2mm}
\end{small}%\vspace{-1mm}
\captionof{figure}{Estimation errors for testing queries across three mapping strategies.} \label{fig:assumption}%\vspace{-3mm} 
\end{figure*}

\spara{Accuracy/cost vs. \# clustering} To measure the sensitivity of our approach to the underlying number of clusters in historical queries, we evaluate \ours on the dataset Overruling with $k$ clusters by varying $k\in\{2,4,8,16,32,64\}$. The accuracy/cost results across different cluster numbers are reported in Figure~\ref{fig:cluster}. As shown, the performance of accuracy/cost is stable w.r.t. the variations of the underlying \ravi{number of} clusters. This is because the queries within Overruling have high semantic similarity, therefore insensitive to the clustering variation. To further verify our findings, we conduct the above analysis on Overruling and compute the success probability estimation errors $\epsilon_{\textrm{our}}$, which remain within the narrow range of $[0.07, 0.09]$ as the cluster number $k$ varies from $1$ to $128$.

\begin{figure*}[!t]
\centering
\begin{small}
\begin{minipage}[t]{0.32\textwidth}
\centering
\begin{small}
\hspace{-1mm}
\begin{tikzpicture}
\begin{customlegend}[legend columns=3,legend style={align=center,draw=none,column sep=1ex, font=\scriptsize},
        legend entries={$k=2$, $k=4$, $k=8$, $k=16$, $k=32$, $k=64$}]
        \addlegendimage{mark=asterisk,red}
        \addlegendimage{mark=oplus,cyan}
        \addlegendimage{mark=triangle,blue}
        \addlegendimage{mark=triangle,cyan}
        \addlegendimage{mark=diamond,orange}
        \addlegendimage{mark=triangle,orange}
        \end{customlegend}
\end{tikzpicture}

\begin{tikzpicture}
\begin{axis}[
    height=\columnwidth/1.5,
    width=\columnwidth,            
    enlarge x limits=true,
    ymin=0.9, ymax=1.0,
    xmin=0.5, xmax=100.5,         
    ylabel={{\em Accuracy}},
    xlabel={\em Cost ($\times 10^{-5}$)},
    xlabel style={yshift=0.1cm},
    ylabel style={yshift=-0.1cm},    
    xtick={1,5,10,50,100},
    xticklabels={1,5,10,50,100},
    xmode=log,
    log basis x={10},
    xticklabel style={font=\scriptsize},
    %every axis y label/.style={at={(current axis.north west)},right=12mm,above=0mm},   
    %legend style={at={(0.05,0.95)},anchor=north west,font=\scriptsize},
    %legend columns=3,
    ]
    \addplot[mark size=1pt, mark=asterisk,red] coordinates{
    (0.477, 0.951388889)
    (2.21,  0.951388889)
    (4.45,  0.953703704)
    (28.0351, 0.953703704)
    (40.9373, 0.953703704)
  }; % k = 2
    \addplot[mark size =1pt, mark=oplus,cyan] coordinates{
    (0.575, 0.953703704)
    (2.28,  0.951388889)
    (4.52,  0.951388889)
    (29.072, 0.953703704)
    (43.262, 0.953703704)
}; % k = 4
    \addplot[mark size =1pt, mark=triangle,blue] coordinates{
    (0.597, 0.956018519)
    (2.43,  0.953703704)
    (4.50,  0.953703704)
    (25.104, 0.953703704)
    (37.9711, 0.956018519)
}; % k = 8
    \addplot[mark size =1pt, mark=triangle,cyan] coordinates {
    (0.589, 0.951388889)
    (2.39,  0.951388889)
    (4.46,  0.953703704)
    (23.469, 0.953703704)
    (37.2377, 0.956018519)
}; % K = 16
    \addplot[mark size =1pt, mark=diamond,orange] coordinates{
    (0.571, 0.953703704)
    (2.39,  0.953703704)
    (4.46,  0.953703704)
    (24.7218, 0.953703704)
    (36.5483, 0.956018519) 
}; % k=32
    \addplot[mark size =1pt, mark=triangle,orange] coordinates{
    (0.542, 0.949074074)
    (2.40,  0.953703704)
    (4.55,  0.956018519)
    (25.2677, 0.951388889)
    (39.1113, 0.951388889)
}; % k=64
\end{axis}
\end{tikzpicture}
%}
\hspace{2mm}
%\vspace{-1mm}%\hspace{2mm}
\end{small}%\vspace{-1mm}
\captionof{figure}{Accuracy vs cost across various cluster numbers on Overruling.} \label{fig:cluster}%\vspace{-3mm} 
\end{minipage}
\hfill
\begin{minipage}[t]{0.32\textwidth}
\centering
\begin{small}
\hspace{-1mm}
\begin{tikzpicture}
\begin{customlegend}[legend columns=3,legend style={align=center,draw=none,column sep=1ex, font=\scriptsize},
        legend entries={$\epsilon=0.01$, $\epsilon=0.05$, $\epsilon=0.1$, $\epsilon=0.2$, $\epsilon=0.4$, $\epsilon=0.5$}]
        \addlegendimage{mark=asterisk,red}
        \addlegendimage{mark=oplus,cyan}
        \addlegendimage{mark=triangle,blue}
        \addlegendimage{mark=triangle,cyan}
        \addlegendimage{mark=diamond,orange}
        \addlegendimage{mark=triangle,orange}
        \end{customlegend}
\end{tikzpicture}

\begin{tikzpicture}
\begin{axis}[
    height=\columnwidth/1.5,
    width=\columnwidth,            
    enlarge x limits=true,
    ymin=0.9, ymax=1,
    xmin=0.5, xmax=100.5,         
    ylabel={{\em Accuracy}},
    xlabel={\em Cost ($\times 10^{-5}$)},
    xlabel style={yshift=0.1cm},
    ylabel style={yshift=-0.1cm},    
    xtick={1,5,10,50,100},
    xticklabels={1,5,10,50,100},
    xmode=log,
    log basis x={10},
    xticklabel style={font=\scriptsize},
    %every axis y label/.style={at={(current axis.north west)},right=12mm,above=0mm},   
    %legend style={at={(0.05,0.95)},anchor=north west,font=\scriptsize},
    %legend columns=3,
    ]
    \addplot[mark size=1pt, mark=asterisk,red] coordinates{
    (0.6,       0.953703704)
    (2.5,       0.956018519)
    (4.51,      0.953703704)
    (25.1762,   0.953703704)
    (38.2308,   0.956018519)
  }; %  = 2
    \addplot[mark size =1pt, mark=oplus,cyan] coordinates{
    (0.604,     0.953703704)
    (2.46,      0.956018519)
    (4.48,      0.953703704)
    (24.9135,   0.953703704)
    (37.579,    0.956018519)
}; % k = 4
    \addplot[mark size =1pt, mark=triangle,blue] coordinates{
    (0.586,     0.953703704)
    (2.41,      0.953703704)
    (4.44,      0.951388889)
    (25.1355,   0.953703704)
    (37.8886,   0.956018519)
}; % k = 8
    \addplot[mark size =1pt, mark=triangle,cyan] coordinates {
    (0.594,     0.953703704)
    (2.41,      0.951388889)
    (4.51,      0.953703704)
    (25.0059,   0.953703704)
    (37.9815,   0.956018519)
}; % K = 16
    \addplot[mark size =1pt, mark=diamond,orange] coordinates{
    (0.579,     0.953703704)
    (2.39,      0.953703704)
    (4.49,      0.953703704)
    (24.8961,   0.953703704)
    (37.9815,   0.956018519)
}; % k=32
    \addplot[mark size =1pt, mark=triangle,orange] coordinates{
    (0.576,     0.951388889)
    (2.39,      0.956018519)
    (4.47,      0.953703704)
    (24.9666,   0.953703704)
    (37.9815,   0.956018519)
}; % k=64
\end{axis}
\end{tikzpicture}
%}
\hspace{2mm}
%\vspace{-1mm}%\hspace{2mm}
\end{small}%\vspace{-1mm}
\captionof{figure}{Accuracy vs cost across various $\epsilon$ values on Overruling.} \label{fig:epsilon}%\vspace{-3mm} 
\end{minipage}
\hfill
\begin{minipage}[t]{0.32\textwidth}
\centering
\begin{small}
\hspace{-1mm}
\begin{tikzpicture}
\begin{customlegend}[legend columns=3,legend style={align=center,draw=none,column sep=1ex, font=\scriptsize},
        legend entries={$\delta=0.01$, $\delta=0.05$, $\delta=0.1$, $\delta=0.2$, $\delta=0.4$, $\delta=0.5$}]
        \addlegendimage{mark=asterisk,red}
        \addlegendimage{mark=oplus,cyan}
        \addlegendimage{mark=triangle,blue}
        \addlegendimage{mark=triangle,cyan}
        \addlegendimage{mark=diamond,orange}
        \addlegendimage{mark=triangle,orange}
        \end{customlegend}
\end{tikzpicture}

\begin{tikzpicture}
\begin{axis}[
    height=\columnwidth/1.5,
    width=\columnwidth,            
    enlarge x limits=true,
    ymin=0.9, ymax=1,
    xmin=0.5, xmax=100.5,         
    ylabel={{\em Accuracy}},
    xlabel={\em Cost ($\times 10^{-5}$)},
    xlabel style={yshift=0.1cm},
    ylabel style={yshift=-0.1cm},    
    xtick={1,5,10,50,100},
    xticklabels={1,5,10,50,100},
    xmode=log,
    log basis x={10},
    xticklabel style={font=\scriptsize},
    %every axis y label/.style={at={(current axis.north west)},right=12mm,above=0mm},   
    %legend style={at={(0.05,0.95)},anchor=north west,font=\scriptsize},
    %legend columns=3,
    ]
    \addplot[mark size=1pt, mark=asterisk,red] coordinates{
    (0.593,     0.953703704)
    (2.43,      0.953703704)
    (4.49,      0.953703704)
    (24.7709,   0.953703704)
    (37.9014,   0.956018519)
  }; %  = 2
    \addplot[mark size =1pt, mark=oplus,cyan] coordinates{
    (0.596,     0.951388889)
    (2.42,      0.953703704)
    (4.50,      0.953703704)
    (25.0401,   0.953703704)
    (37.9871,   0.956018519)    
}; % k = 4
    \addplot[mark size =1pt, mark=triangle,blue] coordinates{
    (0.598,     0.956018519)
    (2.42,      0.951388889)
    (4.50,      0.951388889)
    (25.105,    0.953703704)
    (37.8392,   0.956018519)
}; % k = 8
    \addplot[mark size =1pt, mark=triangle,cyan] coordinates {
    (0.588,     0.953703704)
    (2.42,      0.956018519)
    (4.49,      0.951388889)
    (24.9734,   0.953703704)
    (37.7904,   0.956018519)
}; % K = 16
    \addplot[mark size =1pt, mark=diamond,orange] coordinates{
    (0.596,     0.951388889)
    (2.42,      0.956018519)
    (4.51,      0.953703704)
    (24.9257,   0.953703704)
    (37.8419,   0.956018519)
}; % k=32
    \addplot[mark size =1pt, mark=triangle,orange] coordinates{
    (0.591,     0.958333333)
    (2.40,      0.951388889)
    (4.46,      0.951388889)
    (24.8209,   0.953703704)
    (38.0392,   0.956018519)
}; % k=64
\end{axis}
\end{tikzpicture}
%}
\hspace{2mm}
%\vspace{-1mm}%\hspace{2mm}
\end{small}%\vspace{-1mm}
\captionof{figure}{Accuracy vs cost across various $\delta$ values on Overruling.} \label{fig:delta}%\vspace{-3mm} 
\end{minipage}
\end{small}
\end{figure*}

\spara{Accuracy/cost vs. $\epsilon$ and $\delta$} As defined in Lemma~\ref{lem:montecarlo}, $\epsilon$ controls the relative estimation error and $\delta$ determines the failure probability where $\epsilon, \delta \in (0,1)$. To explore accuracy/cost of \ours in terms of the variation of $\epsilon$ and $\delta$, we set $\epsilon \in \{0.01,0.05,0.1,0.2,0.4,0.5\}$ and $\delta\in\{0.01,0.05,0.1,0.2,0.4,0.5\}$ and report the corresponding accuracy/cost under the same budget setting $B=\{1.0, 5.0, 10, 50, 100\}\times 10^{-5}$ (USD) in Figure~\ref{fig:epsilon} and Figure~\ref{fig:delta}. 
Empirically, $B=1.0 \times 10^{-5}$ USD corresponds to $10^5$ queries per USD, with 2–3 models typically selected per query. In contrast, $B=1.0 \times 10^{-3}$ USD implies 1000 queries per USD, where each query may involve selecting up to 9–10 models. As shown, the performance stays stable across the variation of both $\epsilon$ and $\delta$. We further explore the rationale for the empirical stability below.

\spara{Further exploration of the empirical stability} Above we observe that the accuracy/cost performance of \ours remains stable in terms of the variations of parameters $\epsilon$, and $\delta$. Upon further exploring the underlying reasons for this phenomenon, we have the following explanation. Note that the {\em surrogate function} $\gamma(\cdot)$ in Equation~\eqref{eqn:surrogatefunction} acts as the upper bound of the correctness probability $\PA(\cdot)$. Moreover, $\gamma(\cdot)$ can be calculated {\em exactly}  from input success probabilities without sampling, thus independent of $\epsilon$ and $\delta$. In Algorithm~\ref{alg:surgreedy}, we input $\PA(\cdot)$ and $\gamma(\cdot)$ as the set function to \greedy (Algorithm~\ref{alg:greedy}) and select the sets of LLMs $\S_1$ and $\S_2$ respectively. Then the set with a higher correctness probability is returned. \ravi{In the evaluation, $\S_2$ is equal to $\S_1$ for majority of the queries due to the excellent approximation of $\PA(\cdot)$ by $\gamma(\cdot)$.} Therefore, \ours remains stable when $\epsilon$ and $\delta$ vary. 

To inspect the performance gap between using $\gamma(\cdot)$ and using $\PA(\cdot)$, we inspect those queries for which the resultant sets of LLMs $\S_1$ and $\S_2$ are different from each other. In particular, we observe the largest number of such queries under the budget $B=0.00005$. The accuracy scores corresponding to using  $\PA(\cdot)$ vs $\gamma(\cdot)$  on the five datasets for text classification are reported in Figure~\ref{fig:surrogatevsour}. As shown, function $\PA(\cdot)$ outperforms surrogate set function $\gamma(\cdot)$ across all five datasets. In particular, the performance gains are up to $2.93\%$ and $5.54\%$ on datasets AGNews and Banking77, respectively. This observation shows that the surrogate function $\gamma(\cdot)$ delivers stable and efficient performance for most queries without sampling, whereas our correctness probability function $\PA(\cdot)$ provides high performance for all queries.

\begin{figure*}[!t]
\centering
\begin{small}
{\def\st{0.51}
\def\gap{0.55}
\begin{minipage}[t]{0.32\textwidth}
\centering
\begin{small}
\begin{tikzpicture}
    \begin{customlegend}[        
        legend columns=1,
        area legend,
        legend style={align=center,draw=none,column sep=1ex, font=\footnotesize},
        legend entries={{Correctness probability function $\PA(\cdot)$},{Surrogate set function $\gamma(\cdot)$}}]
        \addlegendimage{,fill=red}
        \addlegendimage{,fill=blue}
    \end{customlegend}
\end{tikzpicture}\vspace{-1mm}

\begin{tikzpicture}
\begin{axis}[
    ybar=2pt,
    height=\columnwidth/1.6,
    width=\columnwidth/1.1,            
    bar width=0.25cm,
    enlarge x limits=true,
    ymin=0.6, ymax=1.1,
    xmin=0.8*\st, xmax=\st+4.2*\gap,         
    ylabel={{\em Accuracy}},
    xtick={{\st},{\st+\gap},{\st+2*\gap},{\st+3*\gap},{\st+4*\gap}},
    xticklabels={Overruling,AGNews,SciQ,Hellaswag,Banking77}, 
    xticklabel style={font=\footnotesize, rotate=-30},
    yticklabel style={font=\small}, 
    ]
    {\addplot[fill=red] coordinates {
    (\st, 0.957086967) (\st+\gap, 0.762283518)(\st+2*\gap, 0.997099525)(\st+3*\gap, 0.924992387)(\st+4*\gap,0.766817345)};
    }{\def\st{0.48}
    \addplot[fill=blue] coordinates {(\st, 0.955191896) (\st+\gap, 0.732971509)(\st+2*\gap, 0.997009515)(\st+3*\gap, 0.90646324)(\st+4*\gap,0.711460085)};}
\end{axis}
\end{tikzpicture}
\end{small}
\captionof{figure}{Accuracy comparison between function $\PA(\cdot)$ and $\gamma(\cdot)$.} \label{fig:surrogatevsour}\vspace{-3mm}
\end{minipage}}\hspace{-1mm}%\vspace{-4mm}
{\def\st{0.515}
\def\gap{0.55}
\begin{minipage}[t]{0.32\textwidth}
\centering
\begin{small}
\begin{tikzpicture}
    \begin{customlegend}[        
        legend columns=2,
        area legend,
        legend style={align=center,draw=none,column sep=1ex, font=\footnotesize},
        legend entries={{Overruling},{Hellaswag}}]
        \addlegendimage{,fill=red}
        \addlegendimage{,fill=blue}
    \end{customlegend}
\end{tikzpicture}\vspace{-1mm}

\begin{tikzpicture}
\begin{axis}[
    ybar=2pt,
    height=\columnwidth/1.6,
    width=\columnwidth/0.9,            
    bar width=0.15cm,
    enlarge x limits=true,
    ymin=0, ymax=1.1,
    xmin=\st, xmax=\st+10.5*\gap,         
    ylabel={{\em Proportion}},
    xtick={{\st},{\st+\gap},{\st+2*\gap},{\st+3*\gap},{\st+4*\gap},{\st+5*\gap},{\st+6*\gap},{\st+7*\gap},{\st+8*\gap},{\st+9*\gap},{\st+10*\gap},{\st+11*\gap}},
    xticklabels={GPT-4o-mini, GPT-4o, Gemini-1.5 Flash, Gemini-1.5 Pro, Gemini-1.0 Pro, Phi-3-mini, Phi-3.5-mini, Phi-3-small, Phi-3-medium, Llama-3 8B, Llama-3 70B, Mixtral-8x7B}, 
    xticklabel style={font=\scriptsize, yshift=5pt, rotate=-70},
    yticklabel style={font=\small}, 
    ]
    {\addplot[fill=red] coordinates {
    (\st, 0.9837963)(\st+1*\gap, 0.0)(\st+2*\gap, 1.0)(\st+3*\gap, 0.00231481)(\st+4*\gap, 0.57638889)(\st+5*\gap, 0.99305556)(\st+6*\gap, 1.0)(\st+7*\gap, 0.95601852)(\st+8*\gap, 0.97222222)(\st+9*\gap, 1.0)(\st+10*\gap, 0.79166667)(\st+11*\gap, 0.88194444)
};
    }{\def\st{0.445}
    \addplot[fill=blue] coordinates {(\st, 0.38179393)(\st+1*\gap, 0.0)(\st+2*\gap, 1.0)(\st+3*\gap, 0.0)(\st+4*\gap, 0.00433478)(\st+5*\gap, 0.53551184)(\st+6*\gap, 0.25141714)(\st+7*\gap, 0.63287763)(\st+8*\gap, 0.87462487)(\st+9*\gap, 1.0)(\st+10*\gap, 0.15238413)(\st+11*\gap, 0.0)
};}
\end{axis}
\end{tikzpicture}
\end{small}
\captionof{figure}{LLM ensemble distribution on Overruling and Hellaswag.} \label{fig:ensemble}%\vspace{-3mm}
\end{minipage}}\hspace{-1mm}
{\def\st{0.505}
\def\gap{0.55}
\begin{minipage}[t]{0.32\textwidth}
\centering
\begin{small}
\begin{tikzpicture}
    \begin{customlegend}[        
        legend columns=2,
        area legend,
        legend style={align=center,draw=none,column sep=1ex, font=\footnotesize},
        legend entries={{Selection time},{Inference time}}]
        \addlegendimage{,fill=red}
        \addlegendimage{,fill=gray!10}
    \end{customlegend}
\end{tikzpicture}\vspace{-1mm}

\begin{tikzpicture}
\begin{axis}[
    ybar stacked,
    height=\columnwidth/1.6,
    width=\columnwidth/0.9,            
    bar width=0.25cm,
    enlarge x limits=true,
    ymin=0, ymax=20,
    %ymode=log,
    xmin=\st, xmax=\st+9*\gap,         
    ylabel={{\em Time (sec)}},
    xtick={{\st},{\st+\gap},{\st+2*\gap},{\st+3*\gap},{\st+4*\gap},{\st+5*\gap},{\st+6*\gap},{\st+7*\gap},{\st+8*\gap},{\st+9*\gap}},
    xticklabels={Overruling,AGNews,SciQ,Hellaswag,Banking77,WDC Products,Abt-Buy,Walmart-Amazon,Amazon-Google,DBLP-Scholar}, 
    xticklabel style={font=\scriptsize, rotate=-60},
    yticklabel style={font=\small}, 
    ]
\addplot[fill=gray!10,draw=black] coordinates {
    (\st, 8.125488426) (\st+\gap, 8.975430921) (\st+2*\gap, 14.26281546)(\st+3*\gap, 19.11728176) (\st+4*\gap, 2.972747421)(\st+5*\gap, 8.875251815980635)(\st+6*\gap, 5.520558506224063)(\st+7*\gap, 5.669246437552392)(\st+8*\gap, 4.570255908720453)(\st+9*\gap, 8.163452648475122)
};
\addplot[fill=red] coordinates {
    (\st, 0.148684) (\st+\gap, 0.286816) (\st+2*\gap, 0.134983)
    (\st+3*\gap, 0.109152) (\st+4*\gap, 0.330097)(\st+5*\gap, 0.120652)(\st+6*\gap, 0.1188)(\st+7*\gap, 0.0914221)(\st+8*\gap, 0.12447)(\st+9*\gap, 0.118133)
};
\end{axis}
\end{tikzpicture}
\end{small}
\captionof{figure}{Running time vs. inference time.} \label{fig:runningtime}\vspace{-3mm}
\end{minipage}}
\end{small}
\end{figure*}

\begin{figure*}
\centering
\begin{small}
\hspace{-1mm}
\begin{tikzpicture}
\begin{customlegend}[legend columns=3,legend style={align=center,draw=none,column sep=1ex, font=\scriptsize},
        legend entries={\ours, Weighted aggregation, Majority vote}]
        \addlegendimage{mark=asterisk,red}
        \addlegendimage{mark=oplus,cyan}
        \addlegendimage{mark=triangle,blue}
        \end{customlegend}
\end{tikzpicture}

\subfloat[{Overruling}]{
\begin{tikzpicture}
\begin{axis}[
    height=\columnwidth/2.5,
    width=\columnwidth/1.6,            
    enlarge x limits=true,
    ymin=0.9, ymax=1.0,
    xmin=0.5, xmax=100.5,         
    ylabel={{\em Accuracy}},
    xlabel={\em Cost ($\times 10^{-5}$)},
    xlabel style={yshift=0.1cm},
    ylabel style={yshift=-0.1cm},    
    xtick={1,5,10,50,100},
    xticklabels={1,5,10,50,100},
    xmode=log,
    log basis x={10},
    xticklabel style={font=\scriptsize},
    %every axis y label/.style={at={(current axis.north west)},right=12mm,above=0mm},   
    %legend style={at={(0.05,0.95)},anchor=north west,font=\scriptsize},
    %legend columns=3,
    ]
    \addplot[only marks, mark size=1pt, mark=asterisk,red] coordinates{
    (0.5868229166666662,0.9513888888888888)
    (2.4283460648148167,0.9537037037037037)
    (4.4832361111111025,0.9537037037037037)
    (24.83767592592593,0.9537037037037037)
    (38.00079050925928,0.9560185185185185)
  }; % our
    \addplot[only marks, mark size =1pt, mark=oplus,cyan] coordinates{
    (0.6166886574074072, 0.9537185185185185)
    (2.48596875, 0.9490740740740741)
    (4.577695601851842, 0.9537037037037037)
    (25.04022569444445, 0.9537037037037037)
    (37.87816666666668, 0.9560185185185185)
}; % weighted
    \addplot[only marks, mark size =1pt, mark=triangle,blue] coordinates{
    (0.6800995370370365, 0.9490740740740741)
    (2.782480324074077, 0.9537037037037037)
    (5.14343402777778, 0.9513888888888888)
    (25.721891203703734, 0.9537037037037037)
    (39.114666666666716, 0.9513888888888888)
}; % majority
\end{axis}
\end{tikzpicture}
}\hspace{2mm}
\subfloat[{AGNews}]{
\begin{tikzpicture}
\begin{axis}[
    height=\columnwidth/2.5,
    width=\columnwidth/1.6,            
    enlarge x limits=true,
    ymin=0.8, ymax=0.9,
    xmin=0.5, xmax=100.5,         
    ylabel={{\em Accuracy}},
    xlabel={\em Cost ($\times 10^{-5}$)},
    xlabel style={yshift=0.1cm},
    ylabel style={yshift=-0.1cm},    
    xtick={1,5,10,50,100},
    xticklabels={1,5,10,50,100},
    xmode=log,
    log basis x={10},
    xticklabel style={font=\scriptsize},
    %every axis y label/.style={at={(current axis.north west)},right=12mm,above=0mm},   
    %legend style={at={(0.05,0.95)},anchor=north west,font=\scriptsize},
    %legend columns=3,
    ]
    \addplot[only marks, mark size=1pt, mark=asterisk,red] coordinates{
    (0.5884641447368416,0.8447368421052631)
    (2.649204605263158,0.8473684210526315)
    (5.244262828947375,0.8486842105263158)
    (28.65560855263157,0.8625)
    (49.76167401315792,0.8644736842105263)
  }; % our
    \addplot[only marks, mark size =1pt, mark=oplus,cyan] coordinates{
    (0.6568101973684204, 0.8460526315789474)
    (2.835899342105263, 0.8414473684210526)
    (5.5792151315789405, 0.8414473684210526)
    (29.47404111842102, 0.85)
    (50.8825266447369, 0.8513157894736842)
}; % weighted
    \addplot[only marks, mark size =1pt, mark=triangle,blue] coordinates{
    (0.7149861842105269, 0.8427631578947369)
    (3.13745657894737, 0.8289473684210527)
    (6.026744736842109, 0.8296052631578947)
    (30.20468059210524, 0.8381578947368421)
    (52.61231151315787, 0.8388157894736842)
}; % majority
\end{axis}
\end{tikzpicture}
}\hspace{2mm}
\subfloat[{Hellaswag}]{
\begin{tikzpicture}
\begin{axis}[
    height=\columnwidth/2.5,
    width=\columnwidth/1.6,            
    enlarge x limits=true,
    ymin=0.82, ymax=0.92,
    xmin=0.5, xmax=100.5,         
    ylabel={{\em Accuracy}},
    xlabel={\em Cost ($\times 10^{-5}$)},
    xlabel style={yshift=0.1cm},
    ylabel style={yshift=-0.1cm},    
    xtick={1,5,10,50,100},
    xticklabels={1,5,10,50,100},
    xmode=log,
    log basis x={10},
    xticklabel style={font=\scriptsize},
    %every axis y label/.style={at={(current axis.north west)},right=12mm,above=0mm},   
    %legend style={at={(0.05,0.95)},anchor=north west,font=\scriptsize},
    %legend columns=3,
    ]
    \addplot[only marks, mark size=1pt, mark=asterisk,red] coordinates{
    (2.823668222740919,0.8579526508836278)
    (3.798212404134714,0.8829609869956653)
    (6.160533344448156,0.8842947649216405)
    (21.502863121040377,0.8942980993664554)
    (38.62696048682907,0.8986328776258753)
  }; % our
    \addplot[only marks, mark size =1pt, mark=oplus,cyan] coordinates{
    (2.8227, 0.85795)
    (3.8003, 0.88196)
    (6.3790, 0.88296)
    (25.0742, 0.88763)
    (41.7526, 0.89330)
}; % weighted
    \addplot[only marks, mark size =1pt, mark=triangle,blue] coordinates{
    (2.837092030676898, 0.8569523174391463)
    (3.993242247415812, 0.8812937645881961)
    (7.46456818939647, 0.8759586528842948)
    (25.49170423474532, 0.8819606535511837)
    (42.66122040680227, 0.8882960986995665)
}; % majority
\end{axis}
\end{tikzpicture}
}\hspace{2mm}
%\vspace{-1mm}%\hspace{2mm}
\end{small}%\vspace{-1mm}
\captionof{figure}{Accuracy vs cost with different aggregation function on Overruling.} \label{fig:aggregation}%\vspace{-3mm} 
\end{figure*}

\spara{Accuracy/cost vs. aggregation variants} Consider a subset of selected models and their responses for a given query. The response aggregation proposed in Section~\ref{sec:aggregation} is based on the success probabilities of the selected models, which are used to calculate the likelihood of each class label. The label with the highest likelihood is returned as the prediction. To evaluate the efficacy of such aggregation design, we develop two variants, namely {\em weighted aggregation} and {\em majority vote}. In particular, weighted aggregation takes the success probability of each model as the weight of the predicted label. The label with the largest weight summation is taken as the final prediction. In contrast, the majority vote \ravi{simply counts the number of times each  label is predicted and returns the one with the largest vote.} The corresponding accuracy/cost results on datasets Overruling, AGNews, and Hellaswag are reported in Figure~\ref{fig:aggregation}. As shown, \ours exhibits a consistent performance with the other two aggregation variants on the dataset Overruling, while it clearly outperforms those aggregation variants on the other two datasets AGNews and Hellaswag. This observation exhibits the effectiveness of our aggregation mechanism and also provides a strong empirical justification for the aggregation design in \ours.

\spara{Diversity of LLM ensemble} To examine the model distribution of the selected LLM ensembles across datasets, we track a vector $\x \in \R^{12}$, where $\x[i] \in [0,1]$ indicates the proportion of test queries for which the $i$-th model is chosen, as indexed according to Table~\ref{tbl:LLMAPI} \ravi{(e.g., 1 = GPT-4o-mini and 12 = Mixtral-8x7B)}. We apply this analysis to datasets Overruling and Hellaswag and obtain the corresponding ensemble distributions depicted in Figure~\ref{fig:ensemble}. As shown, the selected ensemble distribution on Overruling is more balanced among the $12$ LLM candidates than that on Hellaswag. Specifically, model selection on Hellaswag demonstrates a clear preference for Gemini-1.5 Flash,  Llama-3 8B, Phi-3-medium, and Phi-3-small. 
To explain, Overruling has relatively simple queries, resulting in similar performance across models. As a result, \ours distributes selection more evenly among models, excluding only the two most expensive ones (GPT-4o and Gemini-1.5 Pro). In contrast, Hellaswag involves more complex queries, leading to greater performance variation and a stronger preference for models that perform well on this dataset.

%Overruling [0.9837963  0.         1.         0.00231481 0.57638889 0.99305556 1.         0.95601852 0.97222222 1.         0.79166667 0.88194444]
%Hellaswag [0.38179393 0.         1.         0.         0.00433478 0.53551184 0.25141714 0.63287763 0.87462487 1.         0.15238413 0.]

\spara{Running time of LLM selection} As required, we record the running time of the LLM selection component in \ravi{the \ours algorithm} on all five datasets for text classification and five datasets for entity matching by using one budget setting $B=0.001$ (USD). 
Meanwhile, we also report the model API execution time when we apply our selected LLM sets to test queries in Figure~\ref{fig:runningtime}. Note that the API execution time is determined by the configuration of each service provider. As displayed, the model selection time is negligible compared with the inference time. Specifically, for text classification, the selection time of \ours accounts for only $0.5\%\sim 3.2\%$ of the total inference time on Overruling, AGNews, SciQ, and Hellaswag, and $11.1\%$ on Banking77. For entity matching, it takes approximately $1.3\%\sim 2.8\%$ of the total inference time across the five datasets. These results suggest that the selection time is negligible relative to inference and is not a bottleneck for overall efficiency.

}

\clearpage
\bibliographystyle{ACMRefer}
\balance
\bibliography{ref}

\end{sloppy}
\end{document}